\documentclass[fleqn,3p,onecolumn,12pt,nofootinbib,showkeys,showpacs]{revtex4-1}

\newcommand{\versionnumber}{v2.50}

\renewcommand{\today}{KA--TP--35--2011 (\versionnumber)}

\usepackage{graphicx}
\usepackage{bm}
\usepackage{mathrsfs}
\usepackage[latin1]{inputenc}
\usepackage{stmaryrd}
\usepackage{array}
\usepackage{psfrag}
\usepackage{dsfont}
\usepackage{epsfig}
\usepackage{titletoc}
\usepackage{float}   
\usepackage{wrapfig} 
\usepackage{amsmath}
\usepackage[latin1]{inputenc}
\usepackage{amsmath}\usepackage{amssymb,stmaryrd}
\usepackage{mathrsfs}
\usepackage{array}
\usepackage{graphicx}
\usepackage{psfrag}
\usepackage{dsfont}
\usepackage{epsfig}
\usepackage{cancel}
\usepackage{upgreek}
\usepackage{multirow}
\usepackage{color}
\usepackage{hyperref}

\def\be {\begin{equation}}
\def\ee {\end{equation}}
\def\ba {\begin{eqnarray}}
\def\ea {\end{eqnarray}}

\begin{document}


\title{Analysis of the consistency of parity-odd \\
nonbirefringent modified Maxwell theory}

\author{M. Schreck} \email{marco.schreck@kit.edu}
\affiliation{Institute for Theoretical Physics, Karlsruhe Institute of Technology (KIT),\\
         76128 Karlsruhe, Germany}

\begin{abstract}
There exist two deformations of standard electrodynamics that describe
Lorentz symmetry violation in the photon sector: \textit{CPT}-odd Maxwell--Chern--Simons
theory and \textit{CPT}-even modified Maxwell theory. In this article, we focus on the
parity-odd nonbirefringent sector of modified Maxwell theory. It is coupled to a
standard Dirac theory of massive spin-1/2 fermions resulting in a modified
quantum electrodynamics (QED). This theory is discussed with respect to properties
such as microcausality and unitarity, where it turns out that these hold.

Furthermore, \textit{a priori}, the limit of the theory for vanishing Lorentz-violating
parameters seems to be discontinuous. The modified photon polarization vectors
are interweaved with preferred spacetime directions defined by the theory and
one vector even has a longitudinal part. That structure remains in the limit
mentioned. Since it is not clear, whether or not this behavior is a gauge artifact,
the cross section for a physical process --- modified Compton scattering --- is
calculated numerically. Despite the numerical instabilities occurring for scattering
of unpolarized electrons off polarized photons in the second physical polarization
state, it is shown that for Lorentz-violating parameters much smaller than one,
the modified cross sections approach the standard QED results. Analytical
investigations strengthen the numerical computations.

Hence, the theory proves to be consistent, at least with regard to the
investigations performed. This leads to the interesting outcome of the modification
being a well-defined parity-odd extension of QED.
\end{abstract}


\maketitle

\newpage
\setcounter{equation}{0}
\renewcommand{\theequation}{\arabic{section}.\arabic{equation}}
\section{Introduction}
\label{sec:Introduction}

Modern quantum field theories are based on fundamental symmetries. This holds for quantum
electrodynamics (QED)
as well as for the standard model of elementary particle physics. Whenever physicists talk
about symmetries they usually think of gauge invariance or the discrete symmetries charge
conjugation \textit{C}, parity \textit{P}, and time reversal \textit{T}. However, there is one symmetry that often
takes a back seat: Lorentz invariance. This is not surprising, since until now there had
been no convincing experimental evidence for a violation of Lorentz invariance.\footnote{
At the end of September 2011 this seemed to change with the publication of the result by the
OPERA collaboration, which claimed to have discovered Lorentz violation in the neutrino
sector \cite{OPERA:2011zb}. A large number of theoretical models emerged trying to explain
the observed anomaly, for example by Fermi point splitting \cite{Klinkhamer:2011mf},
spontaneous symmetry breaking caused by the existence of a fermionic condensate
\cite{Klinkhamer:2011iz}, or a multiple Lorentz group structure \cite{Schreck:2011ni}.
However, the physics community remained sceptical and articles were published trying to
explain the result by an error source that had not been taken into account
\cite{Contaldi:2011,Besida:2011fi,vanElburg:2011ze}. Unfortunately, at the 25th International
Conference on Neutrino Physics and Astrophysics OPERA announced that their new measurement
yields a deviation of the neutrino velocity from the speed of light, which is consistent with
zero. Now again all laws of nature seem to obey Lorentz invariance.}

However, a violation of other symmetries is part of the everyday life of any high-energy
physicist. For example, violations of \textit{P} and \textit{CP} were measured long ago \cite{Wu:1957,Christenson:1964fg}
and a broken electroweak gauge symmetry with massive $\mathrm{W^{\pm}}$, and $\mathrm{Z^0}$ bosons is
an experimental fact. Why then should Lorentz symmetry and its violation not be of interest?

There exist good theoretical arguments for Lorentz invariance being a symmetry that is restored
at low energies \cite{ChadhaNielsen1983}. At the Planck length the topology of spacetime may be
dynamical, which could lead to it having a foamy structure. The existence of such a spacetime foam
\cite{Wheeler:1957mu,Hawking:1979zw} may define a preferred reference frame --- as is the case
for water in a glass --- and thus violate Lorentz invariance. Since a fundamental
quantum theory of spacetime is still not known, we have to rely on well-established theories such as
the standard model or special relativity for a description of Lorentz violation. By introducing
new parameters that deform these theories it is possible to parameterize Lorentz violation on
the basis of standard physics. One approach is to modify dispersion relations of particles. However,
such a procedure is very \textit{ad hoc} and it is not evident where the modification comes from. Therefore,
a more elementary possibility is to parameterize modifications on the level of Lagrange densities.
A collection of all Lorentz-violating deformations of the standard model that are gauge
invariant
is known as the Lorentz-violating extension of the standard model \cite{ColladayKostelecky1998}.
The minimal version of this extension relies on power-counting renormalizable terms, whereas the
nonminimal version also includes operators of mass dimension $d>4$ (see e.g. the analyses
performed in \cite{Kostelecky:2009zp,Kostelecky:2011gq,Cambiaso:2012vb}).

The theoretical consistency of the standard model itself has been verified by investigations based
on Lorentz-invariant quantum field theory that were performed over decades (see, for example,
Ref.~\cite{JordanPauli1928}). However, it is not entirely clear if a Lorentz-violating theory is
consistent. Some results on certain sectors of the standard model extension already
exist \cite{KosteleckyLehnert2000,AdamKlinkhamer2001,Liberati:2001sd,Mavromatos:2009xg,Casana-etal2009,
Casana-etal2010,Klinkhamer:2010zs,Klinkhamer:2011ez}, but there still remains a lot what we can learn about
Lorentz-violating quantum field theories.
Because of this it is very important to check Lorentz-violating deformations with
respect to fundamental properties such as microcausality and unitarity. Furthermore, it is of
significance whether the modified theory approaches the standard theory for arbitrarily small
deformations. The purpose of this paper is to investigate these questions.

Especially in the case where Lorentz violation resides in the photon sector, it can lead to a variety
of new effects, for example a birefringent vacuum \cite{ColladayKostelecky1998}, new particle
decays \cite{Beall:1970rw,Coleman:1997xq}, and ``aetherlike'' deviations from special relativity,
which are modulated with the rotation of the Earth around the Sun (e.g. Refs.~\cite{Phillips:2000dr,Bear:2000cd}).
From an experimental point of view, photons produce clean signals making the photon sector very important,
in bounding Lorentz-violating parameters.

There exist two gauge-invariant and power-counting renormalizable deformations of the photon
sector: Maxwell--Chern--Simons theory (MCS-theory) \cite{Carroll-etal1990} and modified
Maxwell theory \cite{ColladayKostelecky1998,KosteleckyMewes2002}. Each Lagrangian contains
additional terms besides the Maxwell term of standard electrodynamics. The consistency
of the isotropic and one anisotropic sector of modified Maxwell theory was already shown
in \cite{Klinkhamer:2010zs}. In this article a special sector, that violates parity and
is supposed to show no birefringence, will be investigated.

The paper is organized as follows. In Sec. 2 modified Maxwell theory is presented and
restricted to the parity-odd nonbirefringent case. Additionally, it is coupled to a
standard Dirac theory of massive spin-1/2 fermions, which leads to a theory of modified
QED. In Secs. 3 and 4, we review the nonstandard photon dispersion
relations and the gauge propagator, which are determined from the field equations
\cite{Casana-etal2009,Casana-etal2010}. That completes the current status of research
concerning this special sector of modified Maxwell theory. The successive parts of the
article deal with the main issue, beginning with the deformed polarization vectors, which can
also be obtained from the field equations. After setting up the building blocks we are ready
to discuss unitarity in Sec. 6 and microcausality in Sec. 7. The subsequent two sections are
devoted to the polarization vectors themselves. Since their form is rather uncommon ---
even when considering Lorentz-violating theories --- we make comparisons with MCS-theory and
other sectors of modified Maxwell theory. It will become evident that the polarization
vectors have a property that distinguishes them from the polarization vectors of standard
electrodynamics, even in the limit of vanishing Lorentz violation. To test, whether or not some
residue of the deformation remains in this limit, in Sec. 9 we compute the cross section of the
simplest tree-level process involving external modified photons that is also allowed by standard
QED: Compton scattering. We conclude in the last section. Readers may skip Secs. 4 -- 8 on first
reading.

\section{Modified Maxwell theory}
\label{sec:ModMaxtheory}

\subsection{Action and nonbirefringent Ansatz}
\label{sec:Action-nonbirefringent-Ansatz}

In this article, we focus on modified Maxwell theory~\cite{ChadhaNielsen1983,ColladayKostelecky1998,KosteleckyMewes2002}.
This particular Lorentz-violating theory is characterized by the action
\begin{subequations}\label{eq:action-modified-maxwell-theory}
\begin{eqnarray}
S_{\mathrm{modMax}}&=&\int_{\mathbb{R}^4}\mathrm{d}^4x\,
\mathcal{L}_\text{modMax}(x)\,,\\[2mm]
\mathcal{L}_\text{modMax}(x)&=& -\frac{1}{4}\,
\eta^{\mu\rho}\,\eta^{\nu\sigma}\,F_{\mu\nu}(x)F_{\rho\sigma}(x)
-\frac{1}{4}\,
\kappa^{\mu\nu\varrho\sigma}\,F_{\mu\nu}(x)F_{\varrho\sigma}(x)\,,
\label{eq:L-modified-maxwell-theory}
\end{eqnarray}
\end{subequations}
which involves the field strength tensor $F_{\mu\nu}(x)\equiv\partial_{\mu}A_{\nu}(x)-\partial_{\nu}A_{\mu}(x)$
of the $U(1)$ gauge field $A_{\mu}(x)$. The fields are defined on
Minkowski spacetime with global Cartesian coordinates
$(x^\mu)$ $=$ $(x^0,\boldsymbol{x})$ $=$ $(c\,t,x^1,x^2,x^3)$
and metric $g_{\mu\nu}(x)$ $=$ $\eta_{\mu\nu}$ $\equiv$
$\text{diag}\,(1,\, -1,\, -1,\, -1)\,$. The first term in
Eq. \eqref{eq:L-modified-maxwell-theory} represents the standard
Maxwell term and the second corresponds to a modification of the
standard theory of photons. The fixed background field
$\kappa^{\mu\nu\varrho\sigma}$ selects preferred
directions in spacetime and, therefore, breaks Lorentz
invariance.

The second term in Eq. \eqref{eq:L-modified-maxwell-theory} is
expected to have the same symmetries as the first. These correspond
to the symmetries of the Riemann curvature tensor,
which reduces the number of independent parameters to 20. Furthermore,
a vanishing double trace, $\kappa^{\mu\nu}_{\phantom{\mu\nu}\mu\nu}=0$,
is imposed. A nonvanishing $\kappa^{\mu\nu}_{\phantom{\mu\nu}\mu\nu}$
can be absorbed by a field redefinition \cite{ColladayKostelecky1998}
and does not contribute to physical observables. This additional
condition leads to a remaining number of 19 independent parameters.

Modified Maxwell theory has two distinct parameter sectors that can be
distinguished from each other by the property of birefringence.
The first consists of 10 parameters and leads to birefringent photon
modes at leading-order Lorentz violation. The second is made up of
9 parameters and shows no birefringence, at least to first order with
respect to the parameters. Since the 10 birefringent parameters are
bounded by experiment at the $10^{-32}$ level \cite{Kostelecky:2001mb},
we will restrict our considerations to the nonbirefringent sector,
which can be parameterized by the following
\textit{Ansatz}~\cite{BaileyKostelecky2004}:
\begin{equation}\label{eq:nonbirefringent-Ansatz}
\kappa^{\mu\nu\varrho\sigma}=
\frac{1}{2}\,\Big(
 \eta^{\mu\varrho}\,\widetilde{\kappa}^{\nu\sigma}
-\eta^{\mu\sigma}\,\widetilde{\kappa}^{\nu\varrho}
-\eta^{\nu\varrho}\,\widetilde{\kappa}^{\mu\sigma}
+\eta^{\nu\sigma}\,\widetilde{\kappa}^{\mu\varrho}\Big)\,,
\end{equation}
with a constant symmetric and traceless $4\times 4$ matrix
$\widetilde{\kappa}^{\mu\nu}$. Here and in the following, natural units
are used with $\hbar=c=1$, where $c$ corresponds to the maximal
attainable velocity of the standard Dirac particles, whose action
will be defined in Sec.~\ref{sec:Coupling-to-standard-Dirac-particles}.

There exists a premetric formulation of classical electrodynamics,
that is solely based on the concept of a manifold and does not need a metric.
In this context a tensor density $F$ (electromagnetic field strength) and
pseudotensor densities $\mathcal{H}$, $\mathcal{J}$  (electromagnetic excitation
and electric current) are introduced. Since the resulting field equations for
these quantities are underdetermined, an additional relation between $F$ and
$\mathcal{H}$ has to be imposed, which is governed by the so-called constitutive
four-tensor $\chi$. Modified Maxwell theory emerges as one special case of this
description, namely as the principal part of the constitutive tensor previously
mentioned~\cite{Hehl:2003,Itin:2009aa}. In Eq.~(D.1.80) of the book \cite{Hehl:2003}
the nonbirefringent \textit{Ansatz} of Eq.~\eqref{eq:nonbirefringent-Ansatz} can be
found, as well. Section D.1.6 gives a motivation for it as the simplest --- but not
the most general --- decomposition of the principal part of $\chi$.

Furthermore, note that a special sector of \textit{CPT}-even modified Maxwell theory arises
as a contribution of the one-loop effective action of a \textit{CPT}-odd deformation
involving a spinor field and the photon field \cite{Gomes:2009ch}.

\subsection{Restriction to the parity-odd anisotropic case}
\label{sec:Parity-odd-anisotropic}

The anisotropic case considered concerns the parity-odd
sector of modified Maxwell theory \eqref{eq:action-modified-maxwell-theory}
with the \textit{Ansatz} from Eq.~\eqref{eq:nonbirefringent-Ansatz}.
This case is characterized by one purely timelike
normalized four-vector $\xi^{\mu}$ and one purely spacelike
four-vector $\zeta^{\mu}$ containing three real parameters
$\widetilde{\kappa}^{01}$, $\widetilde{\kappa}^{02}$, and
$\widetilde{\kappa}^{03}$\,:
\begin{subequations}\label{eq:definition-parity-odd-case}
\begin{eqnarray}
\widetilde{\kappa}^{\mu\nu}&=&
\frac{1}{2}\,(\xi^{\mu}\zeta^{\nu}+\zeta^{\mu}\xi^{\nu})
-\frac{1}{4} \, \xi^{\lambda}\zeta_{\lambda}\,\eta^{\mu\nu}
\label{eq:definition-parity-odd-case-widetildekappamunu}
\,,\\[2mm]
(\xi^{\mu})&=&(1,\,0,\,0,\,0)\,,\quad
(\zeta^{\mu})
\equiv  (0,\,2\,\boldsymbol{\zeta})
=
(0,\,2\,\widetilde{\kappa}^{01},\,
   2\,\widetilde{\kappa}^{02},\,
   2\,\widetilde{\kappa}^{03})\,,
\label{eq:definition-parity-odd-case-four-vectors}
\\[2mm]
(\widetilde{\kappa}^{\mu\nu})&=&\begin{pmatrix}
0 & \widetilde{\kappa}^{01} & \widetilde{\kappa}^{02} & \widetilde{\kappa}^{03} \\
\widetilde{\kappa}^{01} & 0 & 0 & 0 \\
\widetilde{\kappa}^{02} & 0 & 0 & 0 \\
\widetilde{\kappa}^{03} & 0 & 0 & 0 \\
\end{pmatrix}\,,
\end{eqnarray}
\end{subequations}
where \eqref{eq:definition-parity-odd-case-widetildekappamunu}
is the most general \textit{Ansatz} for a symmetric and traceless
tensor constructed from two four-vectors.
The second term on the right-hand side
of \eqref{eq:definition-parity-odd-case-widetildekappamunu}
vanishes for the special choice \eqref{eq:definition-parity-odd-case-four-vectors}.

With the replacement rules given in \cite{KlinkhamerRisse2008b},
we can  express our parameters in terms of the Standard Model Extension
(SME) parameters~\cite{KosteleckyMewes2002,BaileyKostelecky2004}:
\begin{subequations}\label{eq:parity-odd-case-SME}
\begin{eqnarray}
\widetilde{\kappa}^{01}&=&-(\widetilde{\kappa}_{\mathrm{o}+})^{(23)}\,,\\
\widetilde{\kappa}^{02}&=&-(\widetilde{\kappa}_{\mathrm{o}+})^{(31)}\,,\\
\widetilde{\kappa}^{03}&=&-(\widetilde{\kappa}_{\mathrm{o}+})^{(12)}\,.
\end{eqnarray}
\end{subequations}
Hence, the case considered here includes only parity-violating coefficients.

This parity-odd case may be of relevance, since it might reflect
the parity-odd low-energy effective photon sector of a quantum theory
of spacetime.
Besides five parameters of the birefringent sector of modified Maxwell
theory, whose coefficients are already strongly bounded, there is only one
alternative parity-odd Lorentz-violating theory for the photon sector, which
is gauge-invariant and power-counting renormalizable: MCS theory
\cite{Carroll-etal1990}. However, the MCS parameters are bounded to lie
below $10^{-42}\,\mathrm{GeV}$
by CMB polarization measurements \cite{Kostelecky:2008ts}.

Since the bounds are not as strong for the parity-odd case of nonbirefringent
modified Maxwell theory defined by Eq. \eqref{eq:definition-parity-odd-case},
a physical understanding of this case is of importance.

\subsection{Coupling to matter: Parity-odd modified QED}
\label{sec:Coupling-to-standard-Dirac-particles}

Modified photons are coupled to matter by the minimal coupling procedure
to standard (Lorentz-invariant) spin-$\textstyle{\frac{1}{2}}$ Dirac
particles with electric charge $e$ and mass $M$. This results in a
parity-odd deformation of QED~\cite{Heitler1954,JauchRohrlich1976,Veltman1994},
which is given by the action
\begin{equation}\label{eq:action-isotropic-modQED} \hspace*{0mm}
S_\text{modQED}^\text{parity-odd}\big[\widetilde{\kappa}^{0m},e,M\big] =
S_\text{modMax}^\text{parity-odd}\big[\widetilde{\kappa}^{0m}\big] +
S^\text{}_\text{Dirac}\big[e,M\big] \,,
\end{equation}
for $m=1$, 2, 3 and with the modified-Maxwell term
\eqref{eq:action-modified-maxwell-theory}--\eqref{eq:definition-parity-odd-case}
for the gauge field $A_\mu(x)$
and the standard Dirac term for the spinor field $\psi(x)$,
\begin{equation}\label{eq:standDirac-action}
S^\text{ }_\text{Dirac}\big[e,M\big] =
\int_{\mathbb{R}^4} \mathrm{d}^4 x \; \overline\psi(x) \Big[
\gamma^\mu \big(\mathrm{i}\,\partial_\mu -e A_\mu(x) \big) -M\Big] \psi(x)\,.
\end{equation}
Equation \eqref{eq:standDirac-action} is to be understood with standard
Dirac matrices $\gamma^\mu$ corresponding to the Minkowski metric
$\eta^{\mu\nu}$.

\section{Dispersion relations}
\label{sec:Dispersion-relations-classical-causality}
\setcounter{equation}{0}

The field equations \cite{ColladayKostelecky1998,KosteleckyMewes2002,BaileyKostelecky2004}
of modified Maxwell theory in momentum space,
\begin{equation}
\label{eq:field-equations-modified-maxwell-theory}
M^{\mu\nu}A_{\nu}=0 \,,\quad
M^{\mu\nu}\equiv
k^{\lambda}k_{\lambda}\,\eta^{\mu\nu}-k^{\mu}k^{\nu}
-2\,\kappa^{\mu\rho\sigma\nu}\,k_{\rho}k_{\sigma}\,,
\end{equation}
lead to the following dispersion relations~\cite{Casana-etal2009}
for the two physical degrees of freedom of
electromagnetic waves (labeled $\lambda=1,2$):
\begin{subequations}\label{eq:dispersion-relation-p-odd-1-2}
\begin{eqnarray}
\omega_1(\mathbf{k})&=&
 \widetilde{\kappa}^{01}\,k_1
+\widetilde{\kappa}^{02}\,k_2
+\widetilde{\kappa}^{03}\,k_3
+\sqrt{|\mathbf{k}|^2+(\widetilde{\kappa}^{01}\,k_1
+\widetilde{\kappa}^{02}\,k_2+\widetilde{\kappa}^{03}\,k_3)^2}\,,
\label{eq:dispersion-relation-p-odd-1}\\[2mm]
\omega_2(\mathbf{k})&=&
 \widetilde{\kappa}^{01}k_1\,
+\widetilde{\kappa}^{02}k_2\,
+\widetilde{\kappa}^{03}k_3\,
+\sqrt{1+(\widetilde{\kappa}^{01})^2
+(\widetilde{\kappa}^{02})^2+(\widetilde{\kappa}^{03})^2}\;|\mathbf{k}|
\label{eq:dispersion-relation-p-odd-2}\,,
\end{eqnarray}
\end{subequations}
for wave vector $\mathbf{k}=(k_1,\,k_2,\,k_3)$ and with
the terms linear in the components $k_m$ explicitly showing
the parity violation. To first order in $\widetilde{\kappa}^{0m}$,
the dispersion relations are equal for both modes, but they
differ at higher order.\footnote{It is evident that the so-called
nonbirefringent \textit{Ansatz} \eqref{eq:nonbirefringent-Ansatz}
is only nonbirefringent to first order in $\widetilde{\kappa}^{\mu\nu}$.
Nevertheless we will still use the term ``nonbirefringent'' in order to
distinguish from the nine-dimensional parameter sector of modified Maxwell
theory, which shows no birefringence at least to first-order Lorentz violation,
from the remaining ten coefficients. In the latter parameter region
birefringent modes emerge already at first order with respect to the
Lorentz-violating parameters \cite{KosteleckyMewes2002}.}
With the modified Coulomb and Amp\`{e}re law it can be shown that the
dispersion relations \eqref{eq:dispersion-relation-p-odd-1-2} indeed
belong to physical photon modes. The procedure given in
\cite{ColladayKostelecky1998} eliminates dispersion relations of
unphysical, i.e. scalar and longitudinal, modes from the field
equations. The two are given by
\begin{equation}
\label{eq:dispersion-relation-unphysical}
\omega_0(\mathbf{k})=\omega_3(\mathbf{k})=|\mathbf{k}|\,,
\end{equation}
where the index ``0'' refers to the scalar and the index ``3''
to the longitudinal degree of freedom of the photon field.

The dispersion relations \eqref{eq:dispersion-relation-p-odd-1-2}
can be cast in a more compact form by defining
components of the wave-vector $\mathbf{k}$ which are parallel or orthogonal
to the background ``three-vector'' $\boldsymbol{\zeta}$:
\begin{equation}
\label{eq:definition-k-orthogonal-k-parallel}
k_{\|}=\mathbf{k}\cdot \widehat{\boldsymbol{\zeta}}\,,\quad
k_{\bot}=|\mathbf{k}-
(\mathbf{k}\cdot\widehat{\boldsymbol{\zeta}})\,
\widehat{\boldsymbol{\zeta}}|\,,\quad
\widehat{\boldsymbol{\zeta}}
\equiv\frac{1}{\sqrt{(\widetilde{\kappa}^{01})^2
+(\widetilde{\kappa}^{02})^2+(\widetilde{\kappa}^{03})^2}}
\begin{pmatrix}
\widetilde{\kappa}^{01} \\
\widetilde{\kappa}^{02} \\
\widetilde{\kappa}^{03} \\
\end{pmatrix}\,,
\end{equation}
where $k_{\parallel}\in (-\infty,\infty)$ and $k_{\bot}\in [0,\infty)$.
By doing so, it is possible to write the dispersion relations
\eqref{eq:dispersion-relation-p-odd-1-2} as follows:
\begin{subequations}\label{eq:dispersion-relation-p-odd-1-2-mathcalE}
\begin{align}
\label{eq:dispersion-relation-p-odd-1-mathcalE}
\omega_1(k_{\bot},k_{\|})=\mathcal{E}\,k_{\|}
+\sqrt{k_{\bot}^2+(1+\mathcal{E}^2)\,k_{\|}^2}\,,
\end{align}
\begin{align}
\label{eq:dispersion-relation-p-odd-2-mathcalE}
\omega_2(k_{\bot},k_{\|})=
\mathcal{E}\,k_{\parallel}+\sqrt{1+\mathcal{E}^2}\,|\mathbf{k}|\,,
\end{align}
where the three Lorentz-violating parameters $\widetilde{\kappa}^{01}$,
$\widetilde{\kappa}^{02}$, and $\widetilde{\kappa}^{03}$ are contained
in the single parameter $\mathcal{E}$ that is defined as
\begin{equation}
\label{eq:definition-parameter-calligraphic-e}
\mathcal{E} \equiv |\boldsymbol{\zeta}| \equiv
\sqrt{(\widetilde{\kappa}^{01})^2+(\widetilde{\kappa}^{02})^2
+(\widetilde{\kappa}^{03})^2}\,.
\end{equation}
\end{subequations}
It is obvious that $\mathcal{E}\in [0,\infty)$, whereas each single
parameter $\widetilde{\kappa}^{01}$, $\widetilde{\kappa}^{02}$, and
$\widetilde{\kappa}^{03}$ can be either positive or negative.
From the first definition of Eq. \eqref{eq:definition-k-orthogonal-k-parallel}
we see that negative parameters $\widetilde{\kappa}^{01}$, $\widetilde{\kappa}^{02}$,
$\widetilde{\kappa}^{03}$ are mimicked by a negative $k_{\parallel}$.

The phase and group velocity \cite{Brillouin1960} of the above two modes
can be cast in the following form for small enough $\mathcal{E}$:
\begin{subequations}
\label{eq:phase-velocity}
\begin{equation}
v_{\mathrm{ph},\,1}
\equiv\frac{\omega_1}{|\mathbf{k}|}
=1+\mathcal{E}\cos\theta
+\frac{\mathcal{E}^2}{2}\cos^2\theta+\mathsf{O}(\mathcal{E}^3)\,,
\end{equation}
\begin{equation}
v_{\mathrm{ph},\,2}\equiv \frac{\omega_2}{|\mathbf{k}|}=1+\mathcal{E}\cos\theta
+\frac{\mathcal{E}^2}{2}+\mathsf{O}(\mathcal{E}^3)\,,
\end{equation}
\end{subequations}
\begin{subequations}
\label{eq:group-velocity}
\begin{equation}
v_{\mathrm{gr},\,1}
\equiv \left|\frac{\partial \omega_1}{\partial \mathbf{k}}\right|
= 1+\mathcal{E}\cos\theta+\frac{\mathcal{E}^2}{2}+\mathsf{O}(\mathcal{E}^3)\,,
\end{equation}
\begin{equation}
v_{\mathrm{gr},\,2}
\equiv \left|\frac{\partial \omega_2}{\partial \mathbf{k}}\right|
= 1+\mathcal{E}\cos\theta+\left(1+\sin^2\theta\right)\frac{\mathcal{E}^2}{2}+\mathsf{O}(\mathcal{E}^3)\,,
\end{equation}
\end{subequations}
where $\theta$ is the angle between the three-momentum $\mathbf{k}$ and the unit
vector $\widehat{\boldsymbol{\zeta}}$: $\cos\theta=\mathbf{k}\cdot \widehat{\boldsymbol{\zeta}}/|\mathbf{k}|$.

To leading order in $\mathcal{E}$, the velocities above are equal:
\begin{equation}
v_{\mathrm{ph},\,1}=v_{\mathrm{ph},\,2}=v_{\mathrm{gr},\,1}=v_{\mathrm{gr},\,2}\,.
\end{equation}
Furthermore, Eqs. \eqref{eq:phase-velocity}, \eqref{eq:group-velocity}
show that both phase and group velocity can be larger than 1. However, what matters
physically is the velocity of signal propagation, which corresponds to the front
velocity \cite{Brillouin1960}:
\begin{equation}
v_{\mathrm{fr}}\equiv\lim_{k\mapsto\infty} v_{\mathrm{ph}}\,.
\label{eq:front-velocity}
\end{equation}
Equation \eqref{eq:front-velocity} can be interpreted as the velocity of the
highest-frequency forerunners of a signal. As can be seen from Eq.
\eqref{eq:phase-velocity}, $v_{\mathrm{ph}}$ and hence also $v_{\mathrm{fr}}$
do not depend on the magnitude of the wave vector, but only on its direction.
For $\mathcal{E}\ll 1$, we obtain $v_{\mathrm{fr},\,1}\simeq v_{\mathrm{fr},\,2}
\equiv v_{\mathrm{fr}}$, where
\begin{subequations}
\label{eq:inequalities-front-velocity}
\begin{eqnarray}
v_{\mathrm{fr}}
&<& 1  \;\;\text{for}\;\;  \pi/2 < \theta < 3\pi/2\,, \\
v_{\mathrm{fr}}
&\geq&    1  \;\;\text{for}\;\;   0\leq \theta\leq \pi/2 \;\vee\; 3\pi/2\leq\theta<2\pi\,.
\end{eqnarray}
\end{subequations}
Observe that, for small enough $\mathcal{E}$, having $v_{\mathrm{fr}}< 1$ or
$v_{\mathrm{fr}}\geq 1$ does not depend on the Lorentz-violating parameters but only
on the direction in which the classical wave propagates. For completeness, we also
give the phase velocities for propagation parallel and orthogonal to
$\widehat{\boldsymbol{\zeta}}$:
\begin{subequations}
\begin{equation}
v_{\mathrm{ph},\|,\,1}=\frac{\omega_1(k_{\bot},k_{\|})}{k_{\|}}\,\Bigg|_{k_{\bot}=0}
=\mathcal{E}\,\mathrm{sgn}(k_{\parallel})+\sqrt{1+\mathcal{E}^2}=v_{\mathrm{ph},\|,\,2}\,,
\end{equation}
\begin{equation}
v_{\mathrm{ph},\bot,\,1}
=\frac{\omega_1(k_{\bot},k_{\|})}{k_{\bot}}\,\Bigg|_{k_{\|}=0}
=1\,,\quad v_{\mathrm{ph},\bot,\,2}=
\frac{\omega_2(k_{\bot},k_{\|})}{k_{\bot}}\,\Bigg|_{k_{\|}=0}=
\sqrt{1+\mathcal{E}^2}\,,
\end{equation}
\end{subequations}
with the sign function
\begin{equation}
\mathrm{sgn}(x)=\left\{\begin{array}{rcl}
1 & \text{for} & x>0\,, \\
0 & \text{for} & x=0\,, \\
-1 & \text{for} & x<0\,. \\
\end{array}
\right.
\end{equation}
Note that the latter results are in agreement with the inequalities
of Eq. \eqref{eq:inequalities-front-velocity}. We conclude that the front
velocity can be larger than 1 for the wave vector pointing in certain
directions. That leads us to the issue of microcausality, which will be
discussed in Sec. \ref{sec:Microcausality-parity-odd}.

\section{Propagator in the Feynman gauge}
\label{sec:Propagator-parity-odd}
\setcounter{equation}{0}
\vspace*{-0mm}

So far, we have investigated the dispersion relations of the classical theory. For a
further analysis, especially concerning the quantum theory, the gauge propagator
will be needed. The propagator is the Green's function of the free field equations
\eqref{eq:field-equations-modified-maxwell-theory} in momentum space. In order to
compute it the gauge has to be fixed. We decide to use the Feynman
gauge~\cite{Veltman1994,ItzyksonZuber1980,PeskinSchroeder1995},
which can be implemented by the gauge-fixing condition
\begin{equation}
\mathcal{L}_{\mathrm{gf}}(x)=
-\frac{1}{2}\big(\partial_{\mu}\,A^{\mu}(x)\big)^2\,.
\label{eq:gauge-fixing-feynman}
\end{equation}
The following \textit{Ansatz} for the propagator turns out to be useful:
\begin{align}\label{eq:propagator-parity-odd-coeff}
\widehat{G}_{\nu\lambda}\,
\big|^{\mathrm{Feynman}}
=-\mathrm{i}\,\Big\{
&+\widehat{a}\,\eta_{\nu\lambda}
+\widehat{b}\,k_{\nu}k_{\lambda}
+\widehat{c}\,\xi_{\nu}\xi_{\lambda}
+\widehat{d}\,(k_{\nu}\xi_{\lambda}+\xi_{\nu}k_{\lambda}) \notag \\
&+\widehat{e}\,\zeta_{\nu}\zeta_{\lambda}
+\widehat{f}\,(k_{\nu}\zeta_{\lambda}+\zeta_{\nu}k_{\lambda})
+\widehat{g}\,(\xi_{\nu}\zeta_{\lambda}+\zeta_{\nu}\xi_{\lambda})
\Big\}\,\widehat{K}_1\,.
\end{align}
The propagator coefficients $\widehat{a}=\widehat{a}(k^0,\mathbf{k})$, $\hdots$,
$\widehat{g}=\widehat{g}(k^0,\mathbf{k})$ and the scalar propagator part
$\widehat{K}_1=\widehat{K}_1(k^0,\mathbf{k})$ follow from the system of equations
$(\widehat{G}^{-1})^{\mu\nu}\widehat{G}_{\nu\lambda}=\mathrm{i}\,\delta^{\mu}_{\phantom{\mu}\lambda}$
with the differential operator
\begin{equation}
(G^{-1})^{\mu\nu}=\eta^{\mu\nu}\partial^2
-2\, \kappa^{\mu\varrho\sigma\nu}\partial_{\varrho}\partial_{\sigma}\,,
\end{equation}
in Feynman gauge transformed to momentum space.
Scalar products $\xi^{\mu}\xi_{\mu}$, $\zeta^{\mu}\zeta_{\mu}$, and $\xi^{\mu}\zeta_{\mu}$
will be kept in the result, in order to gain some insight in the
covariant structure of the functions. However, we remark that,
for the case considered, $\xi^2 \equiv \xi^{\mu}\xi_{\mu}=1$,
$\zeta^2\equiv \zeta^{\mu}\zeta_{\mu}=-4\mathcal{E}^2$,
and $\xi\cdot\zeta\equiv\xi^{\mu}\zeta_{\mu}=0$.

Specifically, the propagator coefficients and the scalar
propagators $\widehat{K}_1$ and $\widehat{K}_2$, where $\widehat{K}_2$
appears in some of these coefficients, are given by
\begin{subequations}\label{eq:propagator-result-parity-odd-scalar-1-2}
\begin{eqnarray}\label{eq:propagator-result-parity-odd-scalar-1}
\widehat{K}_1 &=&
\frac{2}{2\,k\cdot \xi\,k\cdot\zeta+k^2\,\big(2-\xi\cdot\zeta\big)}\,,
\\
\label{eq:propagator-result-parity-odd-scalar-2}
\widehat{K}_2&\equiv&\frac{4}{4\,k\cdot\xi\,k\cdot\zeta
+\xi^2(k\cdot\zeta)^2+\zeta^2(k\cdot\xi)^2+k^2(4-\xi^2\zeta^2)}\,,
\end{eqnarray}
\end{subequations}
\begin{subequations}
\label{eq:propagator-result-parity-odd-coeff}
\begin{equation}
\widehat{a}=1\,,
\end{equation}
\begin{align}
\widehat{b}&=-\frac{1}{4\,k^4}\left\{\Upsilon\,\widehat{K}_2-2\,\chi\,\Big(2\,k\cdot\xi\,k\cdot\zeta
+k^2(2-\xi\cdot\zeta)\Big)\right\}\,,
\label{eq:propagator-result-parity-odd-coeff-b}
\end{align}
\begin{equation}
\widehat{c}=\frac{1}{4}\big[k^2\zeta^2-(k\cdot \zeta)^2\big]\,\widehat{K}_2\,,
\label{eq:propagator-result-parity-odd-coeff-d}
\end{equation}
\begin{equation}
\widehat{d}=\frac{k\cdot\xi\big(2\,(k\cdot \zeta)^2-k^2\zeta^2\big)+2\,k^2\,k\cdot\zeta}{4\,k^2}\,\widehat{K}_2\,,
\end{equation}
\begin{equation}
\widehat{e}=\frac{1}{4}\big[k^2\xi^2-(k\cdot
\xi)^2\big]\,\widehat{K}_2\,,
\end{equation}
\begin{equation}
\widehat{f}=\frac{k\cdot\zeta\big(2\,(k\cdot \xi)^2-k^2\xi^2\big)+2\,k^2\,k\cdot\xi}{4\,k^2}\,\widehat{K}_2\,,
\end{equation}
\begin{eqnarray}\label{eq:Upsilon-def}
\Upsilon&\equiv&-2\,k\cdot\xi\,k\cdot\zeta(2\,k^2-k\cdot\xi\,k\cdot\zeta)+(k\cdot\zeta)^2\big((k\cdot\xi)^2-k^2\xi^2\big) \nonumber \\
&&\quad\,+(k\cdot\xi)^2\big[(k\cdot\zeta)^2-k^2\zeta^2\big]+k^2\big[12\,k\cdot\xi\,k\cdot\zeta+\xi^2(k\cdot\zeta)^2 \nonumber \\
&&\quad\,+\zeta^2(k\cdot\xi)^2+k^2(4-\xi^2\zeta^2)\big]\,,
\end{eqnarray}
\begin{equation}
\widehat{g}=-\frac{1}{4}\big[2\,k^2+k\cdot\xi\,k\cdot\zeta\big]\,\widehat{K}_2\,,
\end{equation}
\end{subequations}
where definition \eqref{eq:Upsilon-def} enters
\eqref{eq:propagator-result-parity-odd-coeff-b}.

The poles of $\widehat{K}_1$ and $\widehat{K}_2$ can be
identified with the dispersion relations obtained in
Sec.~\ref{sec:Dispersion-relations-classical-causality}.
From $\widehat{K}_1(\omega_1,\mathbf{k})^{-1}=0$, that is
\begin{equation}
2\,k\cdot \xi\,k\cdot\zeta+k^2\,
\big(2-\xi\cdot \zeta\big)\,\Big|_{k_0=\omega_1}=0\,,
\label{eq:off-shell-dispersion-relation-1}
\end{equation}
the dispersion relation \eqref{eq:dispersion-relation-p-odd-1}
of the $\lambda=1$ mode is recovered. Similarly, the dispersion
relation \eqref{eq:dispersion-relation-p-odd-2} of the $\lambda=2$
mode follows from $\widehat{K}_2(\omega_2,\mathbf{k})^{-1}=0$, that is
\begin{equation}
4\,k\cdot\xi\,k\cdot\zeta+\xi^2(k\cdot\zeta)^2+\zeta^2(k\cdot\xi)^2+k^2(4-\xi^2\zeta^2)\,\Big|_{k_0=\omega_2}=0\,,
\label{eq:off-shell-dispersion-relation-2}
\end{equation}
The third pole $k^2=0$ corresponds to the
dispersion relation of scalar and longitudinal modes. This is clear
from the fact that this pole appears only in the gauge-dependent
coefficients $\widehat{b}$, $\widehat{d}$, and $\widehat{f}$.
These are multiplied by at least one photon four-momentum and
vanish by the Ward identity,\footnote{assuming $k^2\neq 0$}
if they couple to a conserved current \cite{PeskinSchroeder1995}.
Since the Ward identity results from
gauge invariance, it also holds for modified Maxwell theory, which
is expected to be free of anomalies \cite{ColladayKostelecky1998}.
Because of parity violation the physical poles are asymmetric
with respect to the imaginary $k^0$-axis.

The above result \eqref{eq:propagator-parity-odd-coeff}--
\eqref{eq:propagator-result-parity-odd-coeff} equals the propagator
given in \cite{Casana-etal2010}. Every propagator coefficient, which
contains the scalar propagator $\widehat{K}_2$, is also multiplied by
$\widehat{K}_1$. Hence, both modes appear together throughout the
propagator and the question arises, whether they can be separated.
It can be shown that the propagator can also be written in the following
form:
\begin{equation}
\label{eq:split-propagator}
\widehat{G}_{\mu\nu}(k)\big|^{\mathrm{Feynman}}= \sum_{n=1,2}
\Xi_{\mu\nu}^{(n)}(k^0,\mathbf{k})\,
\Big(-\mathrm{i}\widehat{G}^{(n)}(k)\Big)\,,
\end{equation}
where the tensor structure $\Xi_{\mu\nu}$ is the same for both
parts, hence
\begin{align}
\label{eq:split-propagator-coefficient-functions}
\Xi_{\nu\lambda}^{(1)}=\Xi_{\nu\lambda}^{(2)}=&+\widehat{a}\,\eta_{\nu\lambda}
+\widehat{b}\,k_{\nu}k_{\lambda}
+\widehat{c}\,\xi_{\nu}\xi_{\lambda}
+\widehat{d}\,(k_{\nu}\xi_{\lambda}+\xi_{\nu}k_{\lambda}) \notag \\
&+\widehat{e}\,\zeta_{\nu}\zeta_{\lambda}
+\widehat{f}\,(k_{\nu}\zeta_{\lambda}+\zeta_{\nu}k_{\lambda})
+\widehat{g}\,(\xi_{\nu}\zeta_{\lambda}+\zeta_{\nu}\xi_{\lambda})\,,
\end{align}
with the coefficients $\widehat{a}$, \dots, $\widehat{g}$
from Eq. \eqref{eq:propagator-result-parity-odd-coeff}.
The scalar propagator functions are then given by:
\begin{equation}
\widehat{G}^{(1)}(k)=\frac{4\widehat{K}_1\widehat{K}_2^{-1}}{[(k\cdot\xi)^2-k^2]\zeta^2+(k\cdot\zeta)^2}\,,
\quad \widehat{G}^{(2)}(k)=-\frac{4}{[(k\cdot\xi)^2-k^2]\zeta^2+(k\cdot\zeta)^2}\,.
\label{eq:split-propagator-scalar-parts}
\end{equation}
The first part $\widehat{D}^{(1)}(k)$ contains both
polarization modes encoded in $\widehat{K}_1$ and $\widehat{K}_2$, whereas
the second part does not involve any mode. The denominator
$[(k\cdot\xi)^2-k^2]\zeta^2+(k\cdot\zeta)^2$ that appears in both parts does
not have a zero with respect to $k_0$, hence it contains no dispersion
relation. So it does not seem that the polarization modes can be separated,
such that each propagator part contains exactly one of the modes.

Finally, we can state that the structure of the propagator of parity-odd
nonbirefringent modified Maxwell theory is rather unusual. In the next
section we will compute the polarization vectors.

\section{Polarization vectors}
\label{sec:Polarization-vectors-parity-odd}
\setcounter{equation}{0}
\vspace*{-0mm}

In what follows, the physical (transverse) degrees of freedom will be
labeled with (1) and (2), respectively. For a fixed nonzero
``three-vector'' $\boldsymbol{\zeta}$ and a generic wave vector $\mathbf{k}$,
the polarization vector of the $\lambda=1$ mode reads
\begin{equation}\label{eq:polarization-mode1-parity-violating}
\big(\varepsilon^{(1)\,\mu}\big)
= \frac{1}{\sqrt{N'}}\;
\big(0,\, \boldsymbol{\zeta}\times\mathbf{k}\big)\big/
  \big|\boldsymbol{\zeta}\times\mathbf{k}\big|\,,
\end{equation}
where $N'$ is a normalization factor to be given later.
The polarization vector of the $\lambda=2$
mode is orthogonal to \eqref{eq:polarization-mode1-parity-violating}
and has a longitudinal component. It is given by
\begin{equation}\label{eq:polarization-mode2-parity-violating}
(\varepsilon^{(2)\,\mu})=\frac{1}{\sqrt{N''}}
\frac{1}{\sqrt{|\boldsymbol{\varepsilon}^{(2)}|^2-(\varepsilon^0)^2}}
\left(\varepsilon^0,\,\boldsymbol{\varepsilon}^{(2)}\right)\,,
\end{equation}
with
\begin{subequations}
\label{eq:polarization-mode2-parity-violating-explict}
\begin{equation}
\varepsilon^0=\frac{1}{4}\Big(k^2-(k\cdot\xi)^2\Big)
\Big((k\cdot\zeta)^2-\zeta^2\left[k^2-(k\cdot\xi)^2\right]\Big)\,,
\end{equation}
\begin{align}
\boldsymbol{\varepsilon}^{(2)}&=
\left(2\,|\mathbf{k}|^2\,|\boldsymbol{\zeta}|^2-
\frac{|\mathbf{k}\times (\mathbf{k}\times \boldsymbol{\zeta})|^2}
     {|\mathbf{k}|^2}
+2\,\sqrt{1+|\boldsymbol{\zeta}|^2}\,|\mathbf{k}|
\,(\mathbf{k}\cdot \boldsymbol{\zeta})\right)
\mathbf{k}\times (\mathbf{k}\times \boldsymbol{\zeta}) \notag \\
&\hspace{2cm}\,+\Big(\sqrt{1+|\boldsymbol{\zeta}|^2}\,|\mathbf{k}|
+\mathbf{k}\cdot\boldsymbol{\zeta}\Big)\;
\frac{|\mathbf{k}\times (\mathbf{k}\times\boldsymbol{\zeta})|^2}
     {|\mathbf{k}|^2}\;\mathbf{k}\,.
\end{align}
\end{subequations}
The polarization vector $\varepsilon^{(1)}$ is a solution of the
field equations \eqref{eq:field-equations-modified-maxwell-theory},
when $k^0$ is replaced by $\omega_1(\mathbf{k})$ from
Eq. \eqref{eq:dispersion-relation-p-odd-1}. The polarization
$\varepsilon^{(2)}$ is the corresponding solution for $k^0$
replaced by $\omega_2(\mathbf{k})$ from Eq.
\eqref{eq:dispersion-relation-p-odd-2}. The normalization
factors $N'$ in \eqref{eq:polarization-mode1-parity-violating}
and $N''$ in \eqref{eq:polarization-mode2-parity-violating}
can be computed from the 00--component of the energy-momentum tensor.
Note that the above polarization vectors have been calculated in
the Lorentz gauge, $\partial_{\mu}\,A^{\mu}=0$.

For the Lorentz-violating decay processes considered,
both the $\lambda=1$ and the $\lambda=2$ polarization modes
contribute.
\begin{align}
\label{eq:polarization-sum-parity-violating-1}
\overline{\varepsilon}^{(1)\,\mu}\varepsilon^{(1)\,\nu}&=\frac{1}{N'}\;\Bigl\{-\eta^{\mu\nu}+\widehat{\Gamma}_1k^{\mu}k^{\nu}
+\widehat{\Delta}_1(k^{\mu}\xi^{\nu}+\xi^{\mu}k^{\nu})+\widehat{\Lambda}_1(k^{\mu}\zeta^{\nu}+\zeta^{\mu}k^{\nu}) \notag \\
&\hspace{1.42cm}+\widehat{\Phi}_1\,\xi^{\mu}\xi^{\nu}+\widehat{\Psi}_1\,\zeta^{\mu}\zeta^{\nu}+\widehat{\Theta}_1\,\big(\xi^{\mu}\zeta^{\nu}+\zeta^{\nu}\xi^{\mu}\big)\,
\Bigr\}\,\Big|_{k_0=\omega_1}\,,
\end{align}
with
\begin{subequations}
\label{eq:polarization-sum-coefficients-1}
\begin{equation}
\widehat{\Gamma}_1=\frac{\zeta^2}{Q}\,,\quad\widehat{\Delta}=\frac{\zeta^2\,\big[2k^2+k\cdot\xi\,k\cdot \zeta\big]}{Q(\zeta\cdot k)}\,,\quad\widehat{\Lambda}_1=-\frac{k\cdot\zeta}{Q}\,,
\end{equation}
\begin{equation}
\widehat{\Theta}_1=-\frac{2k^2+k\cdot\xi\,k\cdot\zeta}{Q}\,,\quad \widehat{\Psi}_1=\frac{k^2\,\xi^2-(k\cdot\xi)^2}{Q}\,,\quad \widehat{\Phi}_1=\frac{k^2\,\zeta^2-(k\cdot\zeta)^2}{Q}\,,
\end{equation}
\begin{equation}
N'=\frac{1}{\omega_1}\sqrt{k_{\bot}^2+(1+\mathcal{E}^2)\,k_{\|}^2}\,,\quad Q=\zeta^2\,\big[k^2-(k\cdot\xi)^2\big]-(k\cdot\zeta)^2\,,
\end{equation}
\end{subequations}
where $\omega_1=\omega_1(k_{\bot},k_{\|})$ is given by
\eqref{eq:dispersion-relation-p-odd-1-mathcalE}. The denominator
$Q$ vanishes only for $\widetilde{\kappa}^{01}=\widetilde{\kappa}^{02}
=\widetilde{\kappa}^{03}=0$ or $k_{\bot}=0$.
If the polarization tensor of the $\lambda=1$ mode is contracted
with a gauge-invariant expression using the Ward identity,\footnote{This
means dropping terms that are proportional to at least one external
four-momentum $k^{\mu}$, which we denote by the word ``truncated''}
it can be replaced by $\Pi^{\mu\nu}|_{\lambda=1}$:
\begin{subequations}
\begin{equation}
\overline{\varepsilon}^{(1)\,\mu}\varepsilon^{(1)\,\nu}\mapsto \Pi^{\mu\nu}|_{\lambda=1}\,,
\end{equation}
\begin{align}
\label{eq:polarization-sum-parity-violating-1-truncated}
\Pi^{\mu\nu}|_{\lambda=1}&\equiv\overline{\varepsilon}^{(1)\,\mu}\varepsilon^{(1)\,\nu}
\;\big|^\text{truncated} \notag \\
&=\frac{1}{N'}\;\Bigl\{-\eta^{\mu\nu}
+\widehat{\Phi}_1\,\xi^{\mu}\xi^{\nu}+\widehat{\Psi}_1\,\zeta^{\mu}\zeta^{\nu}+\widehat{\Theta}_1
\big(\xi^{\mu}\zeta^{\nu}+\zeta^{\nu}\xi^{\mu}\big)\Bigr\}
\,\Big|_{k_0=\omega_1}\,.
\end{align}
\end{subequations}

The polarization tensor of the $\lambda=2$ mode is lengthy and
is best written up in terms of $k_{\|}$ and $k_{\bot}$ defined in
\eqref{eq:definition-k-orthogonal-k-parallel}.
\begin{align}
\label{eq:polarization-sum-parity-violating-2}
\overline{\varepsilon}^{(2)\,\mu}\varepsilon^{(2)\,\nu}
&=\frac{1}{N''}\Big\{+\widehat{\Gamma}_2k^{\mu}k^{\nu}+\widehat{\Delta}_2(k^{\mu}\xi^{\nu}
+\xi^{\mu}k^{\nu})+\widehat{\Lambda}_2(k^{\mu}\zeta^{\nu}+\zeta^{\mu}k^{\nu}) \notag \\
&\hspace{1.48cm}+\widehat{\Phi}_2\xi^{\mu}\xi^{\nu}
+\widehat{\Psi}_2\zeta^{\mu}\zeta^{\nu}+\widehat{\Theta}_2(\xi^{\mu}\zeta^{\nu}
+\zeta^{\mu}\xi^{\nu})\Big\}\,\Big|_{k_0=\omega_2}\,,
\end{align}
with
\begin{subequations}
\label{eq:polarization-sum-coefficients-2}
\begin{align}
\widehat{\Gamma}_2=\frac{\mathcal{E}^4}{\mathcal{N}}\Big[\mathcal{E}k_{\parallel}k_{\bot}^2+(2k_{\parallel}^2+k_{\bot}^2)\omega_2\Big]^2\,,
\end{align}
\begin{align}
\widehat{\Delta}_2=\frac{\mathcal{E}^4}{\mathcal{N}}\Big(\mathcal{E}k_{\parallel}k_{\bot}^2+(2k_{\parallel}^2+k_{\bot}^2)\omega_2\Big)\Big\{k_{\bot}^2|\mathbf{k}|^2-\mathcal{E}k_{\parallel}k_{\bot}^2\omega_2-\big[2k_{\parallel}^2+k_{\bot}^2\big]\omega_2^2\Big\}\,,
\end{align}
\begin{align}
\widehat{\Lambda}_2=-\frac{\mathcal{E}^3}{2\mathcal{N}}|\mathbf{k}|^2\Big(2k_{\parallel}\omega_2+\mathcal{E}k_{\bot}^2\Big)\Big[\mathcal{E}k_{\parallel}k_{\bot}^2+(2k_{\parallel}^2+k_{\bot}^2)\omega_2\Big]\,,
\end{align}
\begin{align}
\label{eq:terms-explicit-polarization-sum-2-1}
\widehat{\Phi}_2=\frac{\mathcal{E}^4}{\mathcal{N}}\Big[k_{\bot}^2(k_{\bot}^2-\omega_2^2)-\mathcal{E}k_{\parallel}k_{\bot}^2\omega_2+k_{\parallel}^2(k_{\bot}^2-2\omega_2^2)\Big]^2\,,
\end{align}
\begin{align}
\label{eq:terms-explicit-polarization-sum-2-2}
\widehat{\Psi}_2=\frac{\mathcal{E}^2}{4\mathcal{N}}|\mathbf{k}|^4(\mathcal{E}k_{\bot}^2+2k_{\parallel}\omega_2)^2\,,
\end{align}
\begin{align}
\label{eq:terms-explicit-polarization-sum-2-3}
\widehat{\Theta}_2=\frac{\mathcal{E}^3}{2\mathcal{N}}|\mathbf{k}|^2(\mathcal{E}k_{\bot}^2+2k_{\parallel}\omega_2)\Big[k_{\bot}^2(\omega_2^2-k_{\bot}^2)+\mathcal{E}k_{\parallel}k_{\bot}^2\omega_2-k_{\parallel}^2(k_{\bot}^2-2\omega_2^2)\Big]\,,
\end{align}
\begin{align}
\label{eq:terms-explicit-polarization-sum-2-4}
N''=\frac{\mathcal{E}^4k_{\bot}^2}{2\omega_2^2\mathcal{N}}|\mathbf{k}|^2\Big[&+(4k_{\parallel}^2+k_{\bot}^2)\omega_2^4+4\mathcal{E}k_{\parallel}k_{\bot}^2\omega_2^3+\big(4k_{\parallel}^4+2k_{\bot}^2k_{\parallel}^2+(\mathcal{E}^2-2)k_{\bot}^4\big)\omega_2^2
\notag \\
&+4\mathcal{E}k_{\parallel}k_{\bot}^2|\mathbf{k}|^2\omega_2+k_{\bot}^2|\mathbf{k}|^2(|\mathbf{k}|^2+\mathcal{E}^2k_{\bot}^2)\Big]\,.
\end{align}
where
\begin{equation}
\mathcal{N}=|\boldsymbol{\varepsilon}^{(2)}|^2-(\varepsilon^0)^2\,,
\end{equation}
\end{subequations}
and $\omega_2=\omega_2(k_{\bot},k_{\|})$ is given by
\eqref{eq:dispersion-relation-p-odd-2-mathcalE}. Again, if the
tensor is contracted with a gauge-invariant expression, it can
be replaced by $\Pi^{\mu\nu}|_{\lambda=2}$:
\begin{subequations}
\begin{equation}
\overline{\varepsilon}^{(2)\,\mu}\varepsilon^{(2)\,\nu}\mapsto \Pi^{\mu\nu}|_{\lambda=2}\,,
\end{equation}
\begin{align}
\label{eq:polarization-sum-parity-violating-2-truncated}
\Pi^{\mu\nu}|_{\lambda=2}&\equiv \overline{\varepsilon}^{(2)\,\mu}\varepsilon^{(2)\,\nu}
\;\big|^\text{truncated} \notag \\
&=\frac{1}{N''}\Big\{\widehat{\Phi}_2\xi^{\mu}\xi^{\nu}
+\widehat{\Psi}_2\zeta^{\mu}\zeta^{\nu}+\widehat{\Theta}_2(\xi^{\mu}\zeta^{\nu}
+\zeta^{\mu}\xi^{\nu})\Big\}\,\,\Big|_{k_0=\omega_2}\,.
\end{align}
\end{subequations}
Finally it holds that
\begin{equation}
k_{\mu}\left(\overline{\varepsilon}^{(1)\,\mu}\varepsilon^{(1)\,\nu}\right)(k)=0\,,\quad \lim_{\mathcal{E}\mapsto 0} k_{\mu}\left(\overline{\varepsilon}^{(2)\,\mu}\varepsilon^{(2)\,\nu}\right)(k)=0\,,
\end{equation}
where the second contraction only vanishes for $\mathcal{E}\mapsto 0$ due
to the longitudinal part of $\varepsilon^{(2)\,\mu}$.

The polarization vector \eqref{eq:polarization-mode2-parity-violating}
is normalized to unit length by $\mathcal{N}$. This normalization
factor cancels in $\Pi^{\mu\nu}|_{\lambda=2}$.
Note that the metric tensor $\eta^{\mu\nu}$ does not appear
on the right-hand side of \eqref{eq:polarization-sum-parity-violating-2-truncated},
whereas it does on the right-hand side of Eq.~
\eqref{eq:polarization-sum-parity-violating-1-truncated}.

Furthermore, note that each truncated polarization tensor $\Pi^{\mu\nu}|_{\lambda=1}$
and $\Pi^{\mu\nu}|_{\lambda=2}$ can be written in a covariant form.
This behavior is different from the polarization vectors of standard QED,\footnote{Also
in the isotropic and the parity-even anisotropic sector of modified Maxwell theory the
polarization tensor of one single transversal mode cannot be decomposed covariantly
\cite{Klinkhamer:2010zs}.} where only the whole polarization sum is covariant.

It is now evident that not only is the structure of the photon propagator uncommon,
but the polarization vectors are unusual as well. In the next section we will analyze
how both results are connected.

\section{The optical theorem and unitarity}
\label{sec:Optical-theorem-parity-odd}
\setcounter{equation}{0}
\vspace*{-0mm}

In order to investigate unitarity, the simple test of reflection positivity
used in Ref.~\cite{Klinkhamer:2010zs} for the isotropic case of modified Maxwell
theory cannot be adopted, because there are now essentially two
different scalar propagators, namely $\widehat{K}_1$ and $\widehat{K}_2$
from \eqref{eq:propagator-result-parity-odd-coeff}. Hence, we could either
examine reflection positivity of the full propagator or study the optical
theorem for physical processes involving modified photons. As unitarity of
the S--matrix results in the optical theorem and the latter is directly
related to physical observables, we choose to proceed with the second
approach.

The optical theorem will also show how the modified photon
propagator in Sec.~\ref{sec:Propagator-parity-odd}
is linked to the photon polarizations from the previous section.
The following computations will deal with the physical process that
we already considered for
isotropic modified Maxwell theory \cite{Klinkhamer:2010zs} in the
context of unitarity: annihilation of a left-handed electron
$e_L^-$ and a right-handed positron $e_R^+$ to a modified photon
$\widetilde{\gamma}$. The fermions are considered to be massless
particles, which renders their helicity a physically well-defined
state. Neglecting the axial anomaly, which is of higher order
with respect to the electromagnetic coupling constant, the axial
vector current $j^{\mu}_5=\overline{\psi}\gamma^{\mu}\gamma_5\psi$
is conserved: $\partial_{\mu}j^{\mu}_5=0$. This is the simplest
tree-level process including a modified photon propagator. It has no
threshold and is allowed for both photon modes.
We assume a nonzero Lorentz-violating parameter $\mathcal{E}$.
Furthermore, the four-momenta of the initial electron
and positron are not expected to be collinear.

If the optical theorem holds, the imaginary part of the forward scattering
amplitude $\mathcal{M}(e^{-}_{L}e^{+}_{R}\rightarrow e^{-}_{L}e^{+}_{R})$ is
related to the cross section for the production of a modified photon from a
left-handed electron and a right-handed positron:
\begin{equation}\label{eq:opt-theorem}
\hspace*{-5mm}
2\,\mathrm{Im}\left(\begin{array}{c}
\includegraphics{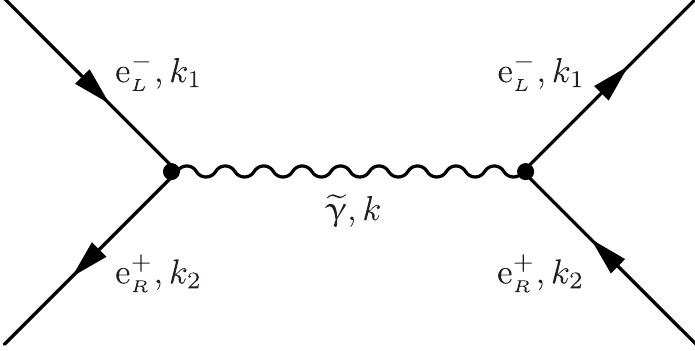}
\end{array}
\right)
\stackrel{?}{=}
\int \mathrm{d}\Pi_1 \left|\begin{array}{c}
\includegraphics{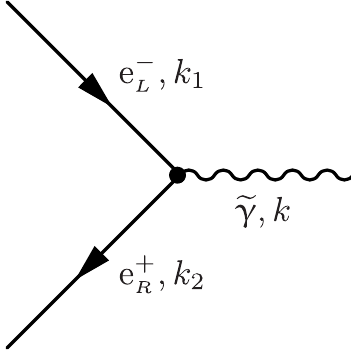}
\end{array}\right|^2\,.
\end{equation}
Herein, $\mathrm{d}\Pi_1$ is the corresponding one-particle phase space element.
By performing an integration over the four-momentum of the virtual photon,
the forward scattering amplitude $\mathcal{M}_{\widehat{1}}\equiv\mathcal{M}(e^{-}_{L}e^{+}_{R}\rightarrow
e^{-}_{L}e^{+}_{R})$ is given by
\begin{align}
\label{eq:forward-scattering-amplitude-optical-theorem}
\mathcal{M}_{\widehat{1}}&=\int \frac{\mathrm{d}^4k}{(2\pi)^4}\,\delta^{(4)}(k_1+k_2-k)\,
e^2\;\overline{u}(k_1)\gamma^{\lambda}\frac{\mathds{1}-\gamma_5}{2}v(k_2)\;
\overline{v}(k_2)\gamma^{\nu}\frac{\mathds{1}-\gamma_5}{2}u(k_1) \notag \displaybreak[0]\\
&\phantom{{}={}\int \frac{\mathrm{d}^4k}{(2\pi)^4}}\,\times\frac{1}{\widehat{K}_1^{-1}+\mathrm{i}\epsilon}\;
\big(+\eta_{\nu\lambda}
+\widehat{b}\,k_{\nu}k_{\lambda}
+\widehat{c}\,\xi_{\nu}\xi_{\lambda}
+\widehat{d}\,(k_{\nu}\xi_{\lambda}+\xi_{\nu}k_{\lambda}) \notag \\
&\phantom{{}={}\int \frac{\mathrm{d}^4k}{(2\pi)^4}\,\times\frac{1}{\widehat{K}_1^{-1}+\mathrm{i}\epsilon}\;\big(}+\widehat{e}\,\zeta_{\nu}\zeta_{\lambda}
+\widehat{f}\,(k_{\nu}\zeta_{\lambda}+\zeta_{\nu}k_{\lambda})
+\widehat{g}\,(\xi_{\nu}\zeta_{\lambda}+\zeta_{\nu}\xi_{\lambda})\big)\,,
\end{align}
with the propagator coefficients $\widehat{b}$, \dots, $\widehat{g}$ from
Eq. \eqref{eq:propagator-result-parity-odd-coeff}. Recall, that the physical
poles have to be treated via Feynman's $\mathrm{i}\epsilon$-prescription.
Hence, the denominator $\widehat{K}_2^{-1}$ from
Eq. \eqref{eq:propagator-result-parity-odd-scalar-2},
which appears in the coefficients $\widehat{b}$, $\widehat{c}$, $\widehat{d}$,
$\widehat{e}$, $\widehat{f}$, and $\widehat{g}$ also has to be replaced by
$\widehat{K}_2^{-1}+\mathrm{i}\epsilon$.

The first contribution to the imaginary part of the matrix element
$\mathcal{M}_{\widehat{1}}$ comes from the physical pole of the scalar
propagator function $\widehat{K}_1$ and corresponds to the dispersion relation
\eqref{eq:dispersion-relation-p-odd-1-mathcalE} of the
$\lambda=1$  polarization
mode. Using the positive and negative photon frequency of the parity-odd
case considered,
\begin{equation}
\label{eq:dispersion-relation-1}
\widetilde{\omega}_{1}^+
\equiv
\mathcal{E}\,k_{\|}+\sqrt{k_{\bot}^2+(1+\mathcal{E}^2)\,k_{\|}^2}\,,\quad
\widetilde{\omega}_{1}^-
\equiv
\mathcal{E}\,k_{\|}-\sqrt{k_{\bot}^2+(1+\mathcal{E}^2)\,k_{\|}^2}\,,
\end{equation}
the scalar part of the propagator is
\begin{equation}
\frac{1}{k\cdot\xi\,k\cdot\zeta+k^2+\mathrm{i}\epsilon}=\frac{1}{(k^0-\widetilde{\omega}_1^++\mathrm{i}\epsilon)(k^0-\widetilde{\omega}_1^--\mathrm{i}\epsilon)}\,.
\end{equation}
The pole with positive real part can be cast in the following form:
\begin{equation}
\frac{1}{k^0-\widetilde{\omega}_{1}^++\mathrm{i}\epsilon}=\mathcal{P}\frac{1}{k^0-\widetilde{\omega}_{1}^+}-\mathrm{i}\pi\,\delta(k^0-\widetilde{\omega}_{1}^+)\,.
\end{equation}
Because of energy conservation only $\widetilde{\omega}_1^+$ and not
$\widetilde{\omega}_1^-$ contributes to the imaginary part. We define
$\widehat{\mathcal{M}}_{\widehat{1}}\equiv \mathcal{M}(e_L^-e_R^+\rightarrow\widetilde{\gamma})$
and obtain:
\begin{align}\label{eq:Optical-theorem-parity-odd}
2\,\mathrm{Im}(\mathcal{M}_{\,\widehat{1}})\big|_{\lambda=1}&=\int
\frac{\mathrm{d}^3k}{(2\pi)^{3}\,2\widetilde{\omega}_{1}^+}\,\delta^{(4)}(k_1+k_2-k) \notag \\
&\hspace{1.5cm}\times e^2\;\overline{u}(k_1)\gamma^{\nu}\frac{\mathds{1}-\gamma_5}{2}v(k_2)\;
\overline{v}(k_2)\gamma^{\mu}\frac{\mathds{1}
-\gamma_5}{2}u(k_1) \notag \\
&\hspace{1.5cm}\times
\frac{1}{N'}\;\big(-\eta_{\mu\nu}-\widehat{c}\,\xi_{\mu}\xi_{\nu}
-\widehat{e}\,\zeta_{\mu}\zeta_{\nu}
-\widehat{g}\,(\xi_{\mu}\zeta_{\nu}+\zeta_{\mu}\xi_{\nu})\big)
\notag \displaybreak[0]\\
&=\int \frac{\mathrm{d}^3k}{(2\pi)^{3}\,2\widetilde{\omega}_1^+}\,\delta^{(4)}(k_1+k_2-k)\,
(\widehat{\mathcal{M}}_{\widehat{1}}^{\,\dagger})^{\nu}(\widehat{\mathcal{M}}_{\widehat{1}})^{\mu}\Big(\Pi_{\mu\nu}|_{\lambda=1}\Big)
\notag \displaybreak[0]\\[1mm]
&=\int
\frac{\mathrm{d}^3k}{(2\pi)^{3}\,2\widetilde{\omega}_1^+}\,\delta^{(4)}(k_1+k_2-k)\left.
|\widehat{\mathcal{M}}_{\widehat{1}}|^2\right|_{\lambda=1}\,,
\end{align}
with
\begin{equation}
\left.\widehat{\mathcal{M}}_{\widehat{1}}\right|_{\lambda=1}\equiv \varepsilon^{(1)}_{\mu}(k)(\widehat{\mathcal{M}}_{\widehat{1}})^{\mu}(k)\,,
\end{equation}
and $k^0$ replaced by the dispersion relation $\widetilde{\omega}_1$ from
Eq. \eqref{eq:dispersion-relation-1}. Furthermore,
\begin{equation}\label{eq:Nprime-parity-odd}
N'=1+\frac{k\cdot\zeta}{2\,\widetilde{\omega}_{1}^+}
=\frac{1}{\widetilde{\omega}_{1}^+}\,
\sqrt{k_{\bot}^2+(1+\mathcal{E}^2)\,k_{\|}^2}\,.
\end{equation}
Using the Ward identity, in the first step of Eq. \eqref{eq:Optical-theorem-parity-odd}
we could eliminate all propagator coefficients that are multiplied by at least one
photon four-momentum. Then we employed the truncated $\lambda=1$ polarization tensor
from Eq. \eqref{eq:polarization-sum-parity-violating-1-truncated}.

The second contribution to the imaginary part of the matrix element comes
from the $\lambda=2$ mode given by the dispersion relation \eqref{eq:dispersion-relation-p-odd-2-mathcalE}.
That mode is contained in $\widehat{K}_2$ from  \eqref{eq:propagator-result-parity-odd-scalar-2}, where
Feynman's $\mathrm{i}\epsilon$-prescription leads to:
\begin{align}
\frac{1}{4\widehat{K}_2^{-1}+\mathrm{i}\epsilon}&=\frac{1}{4\,k\cdot\xi\,k\cdot\zeta+\xi^2(k\cdot\zeta)^2+\zeta^2(k\cdot\xi)^2+k^2(4-\xi^2\zeta^2)+\mathrm{i}\epsilon}
\notag \\
&=\frac{1}{4(k^0-\widetilde{\omega}_2^++\mathrm{i}\epsilon)(k^0-\widetilde{\omega}_2^--\mathrm{i}\epsilon)}\,,
\end{align}
with
\begin{equation}
\label{eq:dispersion-relation-2}
\widetilde{\omega}_2^+\equiv \mathcal{E}k_{\parallel}+\sqrt{1+\mathcal{E}^2}\,|\mathbf{k}|\,,\quad \widetilde{\omega}_2^-\equiv
\mathcal{E}k_{\parallel}-\sqrt{1+\mathcal{E}^2}\,|\mathbf{k}|\,.
\end{equation}
The pole with the positive real part results in the following
contribution to the imaginary part of the matrix element
$\mathcal{M}_{\,\widehat{1}}$:
\begin{equation}
\frac{1}{k^0-\widetilde{\omega}_2^++\mathrm{i}\epsilon}=\mathcal{P}\frac{1}{k^0-\widetilde{\omega}_2^+}-\mathrm{i}\pi\delta(k^0-\widetilde{\omega}_2^+)\,.
\end{equation}
Again, the pole $\widetilde{\omega}_2^-$ with negative real part
does not contribute because of energy conservation.
Using the Ward identity leads to:
\begin{align}
\label{eq:Optical-theorem-parity-odd-2}
2\,\mathrm{Im}(\mathcal{M}_{\widehat{1}})\big|_{\lambda=2}&=\int\frac{\mathrm{d}^3k}{(2\pi)^{3}\,2\widetilde{\omega}_{2}^+}\,\delta^{(4)}(k_1+k_2-k) \notag \\
&\hspace{1cm}\times e^2\;\overline{u}(k_1)\gamma^{\nu}\frac{\mathds{1}-\gamma_5}{2}v(k_2)\;
\overline{v}(k_2)\gamma^{\mu}\frac{\mathds{1}
-\gamma_5}{2}u(k_1)\,\frac{\widehat{K}_1(\widetilde{\omega}_2^+,\mathbf{k})}{2(1-\widetilde{\omega}_2^-/\widetilde{\omega}_2^+)} \notag \\
&\hspace{1cm}\,\times \Big\{\big[(k\cdot \zeta)^2-k^2\zeta^2\big]\xi^{\mu}\xi^{\nu}+\big[(k\cdot
\xi)^2-k^2\xi^2\big]\zeta^{\mu}\zeta^{\nu}\Big. \notag \\
&\hspace{3.20cm}\,\Big.+\big[2\,k^2+k\cdot\xi\,k\cdot\zeta\big](\xi^{\mu}\zeta^{\nu}+\zeta^{\mu}\xi^{\nu})\Big\} \notag \\
&=\int \frac{\mathrm{d}^3k}{(2\pi)^{3}\,2\widetilde{\omega}_2^+}\,\delta^{(4)}(k_1+k_2-k)\,
(\widehat{\mathcal{M}}_{\widehat{1}}^{\,\dagger})^{\nu}(\widehat{\mathcal{M}}_{\widehat{1}})^{\mu}\Big(\Pi_{\mu\nu}|_{\lambda=2}\Big)
\notag \\
&=\int
\frac{\mathrm{d}^3k}{(2\pi)^{3}\,2\widetilde{\omega}_2^+}\,\delta^{(4)}(k_1+k_2-k)\left.
|\widehat{\mathcal{M}}_{\widehat{1}}|^2\right|_{\lambda=2}\,,
\end{align}
where
\begin{equation}
\left.\widehat{\mathcal{M}}_{\widehat{1}}\right|_{\lambda=2}\equiv \varepsilon^{(2)}_{\mu}(k)(\widehat{\mathcal{M}}_{\widehat{1}})^{\mu}(k)\,,
\end{equation}
and $k^0$ is to be replaced by $\widetilde{\omega}_2^+$ from
Eq. \eqref{eq:dispersion-relation-2}. Moreover, we have used that
for $k^0=\widetilde{\omega}_2^+$
\begin{align}
\frac{\widehat{K}_1(\widetilde{\omega}_2^+,\mathbf{k})}
{2(1-\widetilde{\omega}_2^-/\widetilde{\omega}_2^+)}
\Big\{\big[(k\cdot \zeta)^2&-k^2\zeta^2\big]\xi^{\mu}\xi^{\nu}
+\big[(k\cdot \xi)^2-k^2\xi^2\big]\zeta^{\mu}\zeta^{\nu}\Big. \notag \\
&\Big.+\big[2\,k^2+k\cdot\xi\,k\cdot\zeta\big](\xi^{\mu}\zeta^{\nu}
+\zeta^{\mu}\xi^{\nu})\Big\}=\Pi^{\mu\nu}|_{\lambda=2}\,,
\end{align}
with the right-hand side given by
\eqref{eq:polarization-sum-parity-violating-2-truncated}.
Adding the two contributions from Eqs.~\eqref{eq:Optical-theorem-parity-odd}
and \eqref{eq:Optical-theorem-parity-odd-2} leads to
\begin{align}
2\,\mathrm{Im}(\mathcal{M}_{\widehat{1}})&=2\,\mathrm{Im}(\mathcal{M}_{\widehat{1}})\big|_{\lambda=1}+2\,\mathrm{Im}(\mathcal{M}_{\widehat{1}})\big|_{\lambda=2}=
\notag \\
&=\sum_{\lambda=1,2}\int
\frac{\mathrm{d}^3k}{(2\pi)^3\,2\widetilde{\omega}_{\lambda}}\,\delta^{(4)}(k_1+k_2-k)\,\left.|\widehat{\mathcal{M}}_{\widehat{1}}|^2\right|_{\lambda}\,.
\end{align}
But the right-hand side of the previous equation is just the total cross
section of the scattering process. Hence, the optical theorem is valid for
the parity-odd sector of modified Maxwell theory. Furthermore, it reveals
the connection between the modified photon propagator (cf. Eq.
\eqref{eq:forward-scattering-amplitude-optical-theorem}) and the polarization
tensors (cf. penultimate line of Eqs. \eqref{eq:Optical-theorem-parity-odd}
and \eqref{eq:Optical-theorem-parity-odd-2}).
The optical theorem thus provides a good cross check for the obtained results
of Eqs. \eqref{eq:propagator-result-parity-odd-scalar-1-2},
\eqref{eq:polarization-sum-parity-violating-1-truncated}, and
\eqref{eq:polarization-sum-parity-violating-2-truncated}. Since the process
itself only plays a role at the level of providing a valid Ward identity,
the obtained result is consistent with having a unitary theory, at least for
a tree-level process involving conserved currents.

As a final remark we state that the unphysical pole $k^2=0$, which appears in the
propagator coefficients $\widehat{b}$, $\widehat{d}$, and $\widehat{f}$, is
prevented from being reached by energy conservation. Hence it plays no role
in the calculation.

\section{Microcausality}
\label{sec:Microcausality-parity-odd}
\setcounter{equation}{0}
\vspace*{-0mm}

In order to decide whether or not the particular case of parity-odd modified
Maxwell theory considered satisfies the condition of microcausality, we have
to compute the commutator of physical fields at different spacetime points
$y$ and $z$.  The latter can be derived from the commutator of vector
potentials:
\begin{equation}
[A^{\mu}(y),A^{\nu}(z)]=[A^{\mu}(y-z),A^{\nu}(0)]\equiv [A^{\mu}(x),A^{\nu}(0)]=\mathrm{i}\theta^{\mu\nu}\widehat{D}(x)\,,
\end{equation}
where the second step follows from translation invariance.
The tensor structure of this expression is to be put into the function
$\theta^{\mu\nu}$. The causal structure of the commutator is completely
determined by the scalar commutator function $\widehat{D}(x)$, which
corresponds to the scalar part of the Feynman propagator (see, for instance,
Refs.~\cite{AdamKlinkhamer2001} and \cite{Klinkhamer:2010zs}). For this
reason we will restrict our considerations solely to $\widehat{D}(x)$
and forget about the tensor structure. Looking at the propagator
\eqref{eq:propagator-parity-odd-coeff} of Sec. \ref{sec:Propagator-parity-odd}
it is clear that there are two scalar parts, $\widehat{K}_1$ from Eq.
\eqref{eq:propagator-result-parity-odd-scalar-1} and $\widehat{K}_2$ from
Eq. \eqref{eq:propagator-result-parity-odd-scalar-2}, one for each photon
polarization. We begin with $\widehat{K}_1$:
\begin{align}
\widehat{D}_1(x)&=\oint_{C} \frac{\mathrm{d}k_0}{2\pi}
\int\frac{\mathrm{d}^3k}{(2\pi)^3}\;\frac{2}{2(k\cdot\xi)(k\cdot\zeta)
+k^2\,(2-\xi\cdot\zeta)}
\;\exp(\mathrm{i}\,k_0x_0+\mathrm{i}\,\mathbf{k}\cdot\mathbf{x})
\notag \\[1mm]
&=\oint_{C} \frac{\mathrm{d}k_0}{2\pi} \int
\frac{\mathrm{d}^3k}{(2\pi)^3}\;
\frac{1}{k_0^2-k_{\bot}^2-k_{\parallel}^2-2\,\mathcal{E}\,k_0\,k_{\parallel}}
\;\exp(\mathrm{i}\,k_0x_0+\mathrm{i}\,\mathbf{k}\cdot\mathbf{x})
\notag \\[1mm]
&=\oint_{C} \frac{\mathrm{d}k_0}{2\pi} \int
\frac{\mathrm{d}^3k}{(2\pi)^3}\;
\frac{1}{(k_0-\widetilde{\omega}_{1}^+)(k_0-\widetilde{\omega}_{1}^{-})}
\;\exp(\mathrm{i}\,k_0x_0+\mathrm{i}\,\mathbf{k}\cdot\mathbf{x})\,,
\end{align}
where positive and negative energies are defined in Eq.
\eqref{eq:dispersion-relation-1}. These are the poles of
the scalar propagator $\widehat{K}_1$, where $\widetilde{\omega}_{1}^+$
delivers the first contribution to the imaginary part of the forward
scattering amplitude considered in the previous section.

\begin{figure}[t]
\centering
\includegraphics{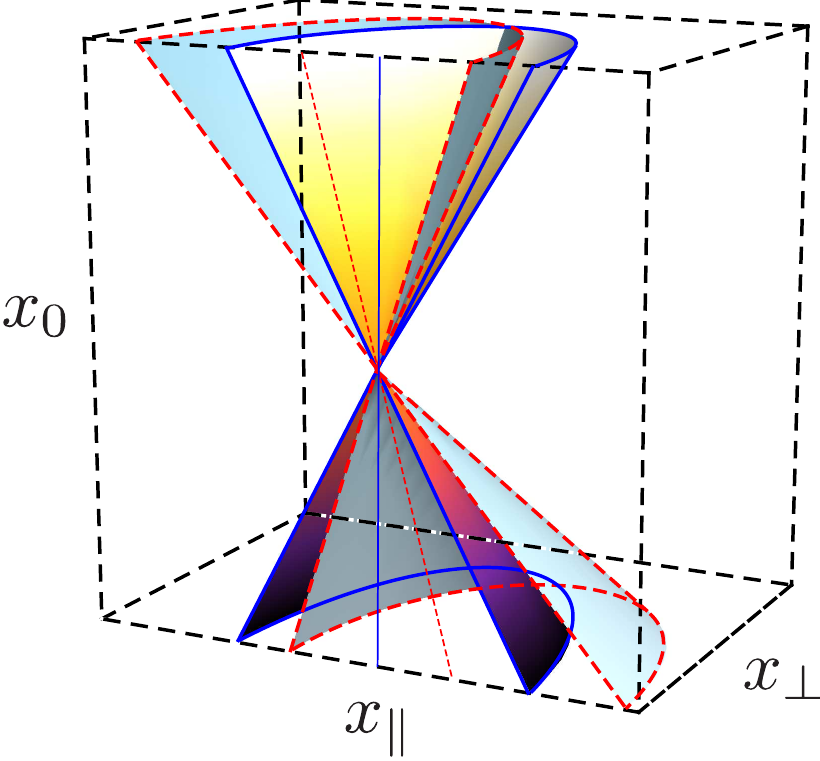}
\caption{Null cone of the standard theory (blue, solid lines) and one of the modified
null cones (red, dashed lines) in configuration space $(x_0,x_{\bot},x_{\parallel})$.
Their rotation axes are shown as well (thin lines).}
\label{fig:modified-null-cones}
\end{figure}

The evaluation of the contour integral gives
\begin{align}
\label{eq:microcausality-computation}
\widehat{D}_1(x)&=\mathrm{i}\int\frac{\mathrm{d}^3k}{(2\pi)^3}\,
\left[\frac{\exp(\mathrm{i}\,\widetilde{\omega}_{1}^+x_0)}{\widetilde{\omega}_{1}^+-\widetilde{\omega}_{1}^-}
+\frac{\exp(\mathrm{i}\,\widetilde{\omega}_{1}^-x_0)}{\widetilde{\omega}_{1}^--\widetilde{\omega}_{1}^+}
\right]
\exp(\mathrm{i}\,\mathbf{k}\cdot\mathbf{x})
\notag \\[1mm]
&=-\int
\frac{\mathrm{d}^3k}{(2\pi)^3}\,
\frac{1}{\sqrt{k_{\bot}^2+(1+\mathcal{E}^2)\,k_{\parallel}^2}}
\;\sin\left(\sqrt{k_{\bot}^2+(1+\mathcal{E}^2)\,k_{\parallel}^2}x_0\right)
\notag \\ &\hspace{3cm}\,\times
\exp\Big(\mathrm{i}\,k_{\bot}x_{\bot}
+\mathrm{i}\,(\mathcal{E}x_0+x_{\parallel})k_{\parallel}\Big)\,.
\end{align}
Substituting $k_{\parallel}=k_{\parallel}'/\sqrt{1+\mathcal{E}^2}$ the
integral results in
\begin{subequations}
\begin{equation}
\label{eq:gauge-field-commutator-spatial-integration}
\widehat{D}_1(x)=-\frac{1}{\sqrt{1+\mathcal{E}^2}}\int
\frac{\mathrm{d}^3k'}{(2\pi)^3}\;
\frac{\sin(|\mathbf{k}'|x_0)}{|\mathbf{k}'|}\;
\exp(\mathrm{i}\,\mathbf{k}'\cdot\mathbf{X})\,,
\end{equation}
with
\begin{equation}
\mathbf{X}\equiv
\begin{pmatrix}
x_{\bot} \\ 0 \\
(\mathcal{E}x_0+x_{\parallel})/\sqrt{1+\mathcal{E}^2}\\
\end{pmatrix}\,.
\end{equation}
\end{subequations}
Hence, Eq. \eqref{eq:gauge-field-commutator-spatial-integration}
is of the same form as an integral that appears in the context
of the standard propagator (see e.g. Eq. (26a) in Ref.~\cite{Heitler1954}).
This leads to the final result:
\begin{align}\label{eq:D-final-parity-odd}
\widehat{D}_1(x)&=-\frac{1}{2\pi\sqrt{1+\mathcal{E}^2}}\;
\mathrm{sgn}(x_0)\;
\delta\Big(x_0^2-2\,\mathcal{E}\,x_0\,x_{\parallel}
-(1+\mathcal{E}^2)\,x_{\bot}^2-x_{\parallel}^2\Big)
\notag \\[1mm]
&=-\frac{1}{2\pi\sqrt{1+\mathcal{E}^2}}\;
\mathrm{sgn}(x_0)\;
\delta\Big((x_0-\mathcal{E}\,x_{\parallel})^2
-(1+\mathcal{E}^2)\,x_{\bot}^2-(1+\mathcal{E}^2)\,x_{\parallel}^2\Big)\,.
\end{align}
Just as for the isotropic case of modified Maxwell theory, whose
consistency was discussed in Ref.~\cite{Klinkhamer:2010zs}, the commutator
function \eqref{eq:D-final-parity-odd} vanishes everywhere except
on the modified null cone
\begin{equation}
\label{eq:modified-null-cone-1}
(x_0-\mathcal{E}\,x_{\parallel})^2
-(1+\mathcal{E}^2)\,x_{\bot}^2-(1+\mathcal{E}^2)\,x_{\parallel}^2=0\,.
\end{equation}
An analogous calculation for the scalar part $\widehat{K}_2$ from
Eq. \eqref{eq:propagator-result-parity-odd-scalar-2} delivers the
following final result for the commutator function $\widehat{D}_2(x)$:
\begin{equation}
\widehat{D}_2(x)=-\frac{1}{2\pi (1+\mathcal{E})^{3/2}}\,\mathrm{sgn}(x_0)\,\delta\Big((x_0-\mathcal{E}\,x_{\parallel})^2-x_{\bot}^2-(1+\mathcal{E}^2)\,x_{\parallel}^2\Big)\,,
\end{equation}
which corresponds to a second modified null cone:
\begin{equation}
\label{eq:modified-null-cone-2}
(x_0-\mathcal{E}\,x_{\parallel})^2-x_{\bot}^2-(1+\mathcal{E}^2)\,x_{\parallel}^2=0\,.
\end{equation}
Both null cones coincide to linear order in $\mathcal{E}$. This is not surprising,
since the theory is birefringent to quadratic order in the Lorentz-violating
parameters. Each of the Eqs. \eqref{eq:modified-null-cone-1} and \eqref{eq:modified-null-cone-2}
corresponds to a null cone, whose rotation axis is different for the past and future
null cone. Neither axes coincides with the time axis, but each is rotated by a
small angle, as shown in Fig. \ref{fig:modified-null-cones}.
Since there are two modes with two different dispersion relations,
one may wonder, if this result is sufficient for taking a decision
about microcausality. For this reason we tried to separate both
modes in Sec. \ref{sec:Propagator-parity-odd} with the result
\eqref{eq:split-propagator} -- \eqref{eq:split-propagator-scalar-parts}.
Therefore, we should investigate $\widehat{G}^{(1)}(k)$ and
$\widehat{G}^{(2)}(k)$ from Eq. \eqref{eq:split-propagator-scalar-parts}:
\begin{align}
\widehat{G}^{(1)}(x)&\equiv \oint_C \frac{\mathrm{d}k_0}{2\pi} \int \frac{\mathrm{d}^3k}{(2\pi)^3} \widehat{G}^{(1)}(k) \notag \\
&=\oint_C \frac{\mathrm{d}k_0}{2\pi} \int \frac{\mathrm{d}^3k}{(2\pi)^3}\,\frac{4\widehat{K}_1\widehat{K}_2^{-1}}{[(k\cdot\xi)^2-k^2]\zeta^2+(k\cdot\zeta)^2}\exp(\mathrm{i}\,k_0x_0+\mathrm{i}\,\mathbf{k}\cdot\mathbf{x}) \notag \\
&=\oint_C \frac{\mathrm{d}k_0}{2\pi} \int \frac{\mathrm{d}^3k}{(2\pi)^3}\,\frac{4\widehat{K}_2^{-1}}{[(k\cdot\xi)^2-k^2]\zeta^2+(k\cdot\zeta)^2}\frac{\exp(\mathrm{i}\,k_0x_0+\mathrm{i}\,\mathbf{k}\cdot\mathbf{x})}{(k_0-\widetilde{\omega}_1^+)(k_0-\widetilde{\omega}_1^-)}\,.
\end{align}
Using
\begin{equation}
\left.\frac{4\widehat{K}_2^{-1}}{[(k\cdot\xi)^2-k^2]\zeta^2+(k\cdot\zeta)^2}\right|_{k_0=\widetilde{\omega}_1^+}=\left.\frac{4\widehat{K}_2^{-1}}{[(k\cdot\xi)^2-k^2]\zeta^2+(k\cdot\zeta)^2}\right|_{k_0=\widetilde{\omega}_1^-}=1\,,
\end{equation}
then leads to the intermediate result of Eq. \eqref{eq:microcausality-computation}
and the rest of the computation is the same. Since $\widehat{G}^{(2)}(k)$ is a constant
function with respect to $k_0$, the evaluation of the contour integral in the complex
$k_0$-plane in
\begin{equation}
\widehat{G}^{(2)}(x)\equiv \oint_C \frac{\mathrm{d}k_0}{2\pi} \int \frac{\mathrm{d}^3k}{(2\pi)^3} \widehat{G}^{(2)}(k)\,,
\end{equation}
will immediately give zero. Hence, the dispersion relation corresponding to
the second mode does not seem to play any role here. The $\lambda=1$ mode
seems to be preferred compared to the $\lambda=2$ mode, what follows
from forcing a parity-odd theory to be nonbirefringent via the \textit{Ansatz}
\eqref{eq:nonbirefringent-Ansatz}. The transversal polarization vectors can
be interpreted as two distinct polarization modes: left- and right-handed.
In a parity-violating theory they are expected to behave differently, for
example with respect to their phase velocity. This would automatically
lead to birefringence, which is suppressed by using Eq. \eqref{eq:nonbirefringent-Ansatz}
as a basis.

The result of Eq. \eqref{eq:D-final-parity-odd} establishes microcausality
for the following parameter domain:
\begin{equation}\label{eq:mathcalE-domain}
\mathcal{E} \equiv |\boldsymbol{\zeta}| \in  [0,\,\infty)\,,
\end{equation}
where $\boldsymbol{\zeta}$ is defined in terms of the SME
parameters by Eqs.~\eqref{eq:definition-parity-odd-case-four-vectors}
and \eqref{eq:parity-odd-case-SME}. Hence, the parity-odd ``nonbirefringent''
sector of modified Maxwell theory is unitary and microcausal for the full
parameter range.

\section{Comparison to other Lorentz-violating theories}
\label{sec:Propagator-polarizations-general-considerations}
\setcounter{equation}{0}

In the previous sections we have seen that both the modified photon propagator and the
polarization vectors have an uncommon structure.
For this reason, we want to have a general look at the photon propagator and polarization
vectors in other Lorentz-violating theories. We start with the photon polarizations of
MCS theory. Besides modified Maxwell theory, MCS theory is another possible example of a
gauge-invariant and power-counting renormalizable theory that violates Lorentz invariance
in the photon sector.
MCS theory is characterized by a mass scale $m_{\scriptscriptstyle{\mathrm{CS}}}$ and a
fixed spacelike\footnote{We assume the four-vector $\zeta^{\mu}$ to be spacelike, since
timelike MCS theory is expected to be nonunitary and noncausal \cite{AdamKlinkhamer2001}.}
``four-vector'' $\zeta^{\mu}$, that plays the role of a background field.
The Chern--Simons mass $m_{\scriptscriptstyle{\mathrm{CS}}}$ gives the amount of
Lorentz violation. MCS theory exhibits two photon modes, which we call `$\oplus$' and
`$\ominus$'. They obey different dispersion relations \cite{Carroll-etal1990}, which results
in birefringence. The polarization vectors follow from the field equations and, in temporal
gauge $A^0=0$, they are given by
\begin{subequations}
\begin{equation}
(\varepsilon^{\oplus\,\mu})=\frac{1}{\sqrt{N'}}\left(0\,,\,-2[(k\cdot \zeta)^2+k^2]\,,\,\mathrm{i}\frac{2\omega}{m_{\mathrm{\scriptscriptstyle{CS}}}}k^2\,,\,2k_{\parallel}k_{\bot}\right)\Bigg|_{\omega=\omega_{\oplus}}\,,
\end{equation}
\begin{equation}
(\varepsilon^{\ominus\,\mu})=\frac{1}{\sqrt{N''}}\left(0\,,\,-2[(k\cdot \zeta)^2+k^2]\,,\,\mathrm{i}\frac{2\omega}{m_{\mathrm{\scriptscriptstyle{CS}}}}k^2\,,\,2k_{\parallel}k_{\bot}\right)\Bigg|_{\omega=\omega_{\ominus}}\,,
\end{equation}
with the normalization constants
\begin{equation}
N'=4\frac{\omega^2k^2}{m^2_{\mathrm{\scriptscriptstyle{CS}}}}[2k^2+m^2_{\mathrm{\scriptscriptstyle{CS}}}\zeta^2]\Big|_{\omega=\omega_{\oplus}}\,,\quad N''=4\frac{\omega^2k^2}{m^2_{\mathrm{\scriptscriptstyle{CS}}}}[2k^2+m^2_{\mathrm{\scriptscriptstyle{CS}}}\zeta^2]\Big|_{\omega=\omega_{\ominus}}\,.
\end{equation}
\end{subequations}
Using the temporal gauge fixing four-vector $(n^{\mu})=(1,0,0,0)$, the polarization tensor for each of
the two modes can be cast in the following form (see \cite{Kaufhold:2005vj} for the truncated versions):
\begin{subequations}
\label{eq:polarization-tensors-mcs-theory}
\begin{equation}
\overline{\varepsilon}^{\oplus\,\mu}(k)\varepsilon^{\oplus\,\nu}(k)=\Pi^{\mu\nu}_{\mathrm{MCS}}\big|_{\omega=\omega_{\oplus}}\,,
\end{equation}
\begin{equation}
\overline{\varepsilon}^{\ominus\,\mu}(k)\varepsilon^{\ominus\,\nu}(k)=\Pi^{\mu\nu}_{\mathrm{MCS}}\big|_{\omega=\omega_{\ominus}}\,,
\end{equation}
where
\begin{align}
\Pi^{\mu\nu}_{\mathrm{MCS}}&=\frac{1}{2k^2+m_{\mathrm{\scriptscriptstyle{CS}}}^2\zeta^2}\left[-\,k^2\eta^{\mu\nu}-\frac{k^2}{(k\cdot n)^2}k^{\mu}k^{\nu}+\frac{k^2}{k\cdot n}(k^{\mu}n^{\nu}+n^{\mu}k^{\nu})\right. \notag \\
&\phantom{{}={}\frac{1}{2k^2+m_{\mathrm{\scriptscriptstyle{CS}}}^2\zeta^2}\Big[}\left.-\,m_{\mathrm{\scriptscriptstyle{CS}}}^2\zeta^{\mu}\zeta^{\nu}-\mathrm{i}m_{\mathrm{\scriptscriptstyle{CS}}}\varepsilon^{\mu\nu\varrho\sigma}\left(\frac{k\cdot\zeta}{k\cdot n}k_{\varrho}n_{\sigma}-\frac{k^2}{k\cdot n}\zeta_{\varrho}n_{\sigma}\right)\right]\,.
\end{align}
\end{subequations}
The polarization sum of standard QED is expected to be recovered for vanishing
$m_{\scriptscriptstyle{\mathrm{CS}}}$. For the truncated polarization sum this is,
indeed, the case:
\begin{equation}
\lim_{m_{\scalebox{0.4}{CS}}\mapsto 0} \left.\Big\{\overline{\varepsilon}^{\oplus\,\mu}(k)\varepsilon^{\oplus\,\nu}(k)+
\overline{\varepsilon}^{\ominus\,\mu}(k)\varepsilon^{\ominus\,\nu}(k)\Big\}\right|^{\text{truncated}}=-\eta^{\mu\nu}\,.
\end{equation}
From
\begin{equation}
\label{eq:polarization-tensors-mcs-theory-limit}
\lim_{m_{\scalebox{0.4}{CS}}\mapsto 0} \overline{\varepsilon}^{\oplus\,\mu}(k)\varepsilon^{\oplus\,\nu}(k)
\,\Big|^{\text{truncated}}=\lim_{m_{\scalebox{0.4}{CS}}\mapsto 0}\overline{\varepsilon}^{\ominus\,\mu}(k)\varepsilon^{\ominus\,\nu}(k)
\,\Big|^{\text{truncated}}=-\frac{\eta^{\mu\nu}}{2}\,,
\end{equation}
it is evident that both modes deliver equal contributions to the polarization
sum. This even holds for nonvanishing $m_{\scriptscriptstyle{\mathrm{CS}}}$. Hence,
the behavior of MCS theory with respect to the polarization modes is completely
different compared to parity-odd nonbirefringent modified Maxwell theory. For
$m_{\scriptscriptstyle{\mathrm{CS}}}\mapsto 0$, there is no residual dependence
from the preferred spacetime direction $\zeta^{\mu}$ in the polarization tensors
of the individual modes, which can be seen from Eq.~\eqref{eq:polarization-tensors-mcs-theory-limit}.

Furthermore, for MCS-theory the photon propagator in the axial gauge has been shown
to be of the following form \cite{AdamKlinkhamer2001}:
\begin{equation}
G_{\mu\nu}(k)\Big|^{\mathrm{axial}}_{\mathrm{MCS}}=-\mathrm{i}\frac{k^2}{\mathscr{P}(k)}\left(\eta_{\mu\nu}+\hdots\right)\,,
\end{equation}
where further terms with the index structure composed of the four-momentum, the preferred
spacelike four-vector $\zeta^{\mu}$, the axial gauge vector and the four-dimensional
Levi-Civita symbol have been omitted.
The denominator $\mathscr{P}(k)$ is a fourth-order polynomial in $k^0$, with its
zeros corresponding to the two different physical dispersion relations. For a special case
of parity-odd `birefringent' modified Maxwell theory\footnote{with nonzero
parity-odd parameters $\kappa^{0213}$, $\kappa^{0123}$
(corresponding to the first two entries of the ten-dimensional vector from Eq.~(8)
in \cite{KosteleckyMewes2002}) plus those related by symmetries and all others set
 to zero} we could show that the propagator in Feynman gauge looks like
\begin{equation}
G_{\mu\nu}(k)\Big|^{\mathrm{Feynman}}_{\substack{\mathrm{birefringent} \\
\mathrm{modMax}\hfill}}=-\mathrm{i}\frac{\mathscr{P}_1(k)}{\mathscr{P}_2(k)}\left(\eta_{\mu\nu}+\hdots\right)\,,
\end{equation}
where $\mathscr{P}_1(k)$ is a second-order polynomial in $k^0$, involving the
Lorentz-violating parameters and $\mathscr{P}_2(k)$ is of fourth order in $k^0$.
The two distinct physical dispersion relations of this birefringent theory follow
from $\mathscr{P}_2(k)=0$. Again, remaining propagator coefficients multiplied
by combinations of the four-momentum and preferred four-vectors have been omitted.

Hence, we see that our result for the propagator for parity-odd nonbirefringent
modified Maxwell theory given by Eqs. \eqref{eq:propagator-parity-odd-coeff} --
\eqref{eq:propagator-result-parity-odd-coeff} is rather unusual. For MCS-theory
and birefringent modified Maxwell theory (at least for the special case examined),
both physical modes emerge as poles of the coefficient before the metric tensor
$\eta_{\mu\nu}$. However, in the case of parity-odd nonbirefringent modified Maxwell
theory, the dispersion relation for the $\lambda=2$ polarization mode is not contained
in the coefficient $\widehat{K}_1$ of Eq. \eqref{eq:propagator-result-parity-odd-scalar-1},
which is multiplied with $\eta_{\mu\nu}$. This peculiarity is also mirrored in the
polarization tensors, where we have shown the interplay in the previous section.

\section{Limit of the polarization tensors for vanishing Lorentz violation}
\label{sec:Limit-polarizations-vanishing-lorentz-violation}
\setcounter{equation}{0}

Taking the limit $\mathcal{E}\mapsto 0$ followed by the limit
$k_{\bot}\mapsto 0$ (see the definition \eqref{eq:definition-k-orthogonal-k-parallel})
for the physical polarization vectors \eqref{eq:polarization-mode1-parity-violating} and
\eqref{eq:polarization-mode2-parity-violating} leads to:
\begin{equation}
\label{eq:polarization-vectors-vanishing-lorentz-violation}
\lim_{\substack{\mathcal{E}\mapsto 0 \\ k_{\bot}\mapsto 0}} (\varepsilon^{(1)\,\mu})=\begin{pmatrix}
0 \\
0 \\
1 \\
0 \\
\end{pmatrix}\,,\quad \lim_{\substack{\mathcal{E}\mapsto 0 \\ k_{\bot}\mapsto 0}} (\varepsilon^{(2)\,\mu})=\begin{pmatrix}
0 \\
1 \\
0 \\
0 \\
\end{pmatrix}\,.
\end{equation}
Taking into account the limit of the four-momentum,
\begin{equation}
\lim_{k_{\bot}\mapsto 0} \begin{pmatrix}
k_{\bot} \\
0 \\
k_{\parallel} \\
\end{pmatrix}=\begin{pmatrix}
0 \\
0 \\
k_{\parallel} \\
\end{pmatrix}\equiv \begin{pmatrix}
0 \\
0 \\
k \\
\end{pmatrix}\,,
\end{equation}
the physical polarization vectors reduce to the standard
transversal QED results.
Note that for both vectors in Eq.
\eqref{eq:polarization-vectors-vanishing-lorentz-violation},
the order in which the limits are taken does not play any role.
As we will see below, this is not the case for the gauge-invariant
parts of the polarization tensors from
Eqs.~\eqref{eq:polarization-sum-parity-violating-1-truncated},
\eqref{eq:polarization-sum-parity-violating-2-truncated},
that is, if the polarization vectors are coupled to conserved
currents. For $\mathcal{E}\mapsto 0$ these tensors result in:
\begin{subequations}
\label{eq:polarization-tensors-physical-limit}
\begin{align}
\lim_{\mathcal{E}\mapsto 0} \Pi^{\mu\nu}|_{\lambda=1}=-\eta^{\mu\nu}-\frac{k_{\parallel}^2}{k_{\bot}^2}\xi^{\mu}\xi^{\nu}-\frac{|\mathbf{k}|^2}{k_{\bot}^2}\widehat{\zeta}^{\mu}\widehat{\zeta}^{\nu}-\frac{|\mathbf{k}|k_{\parallel}}{k_{\bot}^2}(\xi^{\mu}\widehat{\zeta}^{\nu}+\xi^{\nu}\widehat{\zeta}^{\mu})\,,
\end{align}
\begin{align}
\lim_{\mathcal{E}\mapsto 0} \Pi^{\mu\nu}|_{\lambda=2}=\frac{k_{\parallel}^2}{k_{\bot}^2}\xi^{\mu}\xi^{\nu}+\frac{|\mathbf{k}|^2}{k_{\bot}^2}\widehat{\zeta}^{\mu}\widehat{\zeta}^{\nu}+\frac{|\mathbf{k}|k_{\parallel}}{k_{\bot}^2}(\xi^{\mu}\widehat{\zeta}^{\nu}+\xi^{\nu}\widehat{\zeta}^{\mu})\,,
\end{align}
\end{subequations}
with $(\widehat{\zeta}^{\mu})=(0,\widehat{\boldsymbol{\zeta}})$ and
$|\mathbf{k}|=(k_{\scriptscriptstyle{\parallel}}^2+k_{\bot}^2)^{1/2}$.
For completeness, after inserting the explicit four-vectors, we obtain
the following matrices:
\begin{subequations}
\begin{align}
\label{eq:polarization-tensor-1-vanishing-E}
\lim_{\mathcal{E}\mapsto 0} \big(\Pi^{\mu\nu}|_{\lambda=1}\big)&=\frac{1}{k_{\bot}^2}\begin{pmatrix}
-|\mathbf{k}|^2 & 0 & 0 & -|\mathbf{k}|k_{\parallel} \\
0 & k_{\bot}^2 & 0 & 0 \\
0 & 0 & k_{\bot}^2 & 0 \\
-|\mathbf{k}|k_{\parallel} & 0 & 0 & -k_{\parallel}^2 \\
\end{pmatrix}\,,
\end{align}
\begin{equation}
\label{eq:polarization-tensor-2-vanishing-E}
\lim_{\mathcal{E}\mapsto 0} \big(\Pi^{\mu\nu}|_{\lambda=2}\big)=\frac{1}{k_{\bot}^2}\begin{pmatrix}
k_{\parallel}^2 & 0 & 0 & |\mathbf{k}|k_{\parallel} \\
0 & 0 & 0 & 0 \\
0 & 0 & 0 & 0 \\
|\mathbf{k}|k_{\parallel} & 0 & 0 & |\mathbf{k}|^2 \\
\end{pmatrix}\,.
\end{equation}
\end{subequations}
Note that these matrix representations only hold for the special choice
$\widehat{\boldsymbol{\zeta}}=(0,0,1)$. It is evident that
the additional limit $k_{\bot}\mapsto 0$ does not exist for each contribution
$\Pi^{\mu\nu}|_{\lambda=1}$ or $\Pi^{\mu\nu}|_{\lambda=2}$ separately, but only
for the truncated polarization sum $\sum_{\lambda=1,2} \Pi^{\mu\nu}|_{\lambda}$,
which leads to the standard QED result.
For this reason, the polarization vectors are not only deformed ---
unlike for the isotropic case that was examined in Ref.~\cite{Klinkhamer:2010zs}
--- but their structure completely differs from standard QED. Besides that,
no covariant expression exists for each polarization tensor in standard QED,
where only the sum $\sum_{\lambda=1,2} \Pi^{\mu\nu}|_{\lambda}$ can be decomposed
covariantly.

\section{Physical process: Compton scattering with polarized photons}
\label{sec:Compton-scattering}
\setcounter{equation}{0}

\subsection{Description of the process}
\label{subsec:description-compton-process}

The results obtained for the polarization vectors in
Sec.~\ref{sec:Polarization-vectors-parity-odd} together with the observations
that followed forces us to think about the consistency of the modified theory.
The form of the propagator, the polarization vectors and tensors observed in
Secs. \ref{sec:Propagator-parity-odd}, \ref{sec:Polarization-vectors-parity-odd},
and \ref{sec:Limit-polarizations-vanishing-lorentz-violation} reveal the following
uncommon properties:
\begin{itemize}

\item[1)] one of the two physical photon modes seems to be preferred with
respect to the other,
\item[2)] both polarization vectors are interweaved with the spacetime
directions $\xi^{\mu}$ and $\zeta^{\mu}$, even for vanishing Lorentz-violating
parameters,
\item[3)] each physical polarization tensor can be written in covariant form,
\item[4)] and one of the physical polarization vectors has a longitudinal
part.

\end{itemize}
One the one hand, these peculiarities may emerge from the fact that a parity-odd
QED is combined with the claim of being nonbirefringent. Two physical photon
polarizations can be interpreted as two distinct polarization modes:``left-handed''
and ``right-handed''. These are supposed to behave differently because
of parity violation, for example with respect to the phase velocity
of each mode. Hence, birefringence would result from this, which clashes
with the nonbirefringent \textit{Ansatz} of Eq. \eqref{eq:nonbirefringent-Ansatz}.

\begin{figure}[t]
\centering
\includegraphics{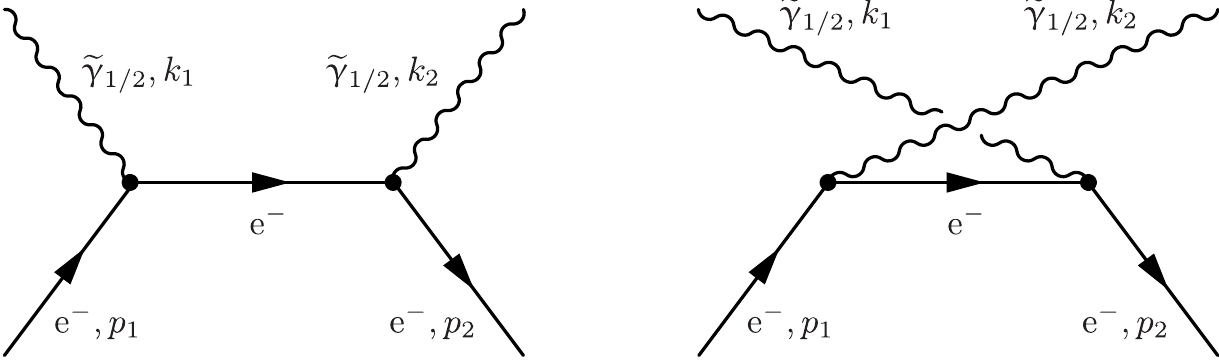}
\caption{Contributions to Compton scattering for polarized modified
photons $\widetilde{\gamma}_{1,2}$, where the subscript refers to the
photon polarization. The photon momenta are denoted as $k_a$ and the
electron momenta as $p_a$, where the label $a=1$, 2 refers to the
initial and final state, respectively.}
\label{fig:compton-scattering}
\end{figure}

On the other hand, the above properties may have emerged from a bad gauge
choice and could possibly be removed by picking a more appropriate gauge.
For this reason a physical process will be considered, whose cross
section does not depend on the gauge. If the mentioned behavior of the
polarization modes is not a gauge artifact, it will show up in the
results for polarized cross sections. The simplest tree-level process
involving external photons, which also occurs in standard QED, is Compton
scattering. We consider an electron scattered off a photon in the
$\lambda=1$ polarization and in the $\lambda=2$ polarization, respectively.
Hence, we want to compute cross sections for the processes
$e^-(p_1)\widetilde{\gamma}_1(k_1)\rightarrow e^-(p_2)\widetilde{\gamma}_1(k_2)$,
$e^-(p_1)\widetilde{\gamma}_1(k_1)\rightarrow e^-(p_2)\widetilde{\gamma}_2(k_2)$,
$e^-(p_1)\widetilde{\gamma}_2(k_1)\rightarrow e^-(p_2)\widetilde{\gamma}_1(k_2)$,
and $e^-(p_1)\widetilde{\gamma}_2(k_1)\rightarrow e^-(p_2)\widetilde{\gamma}_2(k_2)$,
where $\widetilde{\gamma}_{1,2}$ denotes a modified photon in the
$\lambda=1$ or $\lambda=2$ polarization state, respectively. The corresponding Feynman
diagrams are shown in Fig. \ref{fig:compton-scattering}.

For a review of Compton scattering experiments, refer to
Ref.~\cite{Fluegge1958}. Furthermore, Ref.~\cite{Bocquet:2010ke} gives a new bound
on two of the three parameters of parity-odd nonbirefringent modified Maxwell
theory from the study of Compton scattering kinematics at the GRAAL experiment\footnote{
whereas the experiment has been stopped by now} on the European Synchrotron
Radiation Facility (ESRF) at Grenoble in France.

\subsection{Numerical results for polarized Compton scattering cross sections}
\label{subsec:numerical-results-compton-scattering}

We choose special momenta $p_1$, $k_1$ for the initial electron and photon.
The outgoing photon momentum configuration is described in spherical coordinates
with polar angle $\vartheta$ and azimuthal angle $\varphi$. We consider the initial
momentum configuration, for which the electron is at rest:
$(p_1^{\mu})=(m,0,0,0)$, $(k_{1,\lambda}^{\mu})\equiv (\omega_{\lambda}(k),\mathbf{k})$
with $\mathbf{k}=(0,0,|\mathbf{k}|)=(0,0,k_1)$.
\begin{table}[b]
\centering
\setlength{\extrarowheight}{4pt}
\newcolumntype{C}[1]{>{\centering\arraybackslash}m{#1}}
\begin{tabular}{C{0.85cm}||C{2.1cm}|C{2.1cm}|C{2.1cm}|C{2.1cm}|C{2.1cm}C{0.1cm}}
\hline
\multirow{1}{*}{$k_1/m$} & \multirow{1}{*}{$10^{-1}$} & \multirow{1}{*}{$10^{-2}$} & \multirow{1}{*}{$10^{-3}$} & \multirow{1}{*}{$10^{-4}$} & \multirow{1}{*}{$10^{-5}$} \\
\hline
\multirow{1}{*}{$\sigma$} & \multirow{1}{*}{7.048378} & \multirow{1}{*}{8.214276} & \multirow{1}{*}{8.360869} & \multirow{1}{*}{8.375905} & \multirow{1}{*}{8.377413} \\
\hline
\hline
\multirow{1}{*}{$k_1/m$} & \multirow{1}{*}{$10^{-6}$} & \multirow{1}{*}{$10^{-7}$} & \multirow{1}{*}{$10^{-8}$} & \multirow{1}{*}{$10^{-9}$} & \multirow{1}{*}{$10^{-10}$} \\
\hline
\multirow{1}{*}{$\sigma$} & \multirow{1}{*}{8.377564} & \multirow{1}{*}{8.377579} & \multirow{1}{*}{8.377580} & \multirow{1}{*}{8.377580} & \multirow{1}{*}{8.377580} \\
\hline
\end{tabular}
\caption{Unpolarized total Compton scattering cross sections $\sigma$ in standard QED for different values of the initial photon
momentum $k_1$ according to the equation below (5-114) of \cite{ItzyksonZuber1980} or to Eq. (5.81) of \cite{PeskinSchroeder1995}.
The results are given in units of $\alpha^2$ with the fine structure constant $\alpha\equiv e^2/4\pi$. The electron mass is set to
$m=1$.}
\label{tab:Compton-scattering-standard-qed-1}
\end{table}
\begin{table}[t]
\centering
\setlength{\extrarowheight}{4pt}
\newcolumntype{C}[1]{>{\centering\arraybackslash}m{#1}}
\begin{tabular}{C{0.95cm}||C{1.6cm}|C{1.6cm}|C{1.6cm}|C{1.6cm}|C{1.6cm}|C{1.6cm}|C{2.0cm}C{0.1cm}}
\hline
\multirow{1}{*}{$k_1/m$} & \multirow{1}{*}{$\sigma_{11}$} & \multirow{1}{*}{$\sigma_{22}$} & \multirow{1}{*}{$\sigma_{12}$} & \multirow{1}{*}{$\sigma_{21}$} &
\multirow{1}{*}{$\sum_{\lambda'} \sigma_{1\lambda'}$} & \multirow{1}{*}{$\sum_{\lambda'} \sigma_{2\lambda'}$} & \multirow{1}{*}{$\frac{1}{2}\sum_{\lambda,\lambda'}\sigma_{\lambda\lambda'}$} \\
\hline
\hline
\multirow{1}{*}{$10^{-1}$} & \multirow{1}{*}{1.784650} & \multirow{1}{*}{5.263729} & \multirow{1}{*}{5.263729} & \multirow{1}{*}{1.784650} & \multirow{1}{*}{7.048379} & \multirow{1}{*}{7.048379} & \multirow{1}{*}{7.048379} \\
\hline
\multirow{1}{*}{$10^{-2}$} & \multirow{1}{*}{2.053890} & \multirow{1}{*}{6.160386} & \multirow{1}{*}{6.160386} & \multirow{1}{*}{2.053890} & \multirow{1}{*}{8.214277} & \multirow{1}{*}{8.214277} & \multirow{1}{*}{8.214277} \\
\hline
\multirow{1}{*}{$10^{-3}$} & \multirow{1}{*}{2.090221} & \multirow{1}{*}{6.270648} & \multirow{1}{*}{6.270648} & \multirow{1}{*}{2.090221} & \multirow{1}{*}{8.360869} & \multirow{1}{*}{8.360869} & \multirow{1}{*}{8.360869} \\
\hline
\multirow{1}{*}{$10^{-4}$} & \multirow{1}{*}{2.093976} & \multirow{1}{*}{6.281929} & \multirow{1}{*}{6.281929} & \multirow{1}{*}{2.093976} & \multirow{1}{*}{8.375905} & \multirow{1}{*}{8.375905} & \multirow{1}{*}{8.375905} \\
\hline
\multirow{1}{*}{$10^{-5}$} & \multirow{1}{*}{2.094353} & \multirow{1}{*}{6.283060} & \multirow{1}{*}{6.283060} & \multirow{1}{*}{2.094353} & \multirow{1}{*}{8.377413} & \multirow{1}{*}{8.377413} & \multirow{1}{*}{8.377413} \\
\hline
\end{tabular}
\caption{Polarized Compton scattering cross sections $\sigma_{\lambda\lambda'}$ (where $\lambda$ denotes the initial and $\lambda'$
the final photon polarization) of standard QED for different values of the initial photon momentum $k_1$ according to Eq. (11-13)
of \cite{JauchRohrlich1976}. The cross sections are given in units of $\alpha^2$ and the electron mass is set to $m=1$.}
\label{tab:Compton-scattering-standard-qed-2}
\end{table}
\begin{table}[t]
\centering
\setlength{\extrarowheight}{4pt}
\newcolumntype{C}[1]{>{\centering\arraybackslash}m{#1}}
\begin{tabular}{C{1.0cm}||C{1.6cm}|C{1.6cm}|C{1.6cm}|C{1.6cm}|C{1.6cm}|C{1.6cm}|C{1.6cm}C{0.1cm}}
\hline
\multirow{1}{*}{$\widetilde{\kappa}$} & \multirow{1}{*}{$\widetilde{\sigma}_{11}$} & \multirow{1}{*}{$\widetilde{\sigma}_{22}$} & \multirow{1}{*}{$\widetilde{\sigma}_{12}$} & \multirow{1}{*}{$\widetilde{\sigma}_{21}$} & \multirow{1}{*}{$\widetilde{\sigma}_{1X}$} & \multirow{1}{*}{$\widetilde{\sigma}_{2X}$} & \multirow{1}{*}{$\widetilde{\sigma}$} \\
\hline
\hline
\multirow{1}{*}{$10^{-1}$} & \multirow{1}{*}{6.283180} & \multirow{1}{*}{5.526582} & \multirow{1}{*}{2.033397} & \multirow{1}{*}{2.890066} & \multirow{1}{*}{8.316577} & \multirow{1}{*}{8.416647} & \multirow{1}{*}{8.366612} \\
          & \multirow{1}{*}{6.283180} & \multirow{1}{*}{5.526586} & \multirow{1}{*}{2.033399} & \multirow{1}{*}{2.890066} & \multirow{1}{*}{8.316579} & \multirow{1}{*}{8.416652} & \multirow{1}{*}{8.366615} \\
\hline
\multirow{1}{*}{$10^{-2}$} & \multirow{1}{*}{6.283180} & \multirow{1}{*}{6.200243} & \multirow{1}{*}{2.093772} & \multirow{1}{*}{2.177939} & \multirow{1}{*}{8.376952} & \multirow{1}{*}{8.378182} & \multirow{1}{*}{8.377567} \\
          & \multirow{1}{*}{6.283180} & \multirow{1}{*}{6.200245} & \multirow{1}{*}{2.093772} & \multirow{1}{*}{2.177939} & \multirow{1}{*}{8.376952} & \multirow{1}{*}{8.378184} & \multirow{1}{*}{8.377568} \\
\hline
\multirow{1}{*}{$10^{-4}$} & \multirow{1}{*}{6.283180} & \multirow{1}{*}{6.282133} & \multirow{1}{*}{2.094338} & \multirow{1}{*}{2.095235} & \multirow{1}{*}{8.377518} & \multirow{1}{*}{8.377367} & \multirow{1}{*}{8.377443} \\
          & \multirow{1}{*}{6.283180} & \multirow{1}{*}{6.282346} & \multirow{1}{*}{2.094400} & \multirow{1}{*}{2.095235} & \multirow{1}{*}{8.377580} & \multirow{1}{*}{8.377581} & \multirow{1}{*}{8.377581} \\
\hline
\multirow{1}{*}{$10^{-8}$} & \multirow{1}{*}{6.283175} & \multirow{1}{*}{6.282743} & \multirow{1}{*}{2.094270} & \multirow{1}{*}{2.094397} & \multirow{1}{*}{8.377444} & \multirow{1}{*}{8.377140} & \multirow{1}{*}{8.377292} \\
          & \multirow{1}{*}{6.283180} & \multirow{1}{*}{6.283184} & \multirow{1}{*}{2.094400} & \multirow{1}{*}{2.094397} & \multirow{1}{*}{8.377580} & \multirow{1}{*}{8.377581} & \multirow{1}{*}{8.377580} \\
\hline
\multirow{1}{*}{$10^{-16}$} & \multirow{1}{*}{6.283180} & \multirow{1}{*}{6.283258} & \multirow{1}{*}{2.094400} & \multirow{1}{*}{2.094397} & \multirow{1}{*}{8.377581} & \multirow{1}{*}{8.377655} & \multirow{1}{*}{8.377618} \\
           & \multirow{1}{*}{6.283180} & \multirow{1}{*}{6.283184} & \multirow{1}{*}{2.094400} & \multirow{1}{*}{2.094397} & \multirow{1}{*}{8.377581} & \multirow{1}{*}{8.377581} & \multirow{1}{*}{8.377581} \\
\hline
\end{tabular}
\caption{Compton scattering cross sections $\widetilde{\sigma}_{\lambda\lambda'}$
for polarized modified photons, where $\lambda$ is the initial and
$\lambda'$ the final photon mode. The sixth and seventh columns give the combinations
defined by Eq. \eqref{eq:modified-compton-cross-sections-gauge-invariant-sums} and
the eighth column lists the sum of the four cross sections $\widetilde{\sigma}_{\lambda\lambda'}$,
which is averaged over the initial photon polarizations and corresponds to the cross
section of unpolarized modified Compton scattering from Eq. \eqref{eq:total-modified-cross-section}.
The electrons are assumed to be unpolarized. All results are given in units of $\alpha^2$
and $k_1=10^{-10}m$ is used. Moreover, we set the electron mass $m=1$. The
Lorentz-violating parameter $\widetilde{\kappa}$ can be found in the first column. For each
Lorentz-violating parameter we give both the results that follow from Eq.
\eqref{eq:matrix-element-square-x} (first row) and from Eq. \eqref{eq:matrix-element-square-hat-x}
(second row), respectively.}
\label{tab:Compton-scattering-results1}
\end{table}
\begin{table}[h]
\centering
\setlength{\extrarowheight}{4pt}
\newcolumntype{C}[1]{>{\centering\arraybackslash}m{#1}}
\begin{tabular}{C{0.95cm}||C{1.6cm}|C{1.6cm}|C{1.6cm}|C{1.6cm}|C{1.6cm}|C{1.6cm}|C{1.6cm}C{0.1cm}}
\hline
\multirow{1}{*}{$k_1/m$} & \multirow{1}{*}{$\widetilde{\sigma}_{11}$} & \multirow{1}{*}{$\widetilde{\sigma}_{22}$} & \multirow{1}{*}{$\widetilde{\sigma}_{12}$} & \multirow{1}{*}{$\widetilde{\sigma}_{21}$} &
\multirow{1}{*}{$\widetilde{\sigma}_{1X}$} & \multirow{1}{*}{$\widetilde{\sigma}_{2X}$} & \multirow{1}{*}{$\widetilde{\sigma}$} \\
\hline
\hline
\multirow{1}{*}{$10^{-1}$} & \multirow{1}{*}{5.278215} & \multirow{1}{*}{5.280137} & \multirow{1}{*}{1.770163} & \multirow{1}{*}{1.768241} & \multirow{1}{*}{7.048378} & \multirow{1}{*}{7.048378} & \multirow{1}{*}{7.048378} \\
          & \multirow{1}{*}{5.278215} & \multirow{1}{*}{5.280137} & \multirow{1}{*}{1.770163} & \multirow{1}{*}{1.768241} & \multirow{1}{*}{7.048378} & \multirow{1}{*}{7.048378} & \multirow{1}{*}{7.048378} \\
\hline
\multirow{1}{*}{$10^{-2}$} & \multirow{1}{*}{6.160582} & \multirow{1}{*}{6.160613} & \multirow{1}{*}{2.053694} & \multirow{1}{*}{2.053664} & \multirow{1}{*}{8.214277} & \multirow{1}{*}{8.214277} & \multirow{1}{*}{8.214277} \\
          & \multirow{1}{*}{6.160582} & \multirow{1}{*}{6.160613} & \multirow{1}{*}{2.053694} & \multirow{1}{*}{2.053664} & \multirow{1}{*}{8.214277} & \multirow{1}{*}{8.214277} & \multirow{1}{*}{8.214277} \\
\hline
\multirow{1}{*}{$10^{-3}$} & \multirow{1}{*}{6.270645} & \multirow{1}{*}{6.270649} & \multirow{1}{*}{2.090224} & \multirow{1}{*}{2.090220} &
\multirow{1}{*}{8.360869} & \multirow{1}{*}{8.360869} & \multirow{1}{*}{8.360869} \\
          & \multirow{1}{*}{6.270645} & \multirow{1}{*}{6.270649} & \multirow{1}{*}{2.090224} & \multirow{1}{*}{2.090220} & \multirow{1}{*}{8.360869} &
          \multirow{1}{*}{8.360869} & \multirow{1}{*}{8.360869} \\
\hline
\multirow{1}{*}{$10^{-4}$} & \multirow{1}{*}{6.281924} & \multirow{1}{*}{6.281927} & \multirow{1}{*}{2.093982} & \multirow{1}{*}{2.093978} &
\multirow{1}{*}{8.375905} & \multirow{1}{*}{8.375905} & \multirow{1}{*}{8.375905} \\
          & \multirow{1}{*}{6.281924} & \multirow{1}{*}{6.281927} & \multirow{1}{*}{2.093982} & \multirow{1}{*}{2.093978} & \multirow{1}{*}{8.375905} &
          \multirow{1}{*}{8.375905} & \multirow{1}{*}{8.375905} \\
\hline
\multirow{1}{*}{$10^{-5}$} & \multirow{1}{*}{6.283054} & \multirow{1}{*}{6.283058} & \multirow{1}{*}{2.094359} & \multirow{1}{*}{2.094355} & \multirow{1}{*}{8.377413} & \multirow{1}{*}{8.377413} & \multirow{1}{*}{8.377413} \\
          & \multirow{1}{*}{6.283054} & \multirow{1}{*}{6.283058} & \multirow{1}{*}{2.094359} & \multirow{1}{*}{2.094355} & \multirow{1}{*}{8.377413} & \multirow{1}{*}{8.377413} & \multirow{1}{*}{8.377413} \\
\hline
\end{tabular}
\caption{Modified polarized Compton scattering cross sections
$\widetilde{\sigma}_{\lambda\lambda'}$ in units of $\alpha^2$
for fixed Lorentz-violating parameter $\widetilde{\kappa}=10^{-16}$,
where $\lambda$ is the initial
and $\lambda'$ the final photon polarization. The cross sections are
computed for different values of the initial photon momentum $k_1$.
As before, for each $k_1$ the first row gives the results that follow
by using $X_{\lambda\lambda'}$ from Eq. \eqref{eq:matrix-element-square-x}.
The second row delivers the corresponding result $\widehat{X}_{\lambda\lambda'}$
from Eq. \eqref{eq:matrix-element-square-hat-x}.
}
\label{tab:Compton-scattering-results2}
\end{table}
\begin{table}[h]
\centering
\setlength{\extrarowheight}{4pt}
\newcolumntype{C}[1]{>{\centering\arraybackslash}m{#1}}
\begin{tabular}{C{0.6cm}|C{0.6cm}|C{0.6cm}||C{1.5cm}|C{1.5cm}|C{1.5cm}|C{1.5cm}|C{1.5cm}|C{1.5cm}|C{1.5cm}C{0.1cm}}
\hline
\multirow{1}{*}{$\widetilde{\kappa}^{01}$} & \multirow{1}{*}{$\widetilde{\kappa}^{02}$} & \multirow{1}{*}{$\widetilde{\kappa}^{03}$} & \multirow{1}{*}{$\widetilde{\sigma}_{11}$} & \multirow{1}{*}{$\widetilde{\sigma}_{22}$} & \multirow{1}{*}{$\widetilde{\sigma}_{12}$} & \multirow{1}{*}{$\widetilde{\sigma}_{21}$} & \multirow{1}{*}{$\widetilde{\sigma}_{1X}$} & \multirow{1}{*}{$\widetilde{\sigma}_{2X}$} & \multirow{1}{*}{\!\!$\widetilde{\sigma}$} \\
\hline
\hline
\multirow{1}{*}{1} & \multirow{1}{*}{1} & \multirow{1}{*}{2} & \multirow{1}{*}{6.283177} & \multirow{1}{*}{4.188794} & \multirow{1}{*}{2.094403} & \multirow{1}{*}{4.188795} & \multirow{1}{*}{8.377581} & \multirow{1}{*}{8.377590} & \multirow{1}{*}{8.377585} \\
\hline
\multirow{1}{*}{1} & \multirow{1}{*}{2} & \multirow{1}{*}{1} & \multirow{1}{*}{6.281280} & \multirow{1}{*}{7.330112} & \multirow{1}{*}{2.096301} & \multirow{1}{*}{1.047515} & \multirow{1}{*}{8.377581} & \multirow{1}{*}{8.377627} & \multirow{1}{*}{8.377604} \\
   &    &    & \multirow{1}{*}{6.281280} & \multirow{1}{*}{7.330065} & \multirow{1}{*}{2.096301} & \multirow{1}{*}{1.047515} & \multirow{1}{*}{8.377581} & \multirow{1}{*}{8.377581} & \multirow{1}{*}{8.377581} \\
\hline
\multirow{1}{*}{2} & \multirow{1}{*}{1} & \multirow{1}{*}{1} & \multirow{1}{*}{6.281280} & \multirow{1}{*}{7.330112} & \multirow{1}{*}{2.096301} & \multirow{1}{*}{1.047515} & \multirow{1}{*}{8.377581} & \multirow{1}{*}{8.377627} & \multirow{1}{*}{8.377604} \\
\hline
\multirow{1}{*}{1} & \multirow{1}{*}{1} & \multirow{1}{*}{3} & \multirow{1}{*}{6.282966} & \multirow{1}{*}{3.236618} & \multirow{1}{*}{2.094614} & \multirow{1}{*}{5.140967} & \multirow{1}{*}{8.377581} & \multirow{1}{*}{8.377585} & \multirow{1}{*}{8.377583} \\
\hline
\multirow{1}{*}{1} & \multirow{1}{*}{3} & \multirow{1}{*}{1} & \multirow{1}{*}{6.281874} & \multirow{1}{*}{7.806356} & \multirow{1}{*}{2.095706} & \multirow{1}{*}{0.571318} & \multirow{1}{*}{8.377581} & \multirow{1}{*}{8.377674} & \multirow{1}{*}{8.377627} \\
\hline
\multirow{1}{*}{3} & \multirow{1}{*}{1} & \multirow{1}{*}{1} & \multirow{1}{*}{6.281874} & \multirow{1}{*}{7.806356} & \multirow{1}{*}{2.095706} & \multirow{1}{*}{0.571318} & \multirow{1}{*}{8.377581} & \multirow{1}{*}{8.377674} & \multirow{1}{*}{8.377627} \\
\hline
\multirow{1}{*}{1} & \multirow{1}{*}{1} & \multirow{1}{*}{5} & \multirow{1}{*}{6.283181} & \multirow{1}{*}{2.559814} & \multirow{1}{*}{2.094399} & \multirow{1}{*}{5.817768} & \multirow{1}{*}{8.377581} & \multirow{1}{*}{8.377582} & \multirow{1}{*}{8.377581} \\
\hline
\multirow{1}{*}{1} & \multirow{1}{*}{5} & \multirow{1}{*}{1} & \multirow{1}{*}{6.280175} & \multirow{1}{*}{8.145009} & \multirow{1}{*}{2.097405} & \multirow{1}{*}{0.232822} & \multirow{1}{*}{8.377581} & \multirow{1}{*}{8.377831} & \multirow{1}{*}{8.377706} \\
\hline
\multirow{1}{*}{5} & \multirow{1}{*}{1} & \multirow{1}{*}{1} & \multirow{1}{*}{6.280175} & \multirow{1}{*}{8.145009} & \multirow{1}{*}{2.097405} &
\multirow{1}{*}{0.232822} & \multirow{1}{*}{8.377581} & \multirow{1}{*}{8.377831} & \multirow{1}{*}{8.377706} \\
\hline
\multirow{1}{*}{1} & \multirow{1}{*}{1} & \multirow{1}{*}{10} & \multirow{1}{*}{6.283322} & \multirow{1}{*}{2.217729} & \multirow{1}{*}{2.094258} & \multirow{1}{*}{6.159851} & \multirow{1}{*}{8.377581} & \multirow{1}{*}{8.377581} & \multirow{1}{*}{8.377581} \\
\hline
\multirow{1}{*}{1} & \multirow{1}{*}{10} & \multirow{1}{*}{1} & \multirow{1}{*}{6.285520} & \multirow{1}{*}{8.317100} & \multirow{1}{*}{2.092061} & \multirow{1}{*}{0.061577} & \multirow{1}{*}{8.377581} & \multirow{1}{*}{8.378677} & \multirow{1}{*}{8.378129} \\
\hline
\multirow{1}{*}{10} & \multirow{1}{*}{1} & \multirow{1}{*}{1} & \multirow{1}{*}{6.285520} & \multirow{1}{*}{8.317100} & \multirow{1}{*}{2.092061} & \multirow{1}{*}{0.061577} & \multirow{1}{*}{8.377581} & \multirow{1}{*}{8.378677} & \multirow{1}{*}{8.378129} \\
\hline
\multirow{1}{*}{10} & \multirow{1}{*}{1} & \multirow{1}{*}{10} & \multirow{1}{*}{6.281734} & \multirow{1}{*}{5.235428} & \multirow{1}{*}{2.095847} & \multirow{1}{*}{3.142161} & \multirow{1}{*}{8.377581} & \multirow{1}{*}{8.377590} & \multirow{1}{*}{8.377585} \\
\hline
\end{tabular}
\caption{Cross sections of modified polarized Compton scattering in units of $\alpha^2$
for different values of $\widetilde{\kappa}^{01}$, $\widetilde{\kappa}^{02}$, and
$\widetilde{\kappa}^{03}$, where the latter parameters are given in magnitudes of $10^{-16}$.
In the second row we also give the result that follows $\widehat{X}_{\lambda\lambda'}$.
The photon momentum is $k_1=10^{-10}m$.}
\label{tab:Compton-scattering-results3}
\end{table}

Values for the modified polarized Compton scattering cross sections $\widetilde{\sigma}_{11}$,
$\widetilde{\sigma}_{12}$, $\widetilde{\sigma}_{21}$, and $\widetilde{\sigma}_{22}$ are obtained.
These correspond to the processes $1\mapsto 1$, $1\mapsto 2$, $2\mapsto 1$, and $2\mapsto 2$,
where the numbers give the initial and final photon polarization, respectively.
To form gauge-invariant expressions, the sum over final photon polarizations
has to be performed:
\begin{subequations}
\label{eq:modified-compton-cross-sections-gauge-invariant-sums}
\begin{equation}
\widetilde{\sigma}_{1X}\equiv \sum_{\lambda'=1,2} \widetilde{\sigma}_{1\lambda'}\,,
\end{equation}
\begin{equation}
\widetilde{\sigma}_{2X}\equiv \sum_{\lambda'=1,2} \widetilde{\sigma}_{2\lambda'}\,.
\end{equation}
\end{subequations}
Our calculation is based on the assumption that only the initial photon state
can be prepared, especially its polarization. However, the final photon
polarization can only be measured, if the photon is observed or scattered
at a second electron. Since we consider the final photon as an asymptotic
particle according to the Feynman diagrams in Fig. \ref{fig:compton-scattering},
it is not observed and one has to sum over final photon polarizations
\cite{PeskinSchroeder1995,ItzyksonZuber1980}. Hence, what can be measured in
this context are only the quantities $\widetilde{\sigma}_{1X}$ and $\widetilde{\sigma}_{2X}$, so we
also give them.

Finally, we list the sum of all cross sections, which is averaged over the
initial photon polarizations:
\begin{equation}
\label{eq:total-modified-cross-section}
\widetilde{\sigma}\equiv \frac{1}{2}\sum_{\lambda,\lambda'} \widetilde{\sigma}_{\lambda\lambda'}=
\frac{1}{2}(\widetilde{\sigma}_{11}+\widetilde{\sigma}_{12}
+\widetilde{\sigma}_{21}+\widetilde{\sigma}_{22})=
\frac{1}{2}(\widetilde{\sigma}_{1X}+\widetilde{\sigma}_{2X})\,.
\end{equation}
For comparison with the modified Compton cross sections, the cross sections for
unpolarized and polarized Compton scattering in standard QED are presented in Table
\ref{tab:Compton-scattering-standard-qed-1} and Table \ref{tab:Compton-scattering-standard-qed-2},
respectively, for different initial photon momenta $k_1$.

An important issue has to be mentioned first: the calculation of the modified cross section
in the parity-odd theory can be performed in two different ways. The first possibility is to
calculate the matrix element squared \textit{\`{a} la} Sec.~(11.1) of Ref.~\cite{JauchRohrlich1976} by directly
using the modified polarization vectors from Eqs. \eqref{eq:polarization-mode1-parity-violating},
\eqref{eq:polarization-mode2-parity-violating}.
For completeness, we give this equation in a compact form:
\begin{subequations}
\begin{align}
\label{eq:matrix-element-square-x}
X_{\lambda\lambda'}&=\frac{1}{4}\mathrm{Tr}\left\{\left[\frac{1}{\upsilon_1}\cancel{\varepsilon}^{(\lambda')}(\cancel{p}_1+\cancel{k}_1+m)\cancel{\varepsilon}^{(\lambda)}
-\frac{1}{\upsilon_2}\cancel{\varepsilon}^{(\lambda)}(\cancel{p}_1-\cancel{k}_2+m)\cancel{\varepsilon}^{(\lambda')}\right](\cancel{p}_1+m)\right. \notag \\
&\phantom{{}={}\frac{1}{4}\mathrm{Tr}}\hspace{-0.125cm}\left.\times\left[\frac{1}{\upsilon_1}
\cancel{\varepsilon}^{(\lambda)}(\cancel{p}_1+\cancel{k}_1+m)\cancel{\varepsilon}^{(\lambda')}-\frac{1}{\upsilon_2}\cancel{\varepsilon}^{(\lambda')}(\cancel{p}_1
-\cancel{k}_2+m)\cancel{\varepsilon}^{(\lambda)}\right](\cancel{p}_2+m)\right\}\,,
\end{align}
where
\begin{equation}
\upsilon_1=2p_1\cdot k_1+k_1^2\,,\quad \upsilon_2=2p_1\cdot k_2-k_2^2\,,\quad p_2=p_1+k_1-k_2\,.
\end{equation}
\end{subequations}
Here, $\lambda$ denotes the initial and $\lambda'$ the final photon polarization.
For standard QED Eq.~\eqref{eq:matrix-element-square-x} results in Eq.~(11-13)
of Ref.~\cite{JauchRohrlich1976} (which we transform to fit our conventions):
\begin{subequations}
\begin{align}
\label{eq:matrix-element-square-x-stand-qed}
X_{\lambda\lambda'}^{\mathrm{QED}}&=\frac{1}{2}\left(\frac{\upsilon}{\upsilon'}+\frac{\upsilon'}{\upsilon}\right)-1 \notag \\
&\phantom{{}=\frac{1}{2}{}}+2\left[\varepsilon^{(\lambda)}\cdot \varepsilon^{(\lambda')}-\frac{2(\varepsilon^{(\lambda)}\cdot p_1)(\varepsilon^{(\lambda')}\cdot p_2)}{\upsilon}+\frac{2(\varepsilon^{(\lambda)}\cdot p_2)(\varepsilon^{(\lambda')}\cdot p_1)}{\upsilon'}\right]^2\,,
\end{align}
\begin{equation}
\upsilon=2p_1\cdot k_1\,,\quad \upsilon'=2p_1\cdot k_2\,.
\end{equation}
\end{subequations}
Alternatively, the computation can be performed with the matrix element squared
that is obtained without the direct use of the polarization vectors, but with the
polarization tensors from Eqs. \eqref{eq:polarization-sum-parity-violating-1},
\eqref{eq:polarization-sum-parity-violating-2}. This expression is lengthy and
we will not give it in full detail. However, we will state it in a formal manner:
\begin{equation}
\label{eq:matrix-element-square-hat-x}
\widehat{X}_{\lambda\lambda'}=\big(|\mathcal{M}(k_1,k_2)|^2\big)^{\mu\nu\varrho\sigma}\Pi_{\mu\varrho}(k_1)|_{\lambda}\Pi_{\nu\sigma}(k_2)|_{\lambda'}\Big|_{\substack{k_1^0=\omega_{\lambda}(\mathbf{k}_1) \\ k_2^0=\omega_{\lambda'}(\mathbf{k}_2)}}\,,
\end{equation}
where $\Pi_{\mu\nu}$ are photon polarization tensors and
$\big(|\mathcal{M}(k_1,k_2)|^2\big)^{\mu\nu\varrho\sigma}$ includes all parts
that do not directly involve the photon: traces of combinations of $\gamma$-matrices,
electron propagators etc. The structure of $\widehat{X}_{\lambda\lambda'}$ is
similar to Eq. (5.81) of Ref.~\cite{PeskinSchroeder1995}. However, the latter
equation gives the sum over all polarizations, whereas $\widehat{X}_{\lambda\lambda'}$
is the amplitude square for a distinct polarization.

For the configurations of Table \ref{tab:Compton-scattering-results1} and
\ref{tab:Compton-scattering-results2} the results are shown for different
Lorentz-violating parameters $\widetilde{\kappa}$, where $\widetilde{\kappa}$
is defined by $\widetilde{\kappa}\equiv \mathcal{E}/\sqrt{3}$ with $\mathcal{E}$
of Eq.~\eqref{eq:definition-parameter-calligraphic-e}. It suffices to give
$\widetilde{\kappa}$, since for both tables the three Lorentz-violating
parameters $\widetilde{\kappa}^{01}$, $\widetilde{\kappa}^{02}$, and
$\widetilde{\kappa}^{03}$ are chosen to be equal. Table
\ref{tab:Compton-scattering-results3} presents results, where
the latter parameters differ from each other.
Compare the obtained results to the classical Thomson cross section, which
follows from the standard QED result --- first obtained by Klein and Nishina ---
in the limit of vanishing initial photon momentum \cite{PeskinSchroeder1995}:
\begin{equation}
\label{eq:result-thomson-scattering-total}
\sigma^{\mathrm{Th}}\equiv\lim_{\substack{|\mathbf{k}_1|\mapsto 0 \\ |\mathbf{p}_1|\mapsto 0}} \sigma\big(e^-(p_1)\gamma(k_1)\rightarrow e^-(p_2)\gamma(k_2)\big)=\frac{8\pi\alpha^2}{3m^2}\approx \frac{8.377580\alpha^2}{m^2}\,.
\end{equation}

From Table \ref{tab:Compton-scattering-results1} we see that for vanishing
Lorentz violation the gauge-invariant contributions from Eq.
\eqref{eq:modified-compton-cross-sections-gauge-invariant-sums} are equal:
\begin{equation}
\lim_{\widetilde{\kappa}\mapsto 0} \widetilde{\sigma}_{1X}=
\lim_{\widetilde{\kappa}\mapsto 0} \widetilde{\sigma}_{2X}\,.
\end{equation}
Furthermore, the sum of all cross sections then corresponds to the Thomson
limit given in Eq. \eqref{eq:result-thomson-scattering-total}. The
results for $\widetilde{\sigma}_{11}$, $\widetilde{\sigma}_{12}$, and
$\widetilde{\sigma}_{21}$ do not depend on whether Eq. \eqref{eq:matrix-element-square-x}
or Eq. \eqref{eq:matrix-element-square-hat-x} is used for the calculation.

From Table \ref{tab:Compton-scattering-results2} it also follows that, for vanishing
Lorentz violation, $\widetilde{\sigma}_{1X}=\widetilde{\sigma}_{2X}$ and,
furthermore, that the averaged sum over all cross sections corresponds to the standard
Klein--Nishina results. Besides, all results are independent of the fact of whether the
calculation is based on $X_{\lambda\lambda'}$ or $\widehat{X}_{\lambda\lambda'}$. This
is also the case for the first selection of parameters in Table
\ref{tab:Compton-scattering-results3}. Furthermore, this table shows that the individual
cross sections $\widetilde{\sigma}_{11}$, $\widetilde{\sigma}_{12}$, $\widetilde{\sigma}_{21}$,
and $\widetilde{\sigma}_{22}$ depend on the direction of $\boldsymbol{\zeta}$, which is
encoded in the choice of $\widetilde{\kappa}^{01}$, $\widetilde{\kappa}^{02}$, and
$\widetilde{\kappa}^{03}$. However, it is evident that the gauge-invariant expressions
defined in Eq. \eqref{eq:modified-compton-cross-sections-gauge-invariant-sums}
are independent of the direction of $\boldsymbol{\zeta}$.

\subsection{Plots of the amplitude squares $X_{\lambda\lambda'}$ and $\widehat{X}_{\lambda\lambda'}$}
\label{subsec:reason-discrepancy}

\begin{figure}[h]
\centering
\begin{minipage}{0.33\textwidth}
\includegraphics[scale=0.40]{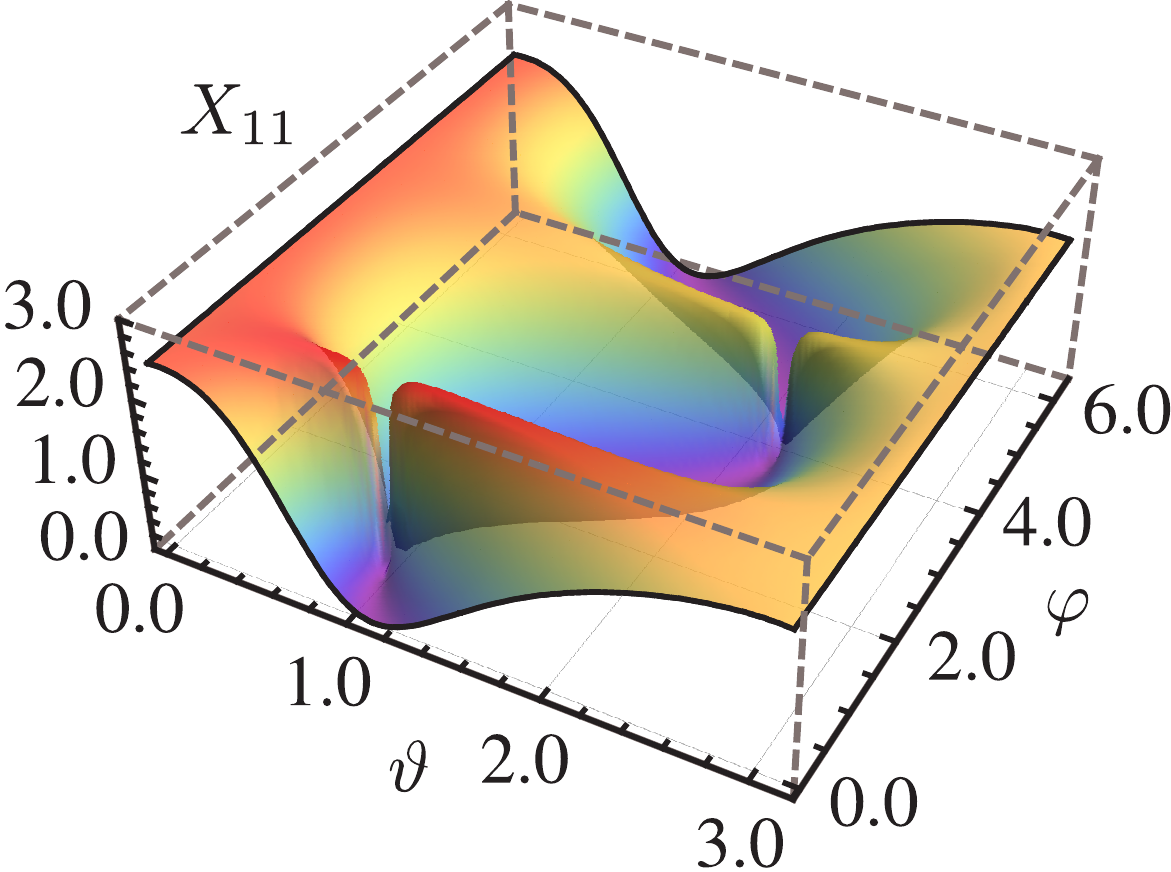}
\end{minipage}\begin{minipage}{0.33\textwidth}
\includegraphics[scale=0.40]{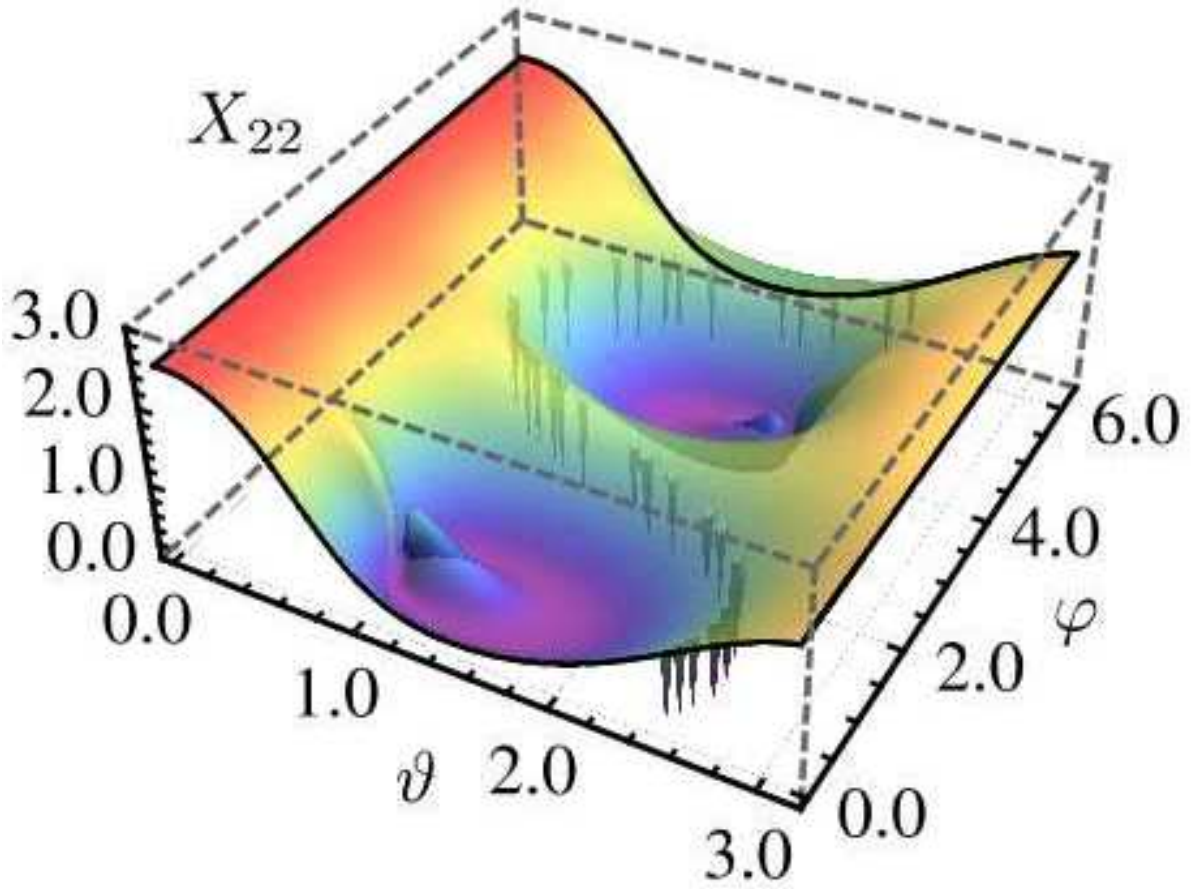}
\end{minipage}\begin{minipage}{0.33\textwidth}
\includegraphics[scale=0.40]{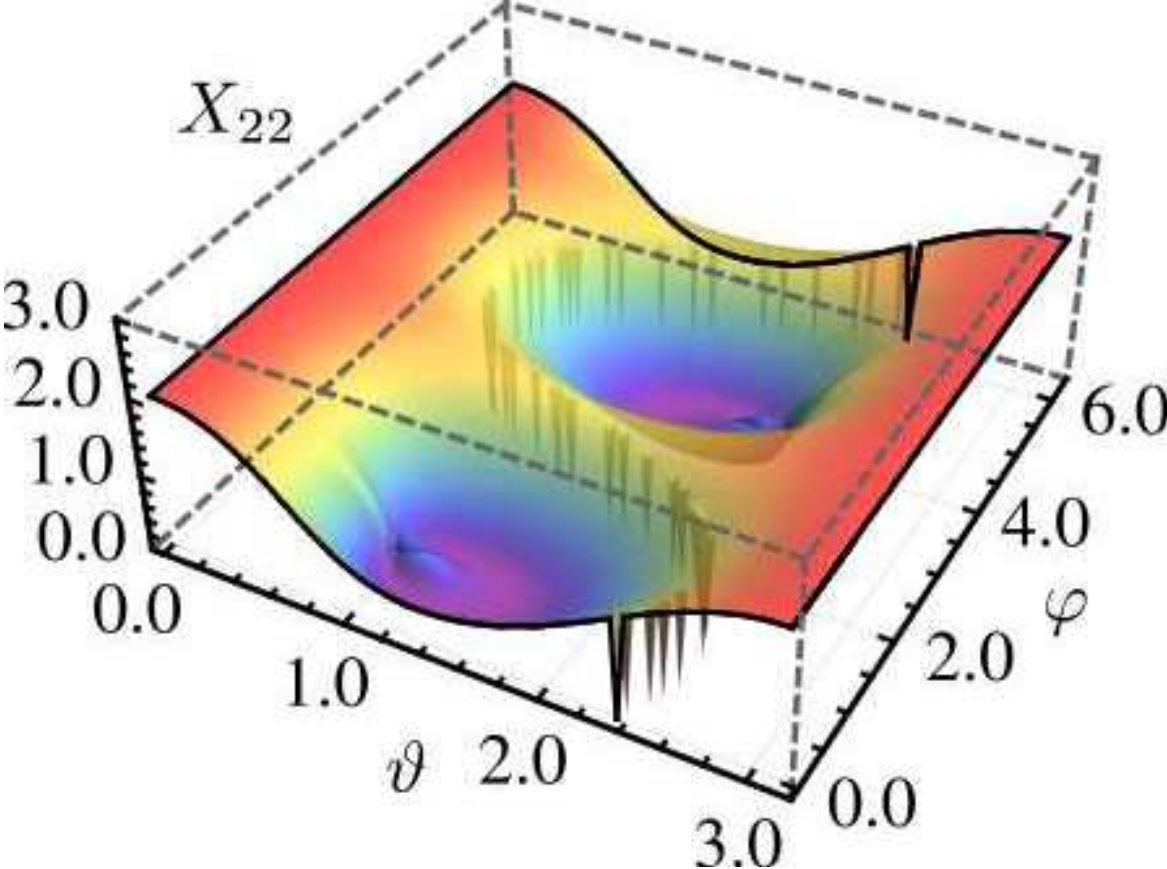}
\end{minipage}
\begin{minipage}{0.33\textwidth}
\includegraphics[scale=0.40]{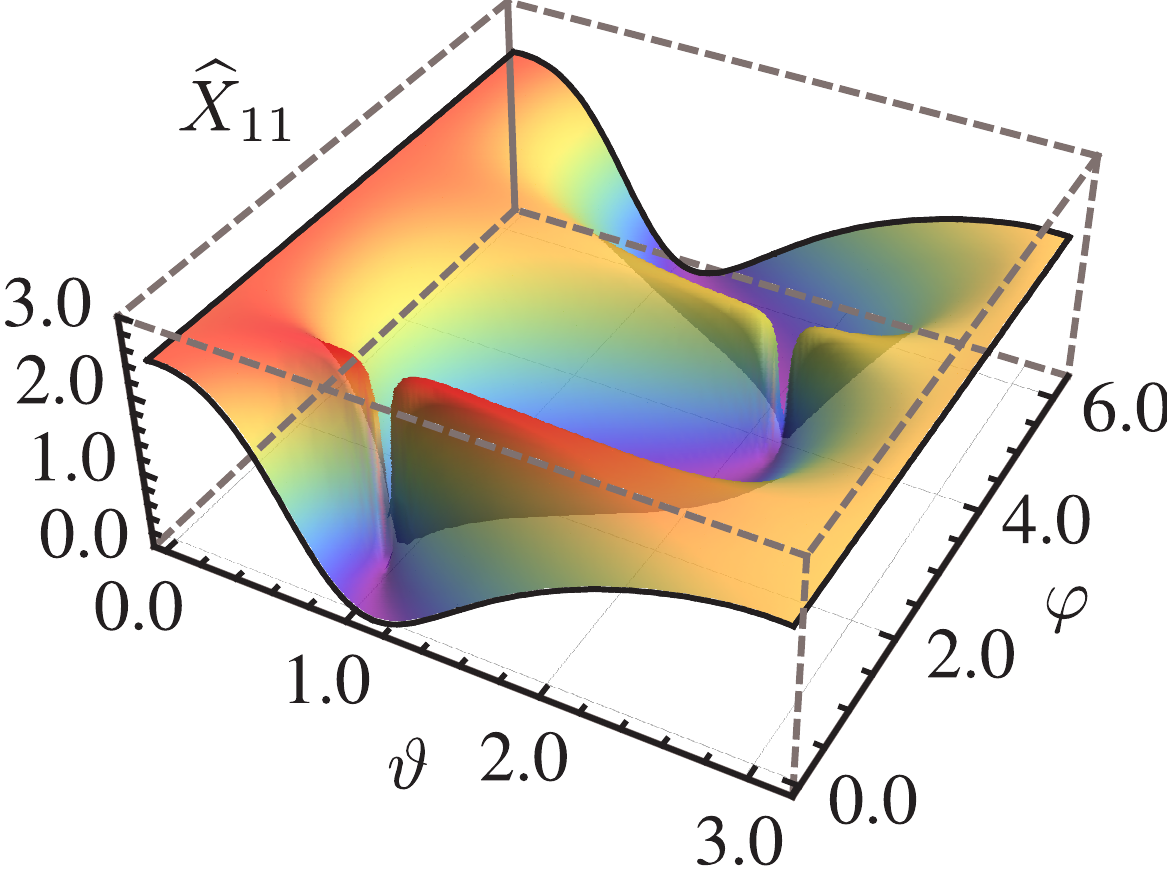}
\end{minipage}\begin{minipage}{0.33\textwidth}
\includegraphics[scale=0.40]{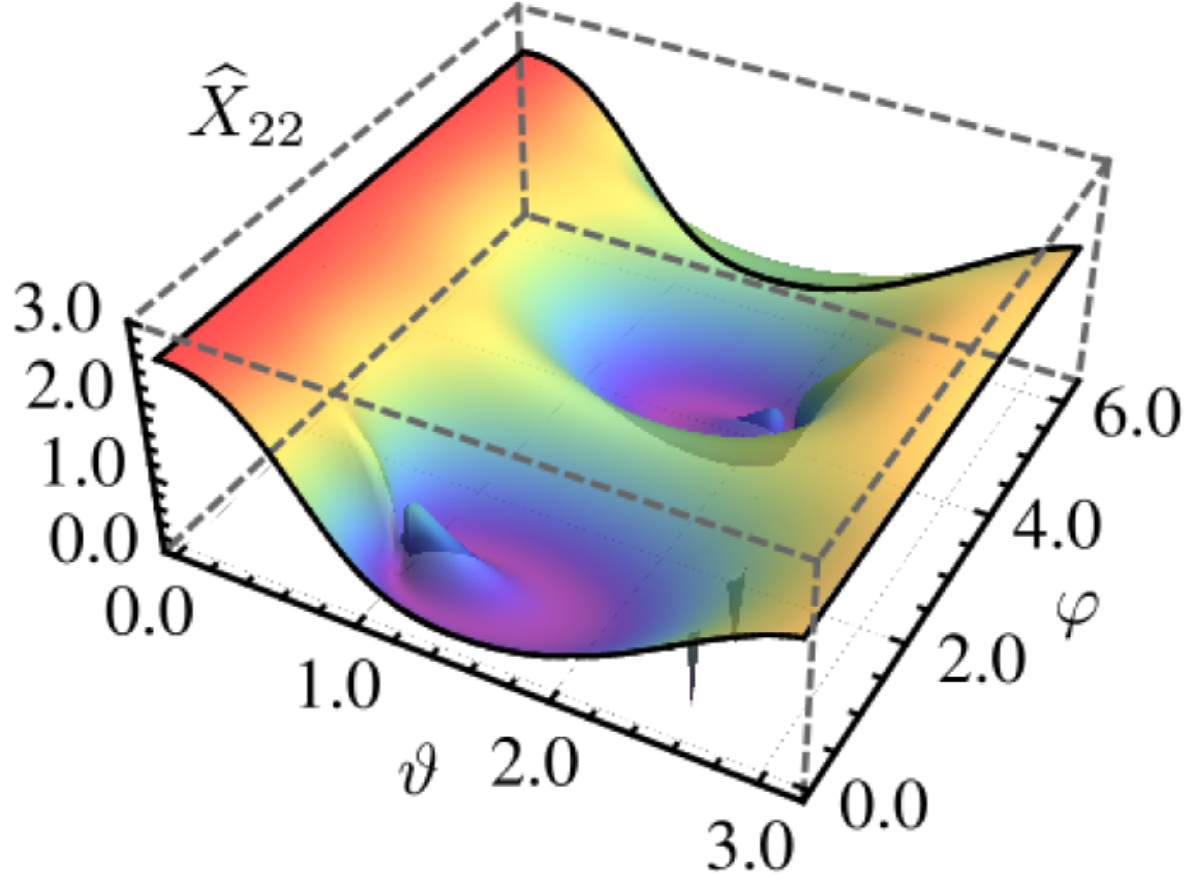}
\end{minipage}\begin{minipage}{0.33\textwidth}
\includegraphics[scale=0.40]{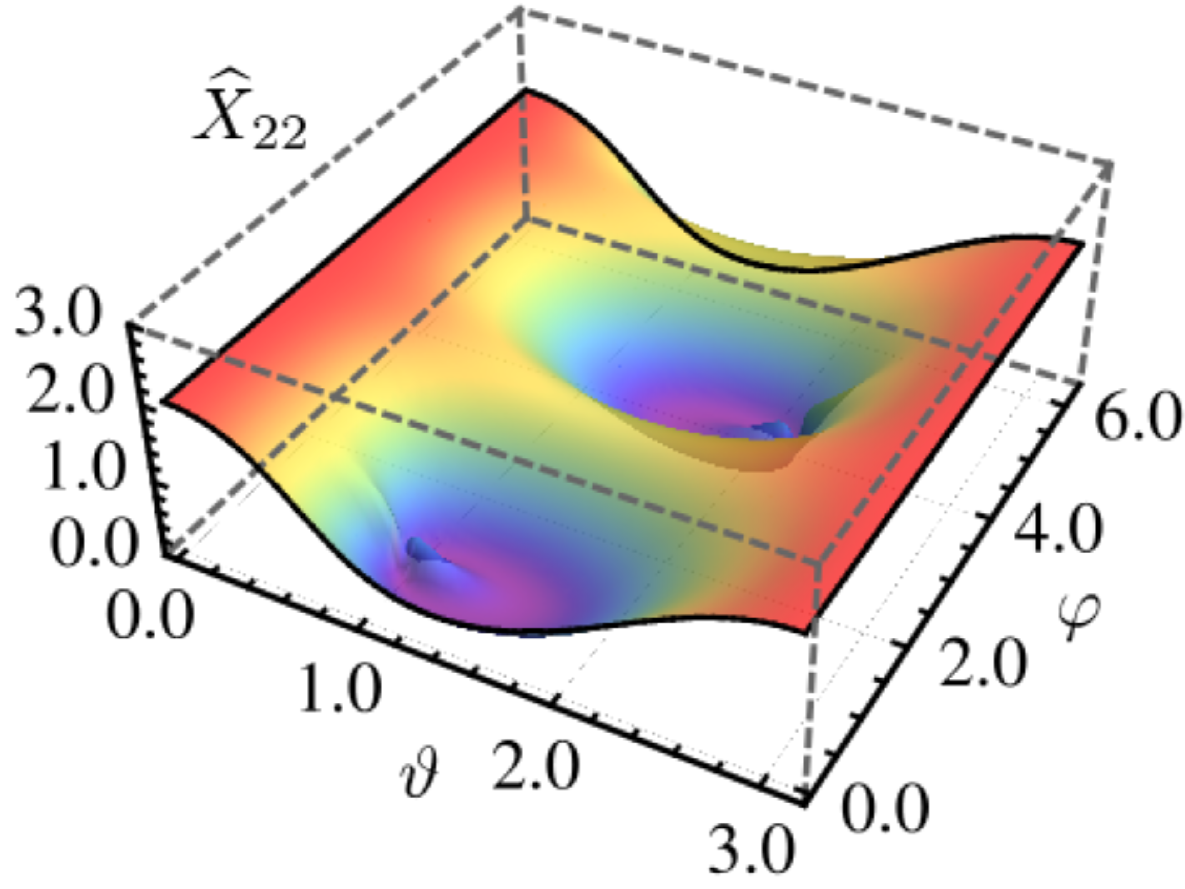}
\end{minipage}
\caption{Matrix element squared dependent on the polar angle $\vartheta$
and the azimuthal angle $\varphi$ for Compton scattering of polarized photons.
The plots in the first row were made by using $X_{\lambda\lambda'}$ from
Eq. \eqref{eq:matrix-element-square-x}. The graphs in the second row illustrate
$\widehat{X}_{\lambda\lambda'}$ from Eq. \eqref{eq:matrix-element-square-hat-x}.
The first column illustrates the corresponding functions for the $1\mapsto 1$ process
with $\widetilde{\kappa}=1/10$. The second and the third demonstrate the $2\mapsto 2$
process, where $\widetilde{\kappa}=1/10$ and $\widetilde{\kappa}=10^{-10}$,
respectively. Here again holds $k_1=10^{-10}m$ and $m=1$.}
\label{fig:plots-matrix-element-square-parameter-choice-3}
\end{figure}

Plotting the matrix element squares $X_{\lambda\lambda'}$ from Eq. \eqref{eq:matrix-element-square-x}
and $\widehat{X}_{\lambda\lambda'}$ from Eq. \eqref{eq:matrix-element-square-hat-x}
for each process $1\mapsto 1$, $1\mapsto 2$, $2\mapsto 1$, and $2\mapsto 2$
leads to a surprise.
We first present graphs of both $X_{\lambda\lambda'}$ and $\widehat{X}_{\lambda\lambda'}$ for
different sets of Lorentz-violating parameters, where the azimuthal angle $\varphi$ is
set to zero.
\begin{figure}[h]
\centering
\begin{minipage}{0.25\textwidth}
\includegraphics[scale=0.30]{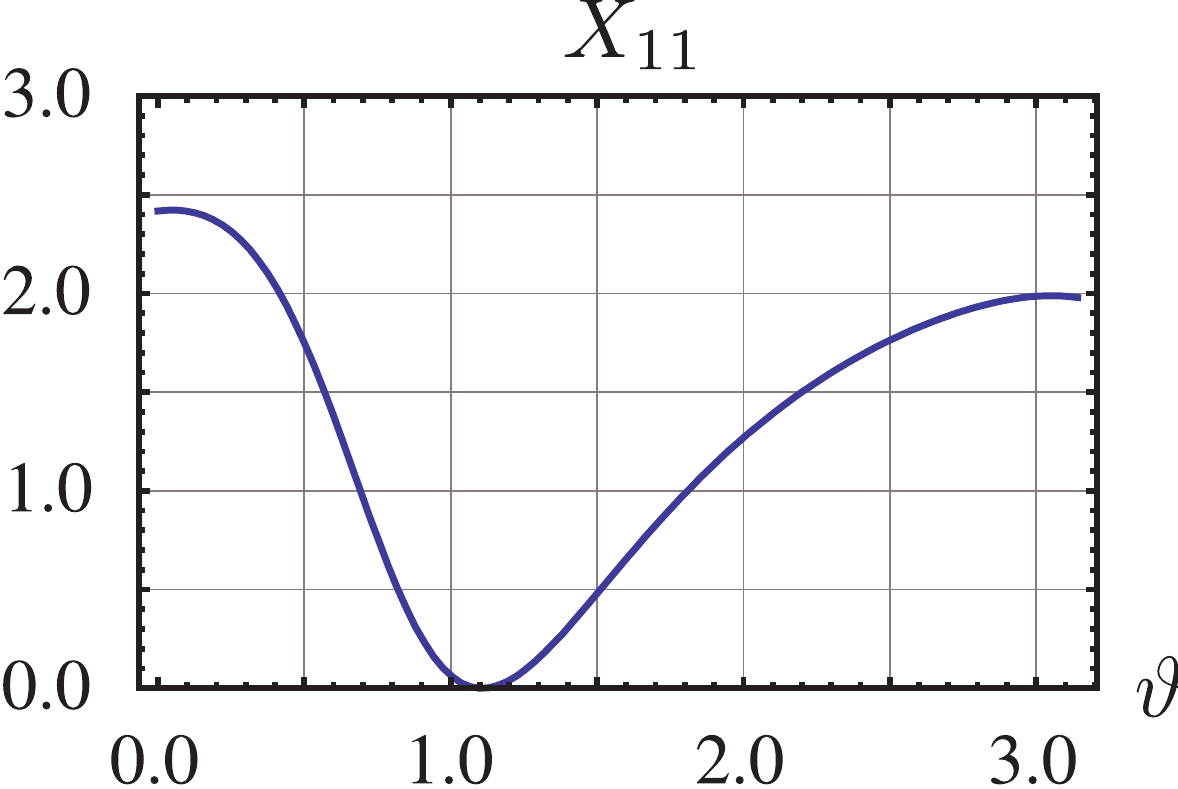}
\end{minipage}\begin{minipage}{0.25\textwidth}
\includegraphics[scale=0.30]{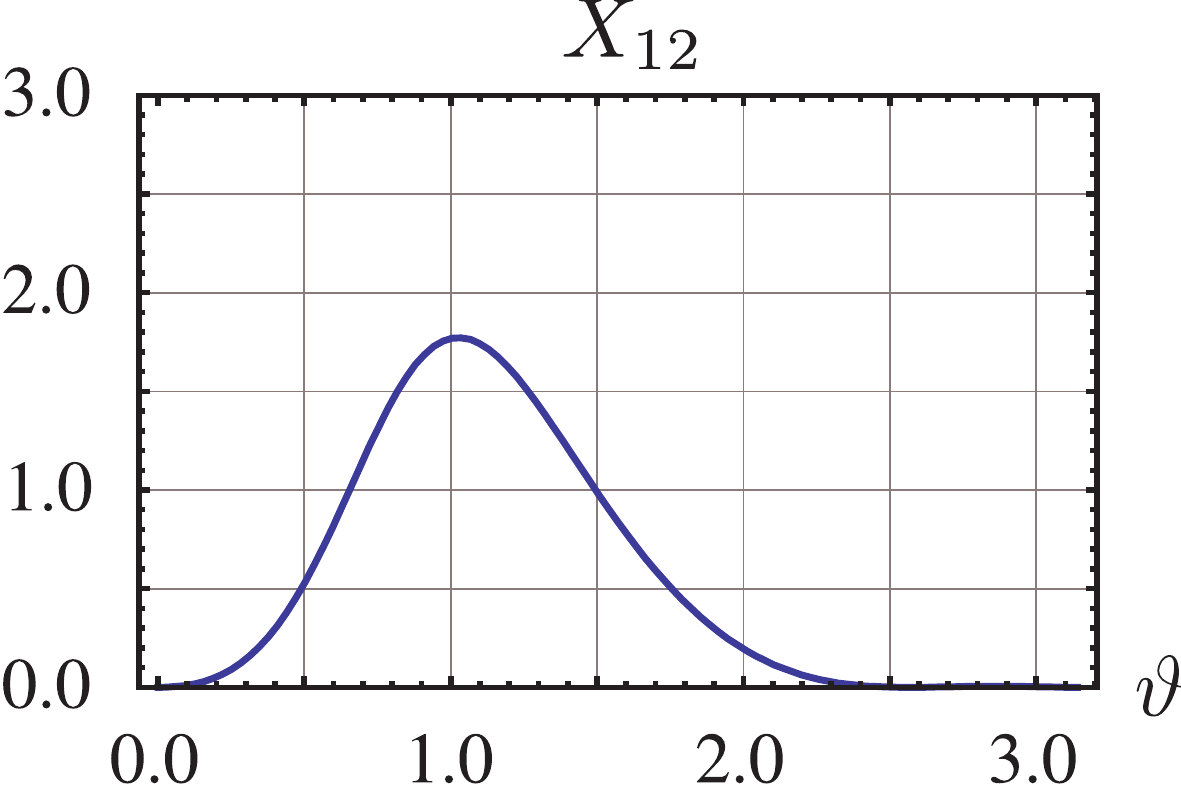}
\end{minipage}\begin{minipage}{0.25\textwidth}
\includegraphics[scale=0.30]{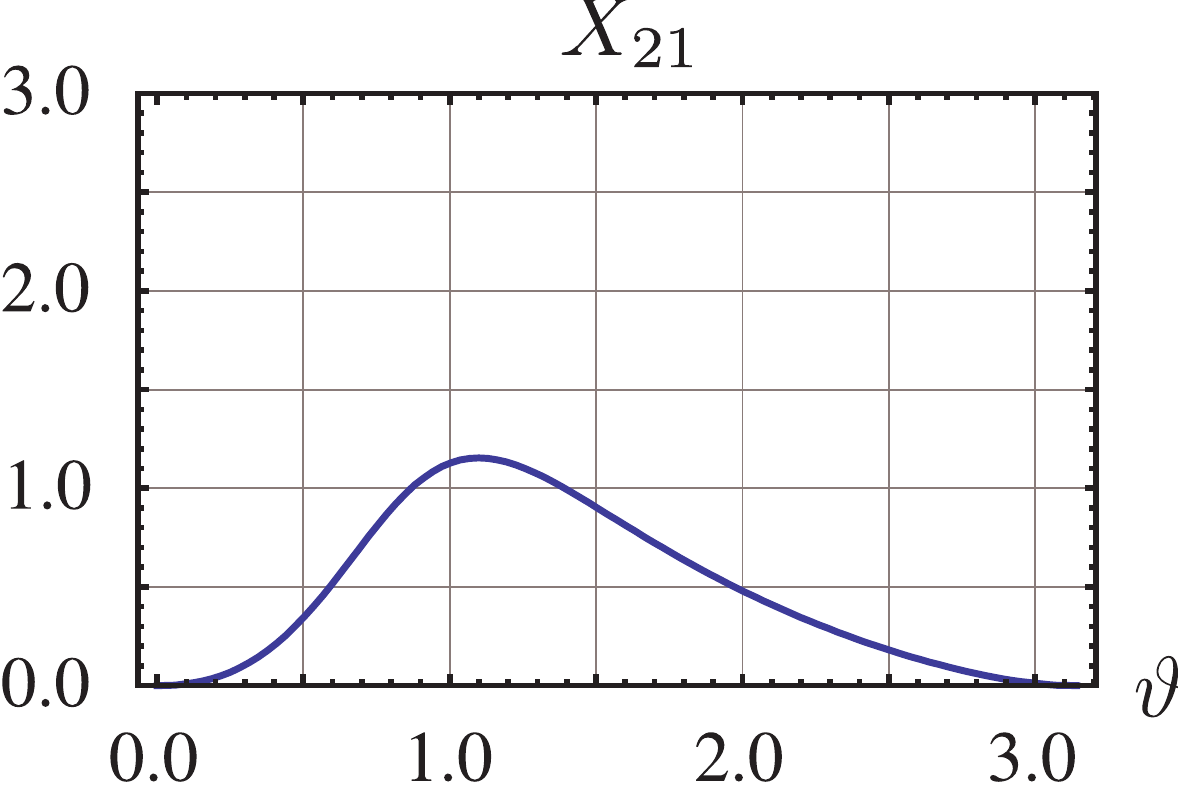}
\end{minipage}\begin{minipage}{0.25\textwidth}
\includegraphics[scale=0.30]{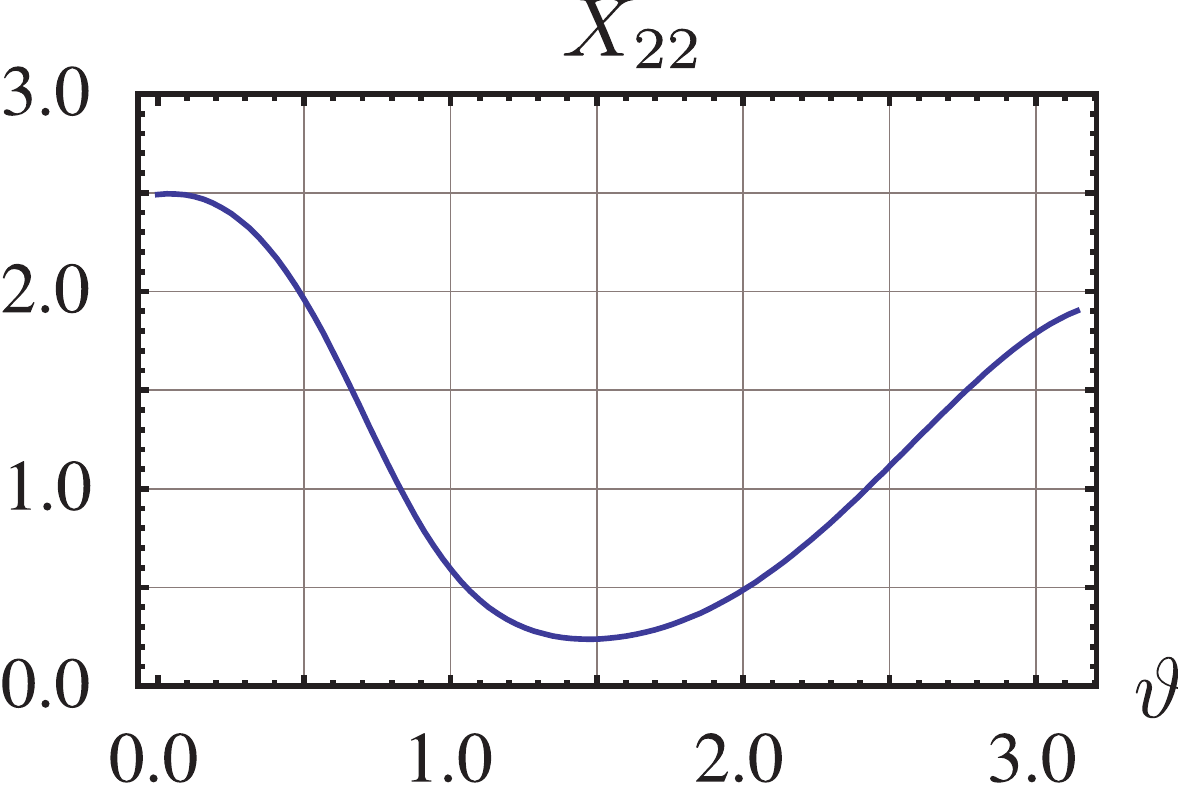}
\end{minipage}
\begin{minipage}{0.25\textwidth}
\includegraphics[scale=0.30]{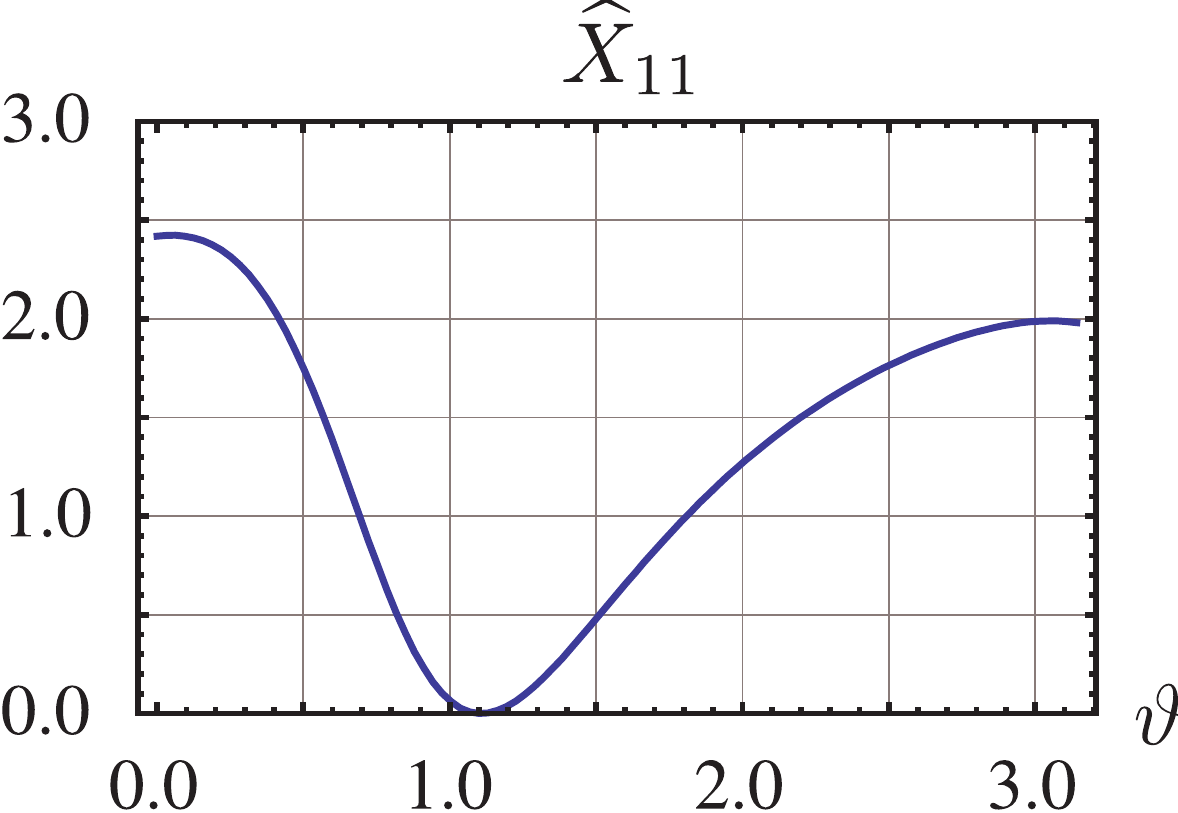}
\end{minipage}\begin{minipage}{0.25\textwidth}
\includegraphics[scale=0.30]{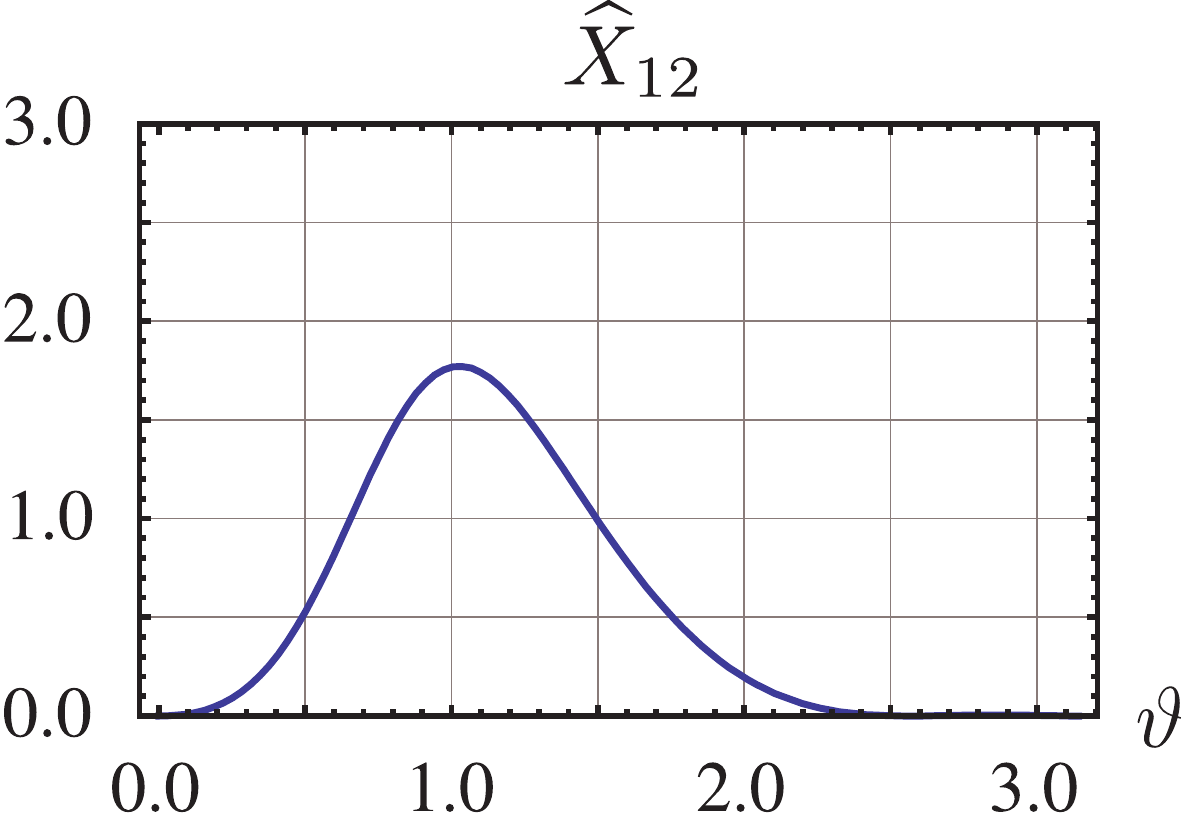}
\end{minipage}\begin{minipage}{0.25\textwidth}
\includegraphics[scale=0.30]{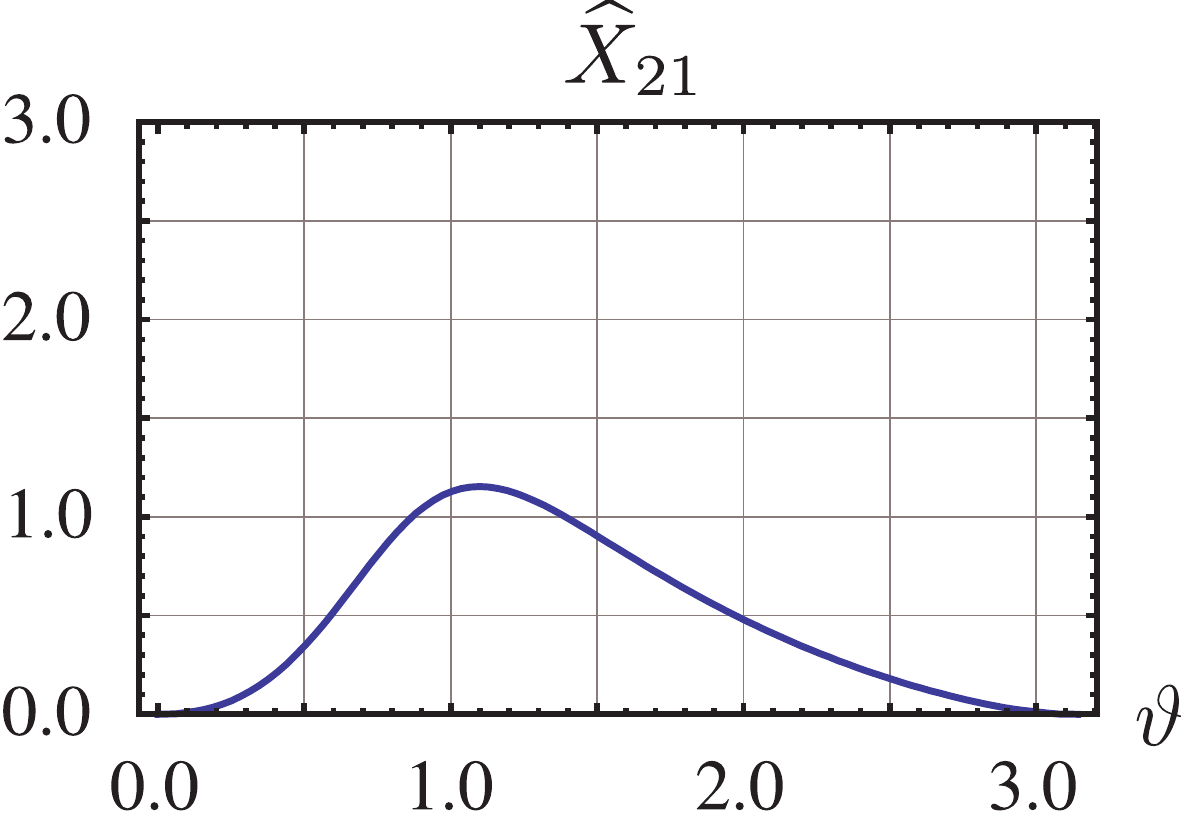}
\end{minipage}\begin{minipage}{0.25\textwidth}
\includegraphics[scale=0.30]{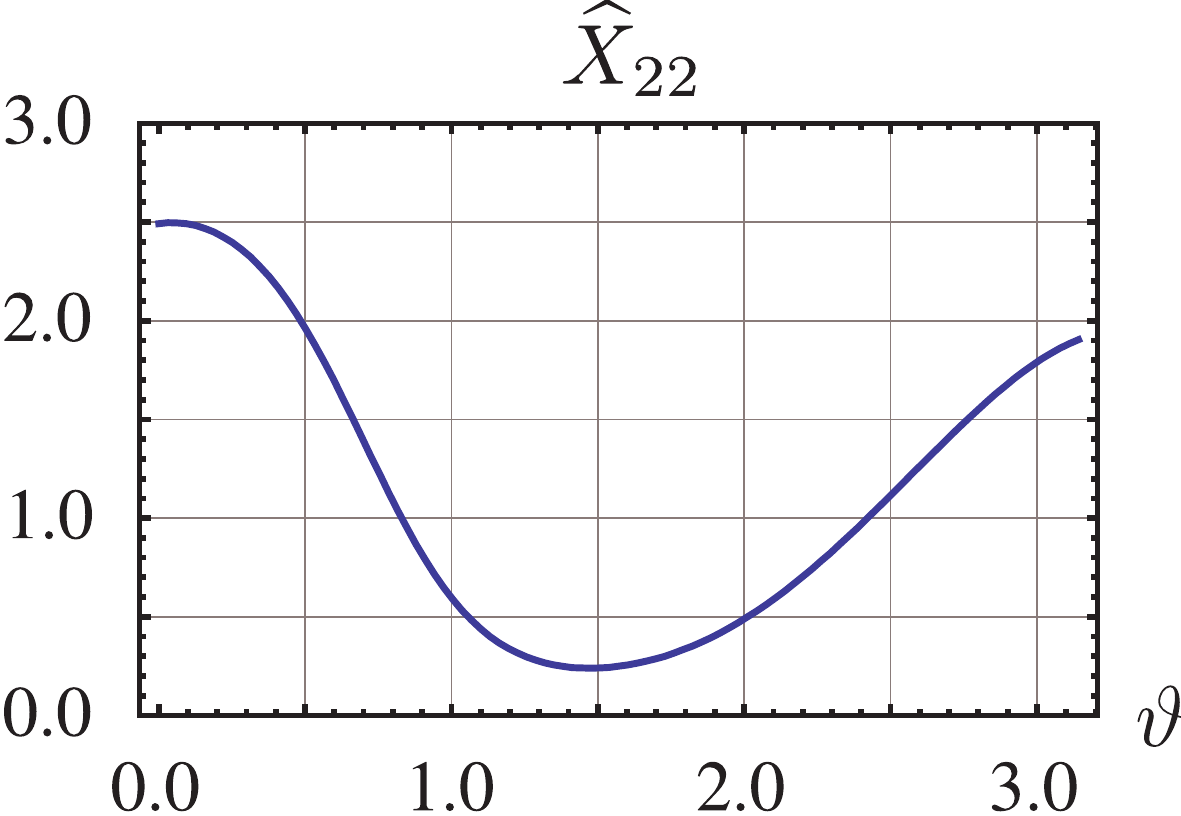}
\end{minipage}
\caption{Matrix element squared for Compton scattering of polarized photons
in the order $1\mapsto 1$, $1\mapsto 2$, $2\mapsto 1$, and $2\mapsto 2$ for
each row. The functions in the first row correspond to Eq.
\eqref{eq:matrix-element-square-x}, where the modified polarization vectors
were plugged in. The panels in the second row show the corresponding function
from Eq. \eqref{eq:matrix-element-square-hat-x} that is obtained by using
polarization tensors instead of the polarization vectors directly. The plots
were made for the special choice $\varphi=0$ and the horizontal axis gives
the polar angle $\vartheta$. The Lorentz-violating parameters are chosen
as $\widetilde{\kappa}^{01}=\widetilde{\kappa}^{02}=\widetilde{\kappa}^{03}=1/10$.
Furthermore, $k_1=10^{-10}m$ and $m=1$.}
\label{fig:plots-matrix-element-square-parameter-choice-1}
\end{figure}
\begin{figure}
\centering
\begin{minipage}{0.25\textwidth}
\includegraphics[scale=0.30]{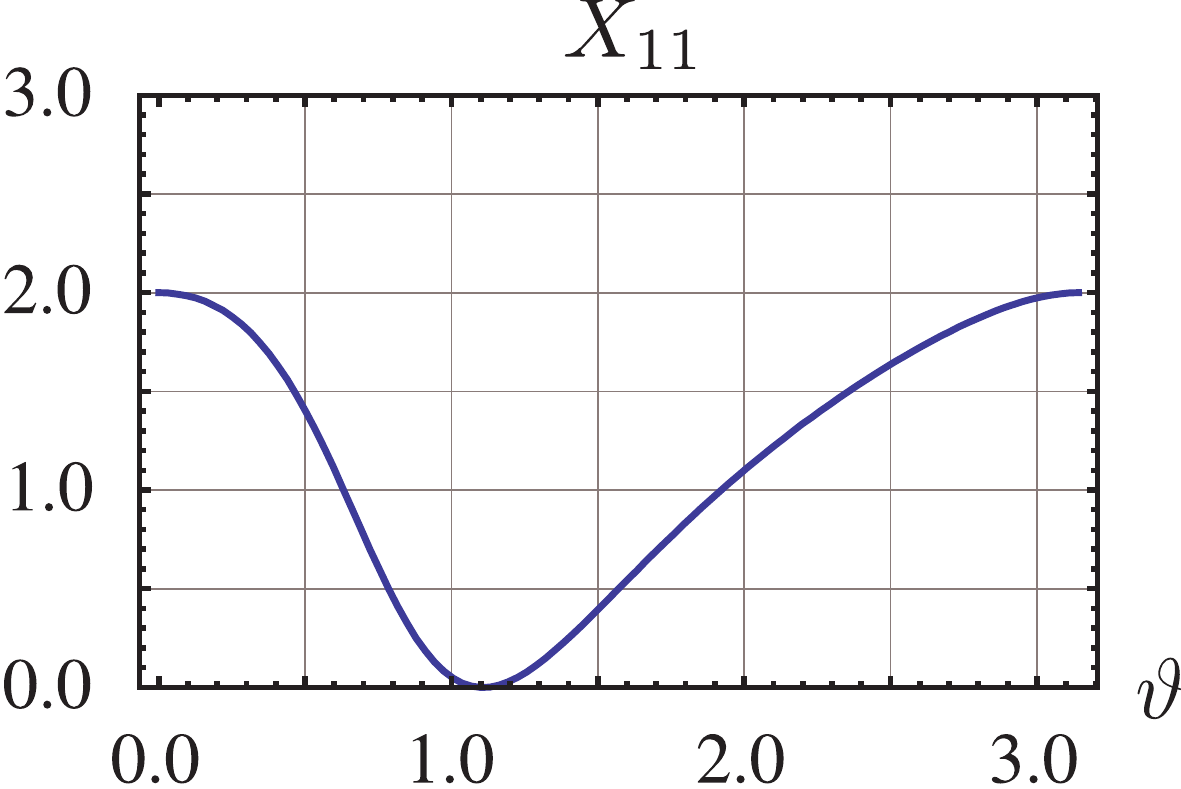}
\end{minipage}\begin{minipage}{0.25\textwidth}
\includegraphics[scale=0.30]{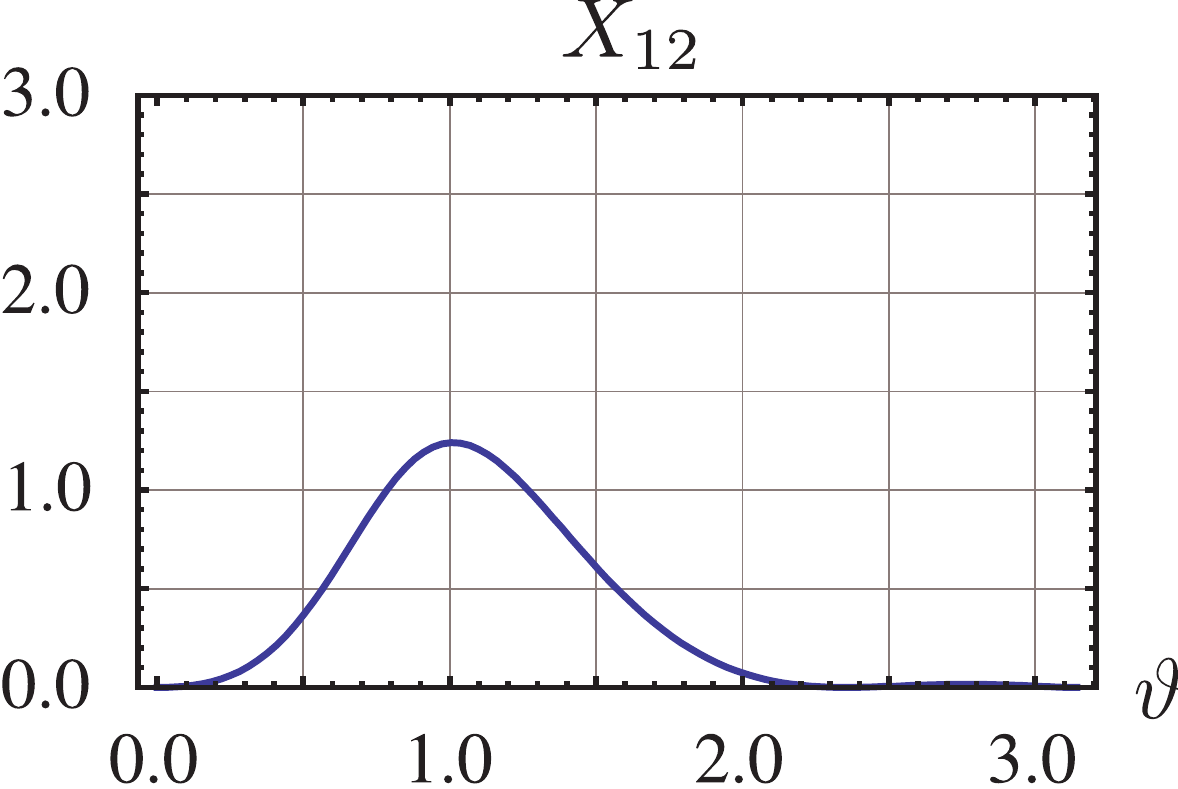}
\end{minipage}\begin{minipage}{0.25\textwidth}
\includegraphics[scale=0.30]{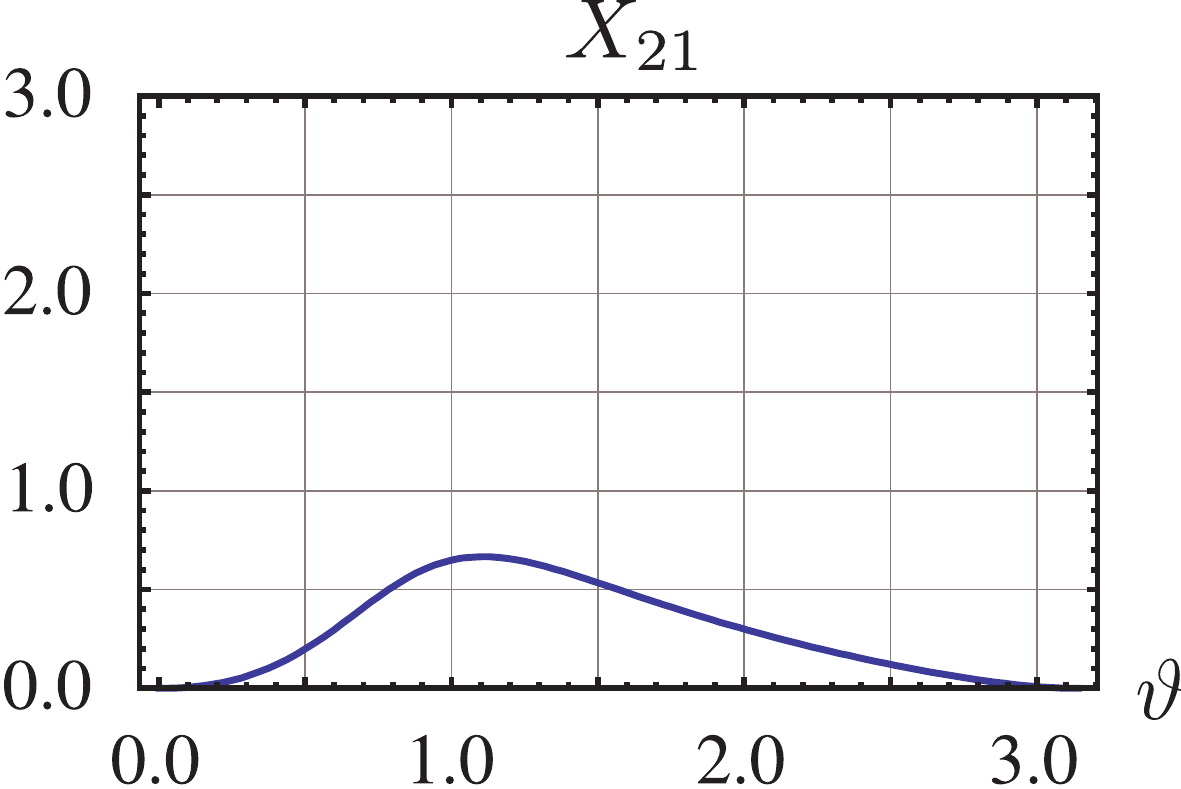}
\end{minipage}\begin{minipage}{0.25\textwidth}
\includegraphics[scale=0.30]{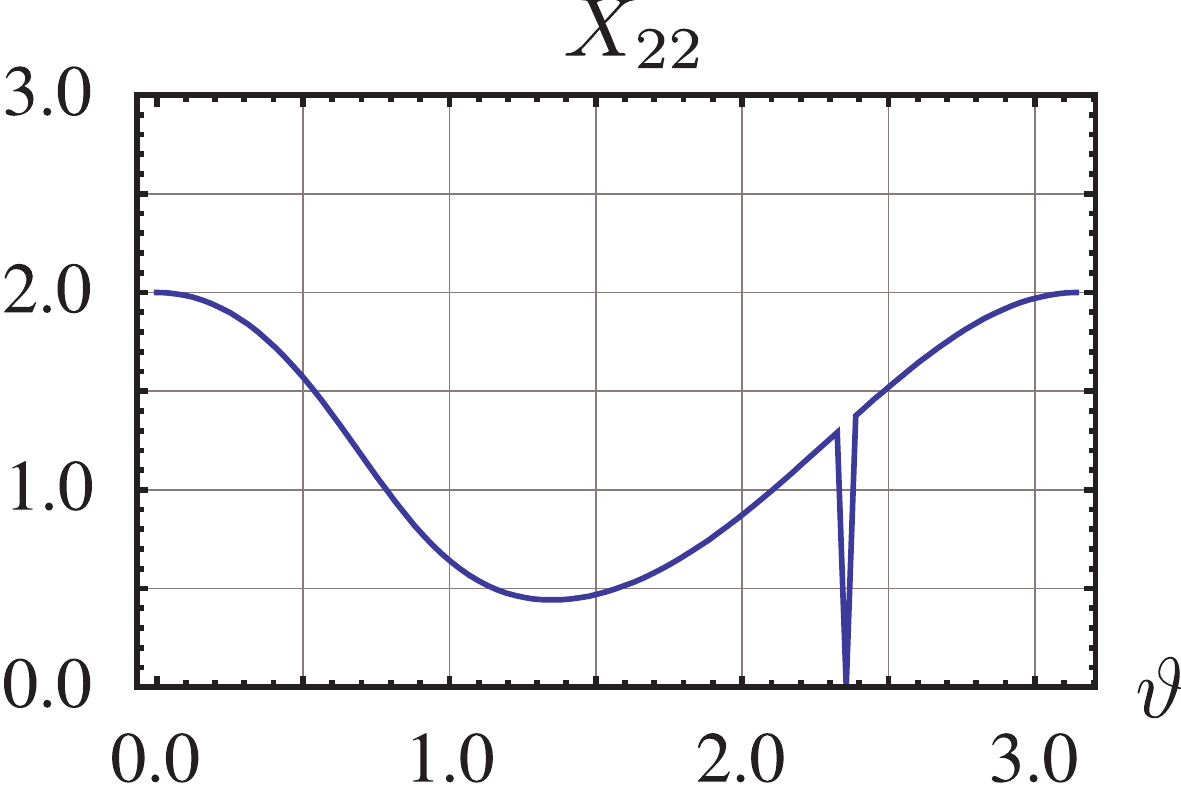}
\end{minipage}
\begin{minipage}{0.25\textwidth}
\includegraphics[scale=0.30]{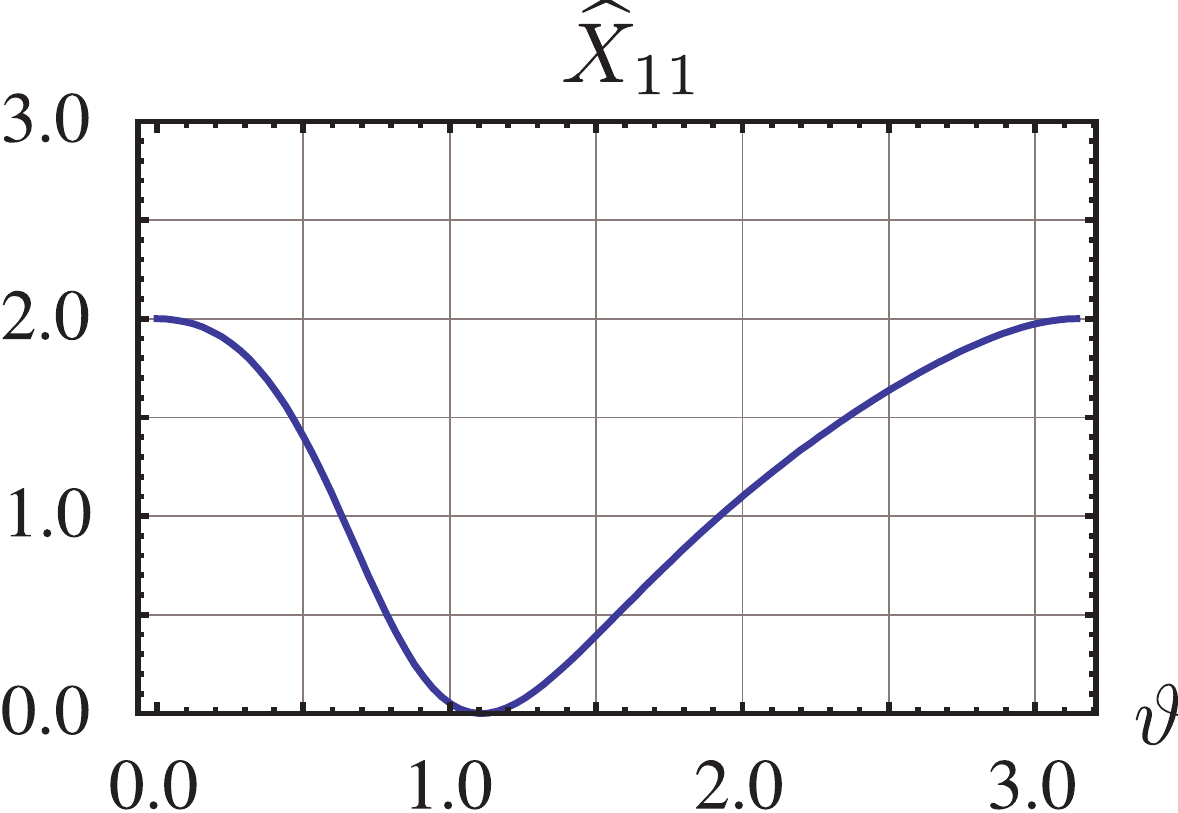}
\end{minipage}\begin{minipage}{0.25\textwidth}
\includegraphics[scale=0.30]{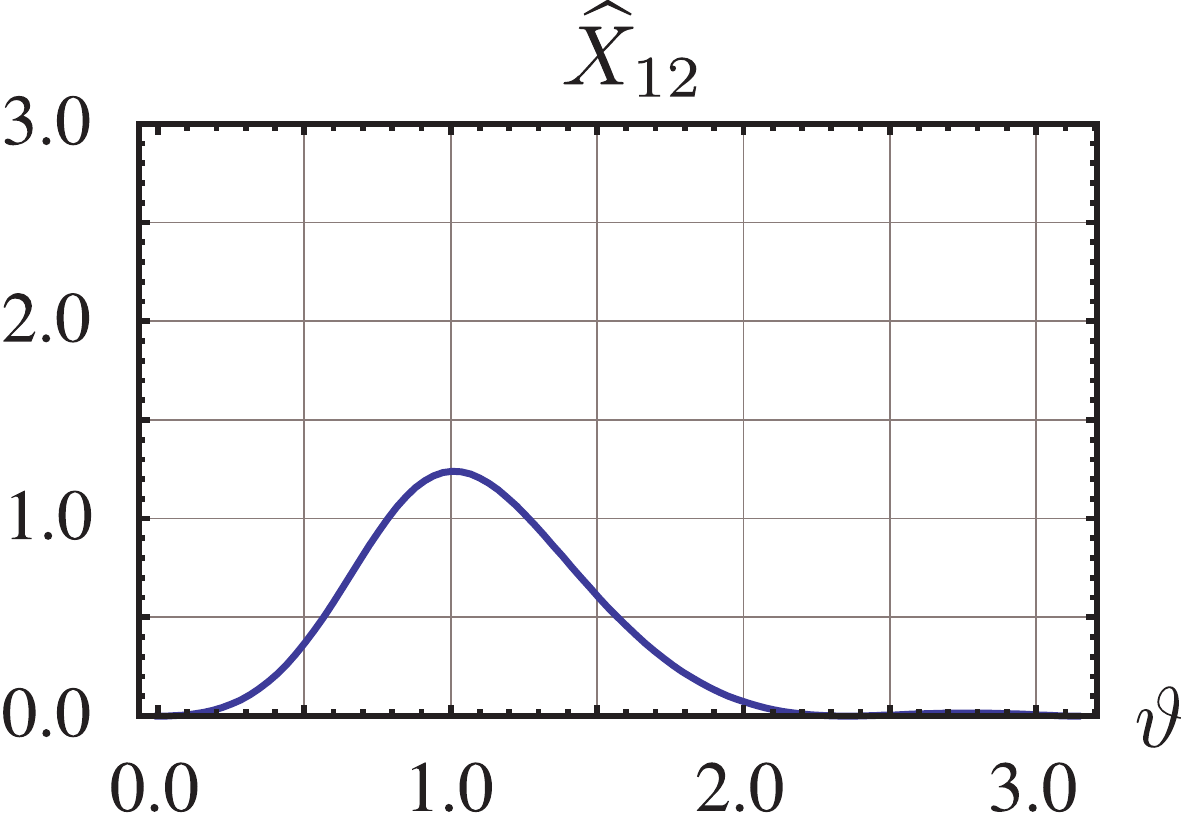}
\end{minipage}\begin{minipage}{0.25\textwidth}
\includegraphics[scale=0.30]{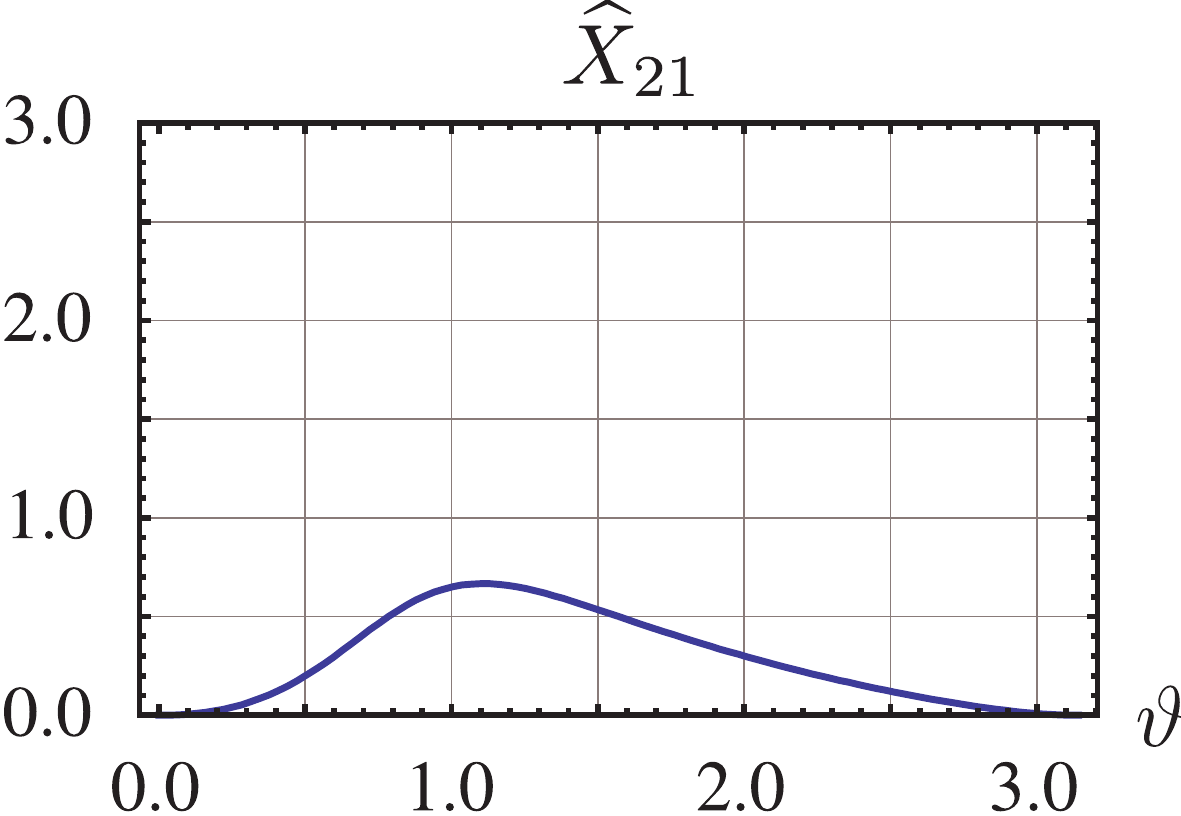}
\end{minipage}\begin{minipage}{0.25\textwidth}
\includegraphics[scale=0.30]{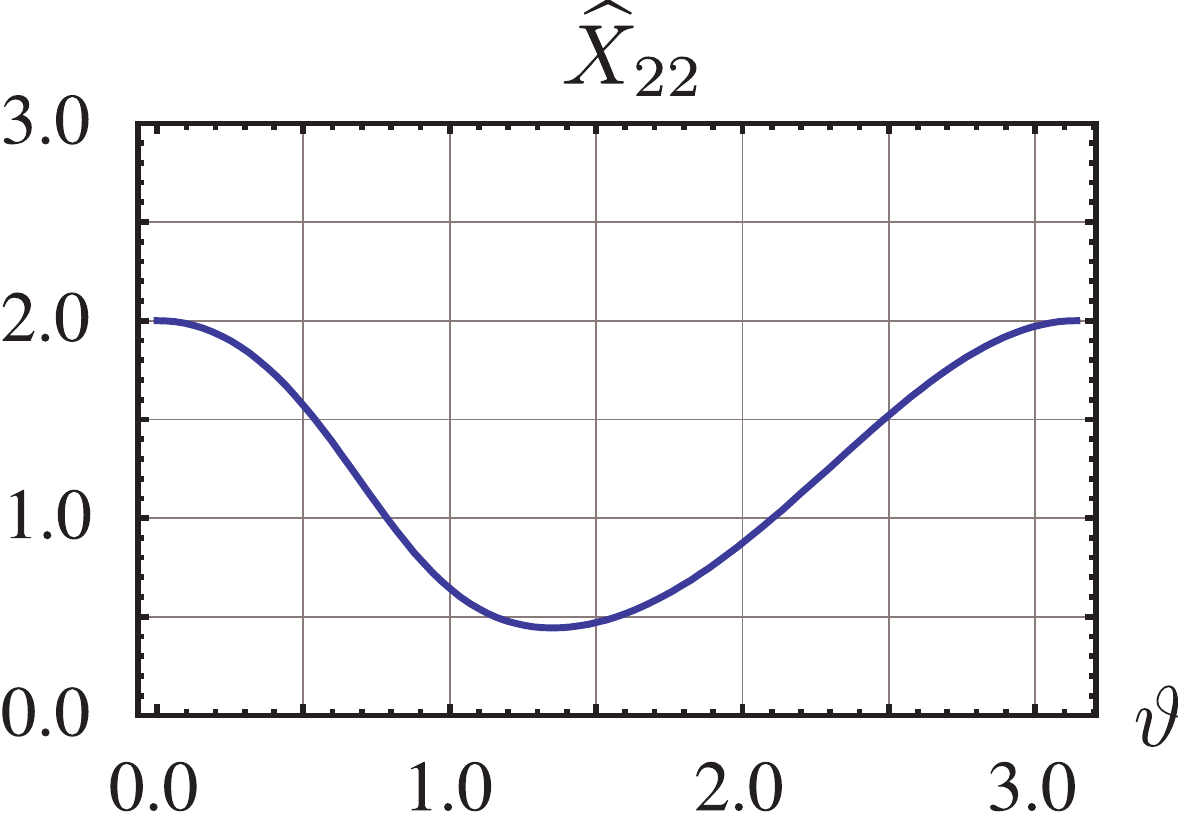}
\end{minipage}
\caption{Same as Fig. \ref{fig:plots-matrix-element-square-parameter-choice-1},
but now for the choice $\widetilde{\kappa}^{01}=\widetilde{\kappa}^{02}=
\widetilde{\kappa}^{03}=10^{-10}$.}
\label{fig:plots-matrix-element-square-parameter-choice-2}
\end{figure}
In Fig. \ref{fig:plots-matrix-element-square-parameter-choice-1}, for which
$\widetilde{\kappa}^{01}=\widetilde{\kappa}^{02}=\widetilde{\kappa}^{03}=1/10$
was inserted, we see that $X_{\lambda\lambda'}$ corresponds to
$\widehat{X}_{\lambda\lambda'}$ for the processes $1\mapsto 1$, $1\mapsto 2$,
$2\mapsto 1$, and $2\mapsto 2$.
The graphs in Fig. \ref{fig:plots-matrix-element-square-parameter-choice-2} for
Lorentz-violating parameters $\widetilde{\kappa}^{01}=\widetilde{\kappa}^{02}=
\widetilde{\kappa}^{03}=10^{-10}$ indicate that for the $2\mapsto 2$ process the
amplitude square $X_{\lambda\lambda'}$ approaches $\widehat{X}_{\lambda\lambda'}$,
but there remains a residue, which appears for $X_{\lambda\lambda'}$ as a narrow
peak at an angle $\vartheta_0\approx 2.35$ (given in arc measure). Finally, in Fig.
\ref{fig:plots-matrix-element-square-parameter-choice-3} we depict
$X_{\lambda\lambda'}$ and $\widehat{X}_{\lambda\lambda'}$ for the $1\mapsto 1$ and the
$2\mapsto 2$ modified Compton scattering as a function of both the polar angle
$\vartheta$ and the azimuthal angle $\varphi$. It is evident that $X_{\lambda\lambda'}$
and $\widehat{X}_{\lambda\lambda'}$ for $1\mapsto 1$ perfectly agree with each other.
This is also the case for the processes $1\mapsto 2$ and $2\mapsto 1$, but we will not
display the corresponding plots here. However, the $2\mapsto 2$ scattering behaves differently.
The matrix element $\widehat{X}_{\lambda\lambda'}$ looks smooth\footnote{Note that the two
small spikes at $(\vartheta,\varphi)\approx (2.45, 0.05)$ and $(\vartheta,\varphi)\approx (2.45,0.12)$,
respectively, probably originate from numerical errors}, whereas for $\widetilde{\kappa}^{01}=
\widetilde{\kappa}^{02}=\widetilde{\kappa}^{03}=1/10$, the amplitude square
$X_{\lambda\lambda'}$ is characterized by a set of sharp peaks. For small Lorentz violation
some of these peaks seem to remain. Whether or not the limit for vanishing Lorentz violation
is influenced by such structures cannot be investigated numerically, but requires analytical
computations.

\subsection{Interpretation}
\label{subsec:interpretation-compton-results}

\subsubsection{Discrepancies between $X_{\lambda\lambda'}$ and $\widehat{X}_{\lambda\lambda'}$ for $2\mapsto 2$ scattering}
\label{subsubsec:discrepancies-x-hat-x-22-scattering}

We already know from Eq. \eqref{eq:polarization-mode2-parity-violating} that the
second polarization vector splits into two contributions: a transverse and a
longitudinal part. For vanishing Lorentz violation it is explicitly true that
\begin{subequations}
\label{eq:polarization-vector-2-transversal-longitudinal-part}
\begin{equation}
(\varepsilon^{(2)\,\mu})=(\varepsilon^0,\widehat{\boldsymbol{\varepsilon}}^{\,(2)}_{\mathrm{transv}}+\varepsilon^0\,\widehat{\mathbf{k}})=(0,\widehat{\boldsymbol{\varepsilon}}^{\,(2)}_{\mathrm{transv}})+\varepsilon^0\,(1,\widehat{\mathbf{k}})
=(0,\widehat{\boldsymbol{\varepsilon}}^{\,(2)}_{\mathrm{transv}})+\frac{\varepsilon^0}{|\mathbf{k}|}\,(k^{\mu})\,,
\end{equation}
\begin{equation}
\widehat{\boldsymbol{\varepsilon}}^{\,(2)}_{\mathrm{transv}}=\widehat{\mathbf{k}}\times(\widehat{\mathbf{k}}\times \widehat{\boldsymbol{\zeta}})\,.
\end{equation}
\end{subequations}
If $\varepsilon^{(2)\,\mu}$ couples to a gauge-invariant quantity, its longitudinal
part will vanish because of the Ward identity, since it is directly proportional to
the four-momentum $k^{\mu}$.

\begin{figure}[t]
\centering
\includegraphics[scale=0.30]{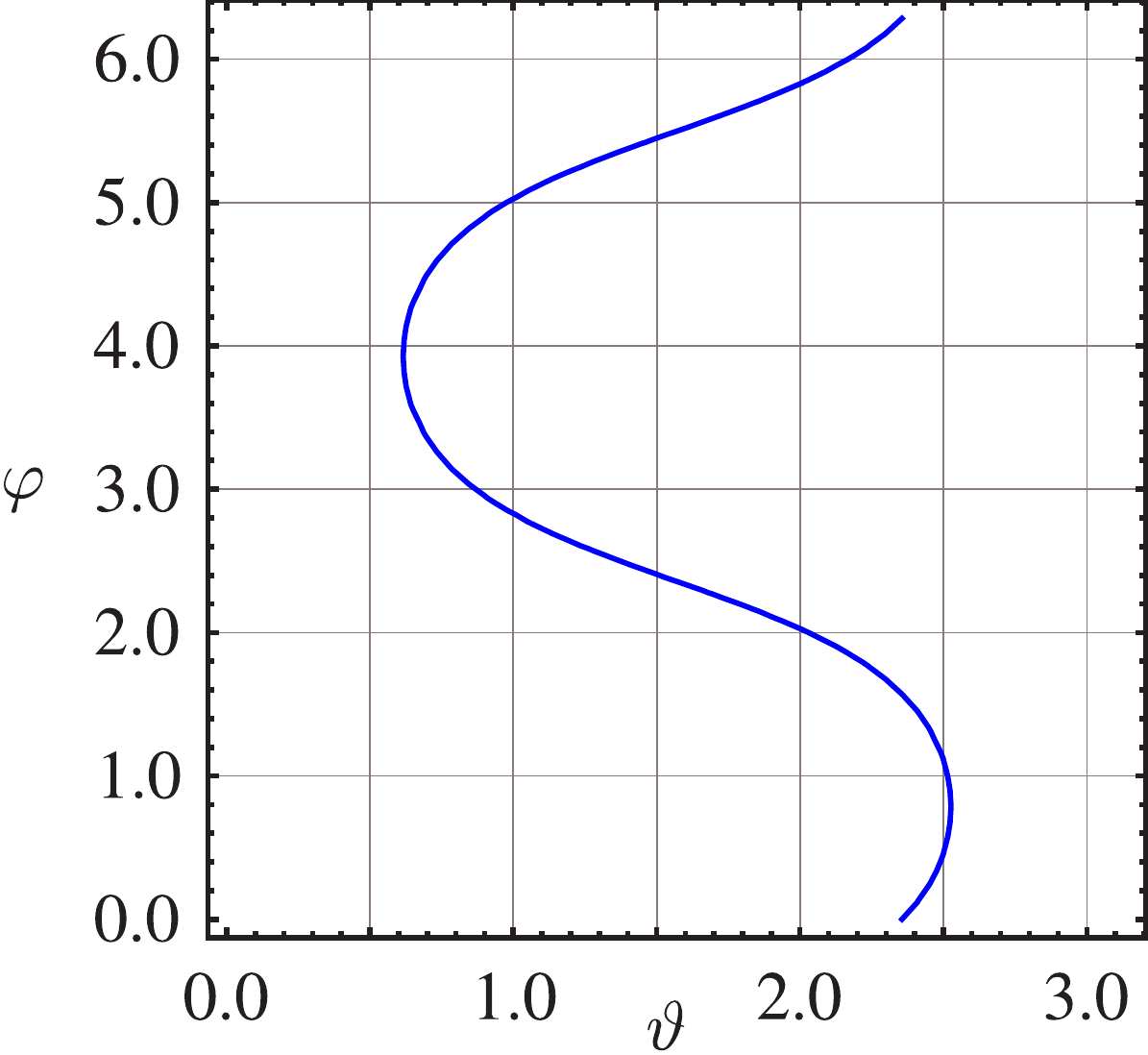}
\caption{Contour of angles $(\varphi_0,\vartheta_0)$, for which the normalization
factor $N''$ vanishes. The peaks in Fig. \ref{fig:plots-matrix-element-square-parameter-choice-3},
lie on this contour.}
\label{fig:zero-contour}
\end{figure}

The weird structures appearing in the matrix element squared $X_{22}$, which were
discussed in the last section, originate from the longitudinal part of
$\varepsilon^{(2)\,\mu}$.
As mentioned, from Eq.~\eqref{eq:polarization-vector-2-transversal-longitudinal-part}
it follows that in the limit of zero Lorentz violation the longitudinal part
vanishes by the Ward identity when contracted with physical quantities.
However, this only holds if the prefactor $\varepsilon^0$ is not zero.
Otherwise, we run into a ``$0/0$'' situation, which is mathematically
not defined. Now, the physical phase space of the process contains a sector, for
which $|k_{\parallel}|$ becomes arbitrarily small. This sector is characterized
by two angles $(\varphi_0,\vartheta_0)$, where for $\varphi_0=0$,
$\vartheta_0\approx 2.35$. This is depicted in Fig.
\ref{fig:zero-contour}.
For this special case the normalization factor $N''$ and, therefore, the prefactor
$\varepsilon^0$ can become arbitrarily small.
This destroys the applicability of the Ward identity and shows up as peaks in
$X_{22}$ of Figs. \ref{fig:plots-matrix-element-square-parameter-choice-2} and
\ref{fig:plots-matrix-element-square-parameter-choice-3}.

Now we would like to analytically investigate the limit $\mathcal{E}\mapsto 0$ of the
second polarization vector, with its transversal part subtracted. We distinguish
between two cases, $k_{\parallel}\sim \mathcal{E}k_{\bot}$ and $k_{\parallel}\gg \mathcal{E}k_{\bot}$.
The first represents the phase space sector for Compton scattering, for which $k_{\parallel}$
becomes arbitrarily small. We begin with the zeroth component of Eq.
\eqref{eq:polarization-mode2-parity-violating-explict}:
\begin{align}
\sqrt{N''}\,\varepsilon^{0}&=\frac{1}{4}\big(k^2-(k\cdot\xi)^2\big)\Big((k\cdot \zeta)^2-\zeta^2\big(k^2-(k\cdot\xi)^2\big)\Big) \notag \\
&\sim \left\{\begin{array}{lcl}
\mathcal{E}^2|\mathbf{k}|^2(k_{\parallel}^2+|\mathbf{k}|^2) & \text{for} & k_{\parallel}\gg \mathcal{E}k_{\bot}\,, \\
\mathcal{E}^2k_{\bot}^4 & \text{for} & k_{\parallel}\sim \mathcal{E}k_{\bot}\,. \\
\end{array}
\right.
\end{align}
The longitudinal part $\boldsymbol{\varepsilon}_{\mathrm{long}}$, which can be extracted from Eq.
\eqref{eq:polarization-mode2-parity-violating-explict} as well, results in:
\begin{align}
\sqrt{N''}\,\boldsymbol{\varepsilon}_{\mathrm{long}}|^{\mathcal{E}\mapsto 0}&\simeq \big(\sqrt{|\mathbf{k}|^2}+\mathbf{k}\cdot \boldsymbol{\zeta}\big)\big(\mathbf{k}\times (\mathbf{k}\times\boldsymbol{\zeta})\big)^2\frac{\mathbf{k}}{|\mathbf{k}|^2} \notag \displaybreak[0] \\
&=\mathcal{E}^2|\mathbf{k}|^4\big(1+\mathcal{E}\widehat{\mathbf{k}}\cdot\widehat{\boldsymbol{\zeta}}\big)\big(\widehat{\boldsymbol{\zeta}}-\widehat{\mathbf{k}}(\widehat{\mathbf{k}}\cdot\widehat{\boldsymbol{\zeta}})\big)^2\,\widehat{\mathbf{k}} \notag \displaybreak[0]\\
&=\mathcal{E}^2|\mathbf{k}|^4\big(1+\mathcal{E}\widehat{\mathbf{k}}\cdot \widehat{\boldsymbol{\zeta}}\big)\big(1-(\widehat{\mathbf{k}}\cdot\widehat{\boldsymbol{\zeta}})^2\big)\widehat{\mathbf{k}} \notag \displaybreak[0]\\
&=\mathcal{E}^2|\mathbf{k}|^4\left(1+\frac{\mathcal{E}k_{\parallel}}{|\mathbf{k}|}\right)\left(1-\frac{k_{\parallel}^2}{|\mathbf{k}|^2}\right)\widehat{\mathbf{k}} \notag \displaybreak[0]\\
&\sim\left\{\begin{array}{lcl}
\mathcal{E}^2|\mathbf{k}|^2\big(|\mathbf{k}|^2-k_{\parallel}^2\big)\widehat{\mathbf{k}} & \text{for} & k_{\parallel}\gg \mathcal{E}k_{\bot}\,, \displaybreak[0]\\
\mathcal{E}^2k_{\bot}^4\widehat{\mathbf{k}} & \text{for} & k_{\parallel}\sim \mathcal{E}k_{\bot}\,.
\end{array}\right.
\end{align}
The normalization factor from Eq. \eqref{eq:terms-explicit-polarization-sum-2-4} is
\begin{equation}
N''|^{\mathcal{E}\mapsto 0}\simeq \mathcal{E}^2k_{\bot}^2|\mathbf{k}|^2(4k_{\parallel}^4+4k_{\parallel}^2k_{\bot}^2+\mathcal{E}^2k_{\bot}^2)\sim \left\{\begin{array}{lcl}
4\mathcal{E}^2k_{\bot}^2k_{\parallel}^2|\mathbf{k}|^4 & \text{for} & k_{\parallel} \gg \mathcal{E}k_{\bot}\,, \\
\mathcal{E}^6k_{\bot}^8 & \text{for} & k_{\parallel} \sim \mathcal{E}k_{\bot}\,. \\
\end{array}\right.
\end{equation}
Respecting $k_{\parallel}\gg \mathcal{E}k_{\bot}$, we obtain for the second
polarization vector:
\begin{align}
(\varepsilon^{(2)\,\mu}-\varepsilon^{(2)\,\mu}_{\mathrm{transv}})|^{k_{\parallel}\gg\mathcal{E}k_{\bot}}&=\left.\frac{1}{\sqrt{N''}}\begin{pmatrix}
\varepsilon^0 \\
\boldsymbol{\varepsilon}_{\mathrm{long}} \\
\end{pmatrix}\right|^{k_{\parallel}\gg \mathcal{E}k_{\bot}} \sim \frac{\mathcal{E}^2|\mathbf{k}|^2}{\sqrt{4\mathcal{E}^2k_{\bot}^2k_{\parallel}^2|\mathbf{k}|^4}}\begin{pmatrix}
k_{\parallel}^2+|\mathbf{k}|^2 \\
\big(|\mathbf{k}|^2-k_{\parallel}^2\big)\widehat{\mathbf{k}} \\
\end{pmatrix} \notag \\
&=\frac{\mathcal{E}|\mathbf{k}|^2}{2k_{\bot}|k_{\parallel}||\mathbf{k}|^2}\begin{pmatrix}
|\mathbf{k}|^2+k_{\parallel}^2 \\
(|\mathbf{k}|^2-k_{\parallel}^2)\widehat{\mathbf{k}} \\
\end{pmatrix}\,,
\end{align}
which vanishes for $\mathcal{E}\mapsto 0$. In contrast to the latter case,
the result for $k_{\parallel}\sim \mathcal{E}k_{\bot}$ is as follows:
\begin{align}
(\varepsilon^{(2)\,\mu}-\varepsilon^{(2)\,\mu}_{\mathrm{transv}})|^{k_{\parallel}\sim\mathcal{E}k_{\bot}}&=\left.\frac{1}{\sqrt{N''}}\begin{pmatrix}
\varepsilon^0 \\
\boldsymbol{\varepsilon}_{\mathrm{long}} \\
\end{pmatrix}\right|^{k_{\parallel}\sim\mathcal{E}k_{\bot}} \notag \\
&\sim \frac{1}{\sqrt{\mathcal{E}^6k_{\bot}^8}}\begin{pmatrix}
\mathcal{E}^2k_{\bot}^4 \\
\mathcal{E}^2k_{\bot}^4\widehat{\mathbf{k}} \\
\end{pmatrix}=\frac{1}{\mathcal{E}}\begin{pmatrix}
1 \\
\widehat{\mathbf{k}} \\
\end{pmatrix}\,.
\end{align}
The latter diverges in the limit $\mathcal{E}\mapsto 0$.

Hence, it becomes evident that for vanishing Lorentz-violating parameter
$\mathcal{E}$, when $k_{\parallel}$ runs into the phase space sector
where it becomes of the order of $\mathcal{E}k_{\bot}$, a peak emerges.
Its width is then $\sim \mathcal{E}k_{\bot}$ and its height
is $\sim 1/\mathcal{E}$. This leads us undoubtedly to the following
representation of a $\delta$-function as the limit of a function sequence:
\begin{equation}
\label{eq:delta-distribution-function-sequence}
\delta(x)=\lim_{n\mapsto 0} g_n(x)\,,\quad g_n(x)=\left\{\begin{array}{lcl}
2/n & \text{for} & |x|\leq n\,, \\
0 & \text{for} & |x|>n\,. \\
\end{array}
\right.
\end{equation}
The role of the function sequence index $n$ in Eq. \eqref{eq:delta-distribution-function-sequence}
is taken by the Lorentz-violating parameter $\mathcal{E}$ in the polarization vector.
As a result, we finally obtain in the limit $\mathcal{E}\mapsto 0$:
\begin{equation}
\label{eq:polarization-2-longitudinal-part-analytic}
(\varepsilon^{(2)\,\mu}-\varepsilon^{(2)\,\mu}_{\mathrm{transv}})|^{\mathcal{E}\mapsto 0}\sim k_{\bot}\delta(k_{\parallel})\begin{pmatrix}
1 \\
\widehat{\mathbf{k}} \\
\end{pmatrix}\,.
\end{equation}
This analytic result shows, besides the numerically obtained plots in Figs.
\ref{fig:plots-matrix-element-square-parameter-choice-2}
and \ref{fig:plots-matrix-element-square-parameter-choice-3}, that the
longitudinal part of the second polarization vector may still play a role
for vanishing Lorentz-violating parameter. Because of the $\delta$-function,
the Ward identity can perhaps not be applied any more.

Now we want to look at the third term of Eq. \eqref{eq:matrix-element-square-x-stand-qed},
which is enclosed by round brackets. It will be denoted as $X^{(3)}_{\lambda\lambda'}$
in what follows.
We consider the $2\mapsto 2$ scattering process, where for the polarization vector
in the final state we insert only its longitudinal part according to Eq.
\eqref{eq:polarization-2-longitudinal-part-analytic}. Note that the longitudinal
part of the initial state polarization vector vanishes, since $k_{1,\parallel}\neq 0$.
Then we obtain:
\begin{align}
X_{22}^{(3)}&=\varepsilon^{(2)}(k_1)\cdot \varepsilon^{(2)}(k_2)-\frac{(\varepsilon^{(2)}(k_1)\cdot p_1)(\varepsilon^{(2)}(k_2)\cdot p_2)}{p_1\cdot k_1}+\frac{(\varepsilon^{(2)}(k_1)\cdot p_2)(\varepsilon^{(2)}(k_2)\cdot p_1)}{p_1\cdot k_2} \notag \displaybreak[0]\\
&\sim k_{2,\bot}\delta(k_{2,\parallel})\begin{pmatrix}
0 \\
\widehat{\boldsymbol{\varepsilon}}^{(2)}(k_1) \\
\end{pmatrix}\cdot \begin{pmatrix}
1 \\
\widehat{\mathbf{k}}_2 \\
\end{pmatrix} \notag \displaybreak[0]\\
&\quad\,-\frac{k_{2,\bot}\delta(k_{2,\parallel})}{p_1\cdot k_1}\left[\begin{pmatrix}
0 \\
\widehat{\boldsymbol{\varepsilon}}^{(2)}(k_1) \\
\end{pmatrix}\cdot\begin{pmatrix}
p_1^0 \\
\mathbf{p}_1 \\
\end{pmatrix}\right]\left[\begin{pmatrix}
1 \\
\widehat{\mathbf{k}}_2 \\
\end{pmatrix}\cdot \begin{pmatrix}
p_2^0 \\
\mathbf{p}_2 \\
\end{pmatrix}\right] \notag \displaybreak[0]\\
&\quad\,+\frac{k_{2,\bot}\delta(k_{2,\parallel})}{p_1\cdot k_2}\left[\begin{pmatrix}
0 \\
\widehat{\boldsymbol{\varepsilon}}^{(2)}(k_1) \\
\end{pmatrix}\cdot \begin{pmatrix}
p_2^0 \\
\mathbf{p}_2 \\
\end{pmatrix}\right]\cdot \left[\begin{pmatrix}
1 \\
\widehat{\mathbf{k}}_2 \\
\end{pmatrix}\cdot \begin{pmatrix}
p_1^0 \\
\mathbf{p}_1 \\
\end{pmatrix}\right] \notag \displaybreak[0]\\
&=k_{2,\bot}\delta(k_{2,\parallel})\left[\widehat{\boldsymbol{\varepsilon}}^{(2)}(k_1)\cdot \widehat{k}_2-\frac{(\widehat{\boldsymbol{\varepsilon}}^{(2)}(k_1)\cdot \mathbf{p}_1)(p_2\cdot k_2)}{|\mathbf{k}_2|\,p_1\cdot k_1}+\frac{(\widehat{\boldsymbol{\varepsilon}}^{(2)}(k_1)\cdot \mathbf{p}_2)(p_1\cdot k_2)}{|\mathbf{k}_2|\,p_1\cdot k_2}\right] \notag \\
&=k_{2,\bot}\delta(k_{2,\parallel})\left[\widehat{\boldsymbol{\varepsilon}}^{(2)}(k_1)\cdot \widehat{\mathbf{k}}_2-\frac{\widehat{\varepsilon}^{(2)}(k_1)\cdot \mathbf{k}_2}{|\mathbf{k}_2|}\right]=0\,,
\end{align}
where we have used $\widehat{\boldsymbol{\varepsilon}}^{(2)}(k_1)\cdot \mathbf{p}_1=0$.
Hence, the Ward identity does not seem to care about the $\delta$-function.
The contribution from the longitudinal part vanishes anyway. The conclusion
is that the peaks in Fig. \ref{fig:plots-matrix-element-square-parameter-choice-3} are
--- most likely --- numerical artifacts. Besides that, we expect this to hold also for
the peaks in Fig. \ref{fig:plots-matrix-element-square-parameter-choice-2}, where the
Lorentz-violating parameter has the finite value $1/10$.\footnote{At the moment, a neat
analytical proof is not available for finite Lorentz-violating parameter.
However, if the peaks were not a numerical artifact but the cause of a inconsistency of
the theory, we would expect them to scale with increasing Lorentz-violation, which is
obviously not the case.}

\subsubsection{Limit of $\widehat{X}_{\lambda\lambda'}$ for vanishing Lorentz violation}
\label{subsubsec:limit-hat-x-zero-lorentz-violation}

In section \ref{sec:Limit-polarizations-vanishing-lorentz-violation} we have
seen that preferred spacetime directions $\xi^{\mu}$ and $\zeta^{\mu}$ appear
in the polarization tensors $\Pi_{\mu\nu}$ even for vanishing Lorentz violation.
However, since the limit of $\widehat{X}_{\lambda\lambda'}$ for vanishing
Lorentz-violating parameters seems to coincide with the standard QED result, they
obviously do not play a role for physical quantities. The question then arises as
to why this is the case.

We consider an amplitude $\mathcal{M}$, to which one external photon with four-momentum
$k^{\mu}$ and polarization $\lambda$ couples:
$\mathcal{M}=\varepsilon_{\mu}^{(\lambda)}(k)\mathcal{M}^{\mu}(k)$.
In what follows, the term ``matrix element squared'' is understood in the sense of
individual contributions $|\varepsilon_{\mu}^{(\lambda)}(k)\mathcal{M}^{\mu}(k)|^2$.
For a virtual state,\footnote{a state with off-shell external particles}
all polarization vectors, hence also the scalar and the longitudinal ones, contribute
to the polarization-summed matrix element squared --- denoted as $|\mathcal{M}|^2$:
\begin{equation}
|\mathcal{M}|^2\Big|^{\mathrm{unphys}}\equiv\sum_{\lambda=0}^3 |\varepsilon_{\mu}^{(\lambda)}(k)\mathcal{M}^{\mu}(k)|^2\,.
\end{equation}
Evaluating $|\mathcal{M}|^2$ for a real state means that the Ward identity is used.
For standard QED, if $(k^{\mu})=(k,0,0,k)$ is chosen, the Ward identity will result in
\begin{equation}
k^{\mu}\mathcal{M}_{\mu}=k_0\mathcal{M}_0-k_3\mathcal{M}_3=k(\mathcal{M}_0+\mathcal{M}_3)=0\,,
\end{equation}
from which it follows that $\mathcal{M}_0=-\mathcal{M}_3$ or $|\mathcal{M}_0|^2=|\mathcal{M}_3|^2$.
Because of this, the unphysical degrees of freedom cancel each other and what remains
are terms which involve the physical polarization vectors ($\lambda=1$, 2). Since the
latter can be chosen as $(\varepsilon_1^{\mu})=(0,1,0,0)$ and $(\varepsilon_2^{\mu})=(0,0,1,0)$,
we obtain
\begin{equation}
|\mathcal{M}|^2\Big|^{\mathrm{phys}}_{\mathrm{QED}}=\sum_{\lambda=1,2} |\varepsilon_{\mu}^{(\lambda)}(k)\mathcal{M}^{\mu}(k)|^2=|\mathcal{M}_1|^2+|\mathcal{M}_2|^2=\sum_{\lambda=1,2} |\mathcal{M}_{\lambda}|^2\,,
\end{equation}
where `phys' means that the Ward identity has been used.

In order to understand the limits of the polarization tensors from Eq.
\eqref{eq:polarization-tensors-physical-limit} we will perform a similar analysis in
the context of the modified theory. For $(k^{\mu})=(|\mathbf{k}|,\mathbf{k})$ with
$\mathbf{k}=(k_{\bot},0,k_{\parallel})$ the Ward identity reads
\begin{equation}
\label{eq:Ward-identity}
k^{\mu}\mathcal{M}_{\mu}=k_0\mathcal{M}_0-k_1\mathcal{M}_1-k_3\mathcal{M}_3=|\mathbf{k}|\mathcal{M}_0+k_{\bot}\mathcal{M}_1+k_{\parallel}\mathcal{M}_3=0\,,
\end{equation}
and therefore, $\mathcal{M}_1$ can be expressed as follows:
\begin{subequations}
\begin{equation}
\label{eq:matrix-element-m1}
\mathcal{M}_1=-\frac{|\mathbf{k}|}{k_{\bot}}\mathcal{M}_0-\frac{k_{\parallel}}{k_{\bot}}\mathcal{M}_3\,,
\end{equation}
\begin{equation}
\label{eq:matrix-element-m1-square}
|\mathcal{M}_1|^2=\frac{|\mathbf{k}|^2}{k_{\bot}^2}|\mathcal{M}_0|^2+\frac{k_{\scriptscriptstyle{\parallel}}^2}{k_{\bot}^2}|\mathcal{M}_3|^2+\frac{2|\mathbf{k}|k_{\parallel}}{k_{\bot}^2}\mathrm{Re}(\mathcal{M}_0\mathcal{M}_3^{*})\,.
\end{equation}
\end{subequations}
Using the result of Eq. \eqref{eq:matrix-element-m1-square}, the contribution of the
matrix element squared involving the first polarization mode results in:
\begin{align}
|\varepsilon_{\mu}^{(1)}\mathcal{M}^{\mu}|^2\Big|^{\mathrm{phys}}_{\substack{\mathrm{mod} \\ \mathcal{E}\mapsto 0}}&=|\mathcal{M}_1|^2+|\mathcal{M}_2|^2-\frac{1}{k_{\bot}^2}\left\{|\mathbf{k}|^2|\mathcal{M}_0|^2+k_{\scriptscriptstyle{\parallel}}^2|\mathcal{M}_3|^2+2|\mathbf{k}|k_{\parallel}\mathrm{Re}(\mathcal{M}_0\mathcal{M}_3^{*})\right\}\!\Big|^{\mathrm{phys}} \notag \\
&=\frac{1}{k_{\bot}^2}\left\{|\mathbf{k}|^2|\mathcal{M}_0|^2+k_{\scriptscriptstyle{\parallel}}^2|\mathcal{M}_3|^2+2|\mathbf{k}|k_{\parallel}\mathrm{Re}(\mathcal{M}_0\mathcal{M}_3^{*})\right\}+|\mathcal{M}_2|^2 \notag \\
&\hspace{1cm}\,-\frac{1}{k_{\bot}^2}\left\{|\mathbf{k}|^2|\mathcal{M}_0|^2+k_{\scriptscriptstyle{\parallel}}^2|\mathcal{M}_3|^2+2|\mathbf{k}|k_{\parallel}\mathrm{Re}(\mathcal{M}_0\mathcal{M}_3^{*})\right\} \notag \\
&=|\mathcal{M}_2|^2\,,
\end{align}
where the Ward identity has been used in the second step.
Hence, restricting the ``matrix element squared'' to the physical subspace with the
Ward identity guarantees that the additional parts, that depend on the preferred
directions $\xi^{\mu}$ and $\zeta^{\mu}$, cancel.

Now consider the $\lambda=2$ polarization mode. With Eq. \eqref{eq:matrix-element-m1-square}
we obtain:
\begin{align}
|\varepsilon_{\mu}^{(2)}\mathcal{M}^{\mu}|^2\Big|^{\mathrm{phys}}_{\substack{\mathrm{mod} \\ \mathcal{E}\mapsto 0}}&=\frac{k_{\scriptscriptstyle{\parallel}}^2}{k_{\bot}^2}|\mathcal{M}_0|^2+\frac{|\mathbf{k}|^2}{k_{\bot}^2}|\mathcal{M}_3|^2+\frac{2|\mathbf{k}|k_{\parallel}}{k_{\bot}^2}\mathrm{Re}(\mathcal{M}_0\mathcal{M}_3^{*})\Big|^{\mathrm{phys}} \notag \\
&=\frac{k_{\scriptscriptstyle{\parallel}}^2}{k_{\bot}^2}|\mathcal{M}_0|^2+\frac{|\mathbf{k}|^2}{k_{\bot}^2}|\mathcal{M}_3|^2+\Bigl\{|\mathcal{M}_1|^2-\frac{|\mathbf{k}|^2}{k_{\bot}^2}|\mathcal{M}_0|^2-\frac{k_{\scriptscriptstyle{\parallel}}^2}{k_{\bot}^2}|\mathcal{M}_3|^2\Bigr\} \notag \\
&=\frac{k_{\scriptscriptstyle{\parallel}}^2-|\mathbf{k}|^2}{k_{\bot}^2}|\mathcal{M}_0|^2+|\mathcal{M}_1|^2+\frac{|\mathbf{k}|^2-k_{\scriptscriptstyle{\parallel}}^2}{k_{\bot}^2}|\mathcal{M}_3|^2 \notag \\
&=-|\mathcal{M}_0|^2+|\mathcal{M}_1|^2+|\mathcal{M}_3|^2\,.
\end{align}
Setting $k_{\bot}=0$ in Eq. \eqref{eq:Ward-identity} results in
$\mathcal{M}_0=-\mathrm{sgn}(k_{\parallel})\mathcal{M}_3$ and therefore
$|\mathcal{M}_0|^2=|\mathcal{M}_3|^2$. This then leads to
\begin{equation}
|\varepsilon_{\mu}^{(2)}\mathcal{M}^{\mu}|^2\Big|^{\mathrm{phys}}_{\substack{\mathrm{mod} \\ \mathcal{E}\mapsto 0}}=|\mathcal{M}_1|^2\,.
\end{equation}
Hence, we see that by using the Ward identity all contributions depending
on $\xi^{\mu}$ and $\zeta^{\mu}$ also vanish for the second mode. Therefore,
for vanishing Lorentz violation the standard result
\begin{equation}
|\mathcal{M}|^2\Big|^{\mathrm{phys}}_{\substack{\mathrm{mod} \\ \mathcal{E}\mapsto 0}}=\sum_{\lambda=1,2} |\varepsilon_{\mu}^{(\lambda)}\mathcal{M}^{\mu}|^2\Big|^{\mathrm{phys}}_{\substack{\mathrm{mod} \\ \mathcal{E}\mapsto 0}}
=|\mathcal{M}_1|^2+|\mathcal{M}_2|^2\,,
\end{equation}
is recovered.

\section{Discussion and conclusion}
\label{sec:Discussion-parity-odd}
\setcounter{equation}{0}

In this article, a special sector of a \textit{CPT}-even Lorentz-violating modification of QED,
with the characteristics of being parity-odd and nonbirefringent, was examined with respect
to consistency. The deformation of QED is described by one fixed timelike ``four-vector'',
one fixed spacelike ``four-vector'', and three Lorentz-violating parameters.

The nonbirefringent \textit{Ansatz} combined with the parity-violating parameter
choice leads to two distinct physical photon polarization modes. These modes are
characterized by dispersion relations, that differ to quadratic order in the Lorentz-violating
parameters. Hence, the theory is only nonbirefringent to linear order. The dispersion relations
coincide with the formulas previously obtained in Ref.~\cite{Casana-etal2009}. The new most important
results of this article are summarized in the subsequent items:
\begin{itemize}

\item With the optical theorem, unitarity is verified for tree-level processes involving
conserved currents.
\item Microcausality is established for the full range of Lorentz-violating parameters.
Information only propagates along the modified null cones.
\item It has turned out that \textit{covariant} polarization tensors can be constructed
for each photon mode. This is not possible in standard QED, where only the polarization tensor
of the sum of both modes can be written covariantly.
\item The gauge-invariant\footnote{with all terms dropped that involve one or more external photon four-momenta}
polarization tensor of each mode depends on the background field directions. For vanishing Lorentz
violation this dependence remains. It only cancels when considering the sum of both modes, which
leads to the polarization sum of standard QED.
\item The fact that the polarization tensors depend on the background field directions even
for vanishing Lorentz violation, makes us think about the question of whether the limit
of zero Lorentz violation is continuous. In other words, \textit{a priori} it is not clear, whether
or not the modified theory approaches standard QED for vanishing Lorentz violation. This is
the motivation to test the theory via brute force by calculating one special process: Compton
scattering for unpolarized electrons scattered by polarized photons.
\item The cross sections can be computed either by using the modified polarization vectors or
the modified polarization tensors. The upshot is that the results for $1\mapsto 1$, $1\mapsto~2$,
and $2\mapsto 1$ coincide, but a numerical treatment reveals a discrepancy for $2\mapsto 2$ scattering.
\footnote{Here, the numbers indicate the photon polarizations.} The Ward identity is shown to cure
the polarization vectors and tensors from their bad behavior for vanishing Lorentz violation,
at least for the first three processes.
However, if the matrix element squared is computed for the fourth process by using the modified
polarization vectors, there exists a phase space sector, for which the longitudinal part of the
second polarization vector
is proportional to a $\delta$-function. This could be shown by an analytic investigation.
It could also be proven analytically that the Ward identity can cancel this contribution,
nevertheless.

\end{itemize}
To conclude, the parity-odd ``nonbirefringent'' sector of modified Maxwell theory seems --- with
regard to the performed investigations --- to be consistent. Further steps in the context of
consistency of Lorentz-violating quantum field theories may involve the analysis of unitarity
at one-loop level, where the Lorentz-violating structure is treated in an exact way. Especially
for this parity-odd theory it would be interesting to know if its consistency is inherited to
higher orders of perturbation theory. However, this is beyond the scope of this article.

In light of the consistency of this Lorentz-violating extension at tree level, nature decides on
the values of the Lorentz-violating parameters. Therefore, they have to be measured with experiments.
For a summary of the current experimental status we refer to Ref.~\cite{Exirifard:2010xm} and references
therein. The latter article also gives new experimental bounds on the parity-odd parameters.

\newpage
\begin{appendix}
\numberwithin{equation}{section}

\section{Technical details concerning the calculation of the Compton cross sections}
\label{subsec:technical-details-calculation}

We compute the cross section in two different manners. The first possibility is
to follow Sec.~11.1 of Ref.~\cite{JauchRohrlich1976}, which gives the matrix
element squared for Compton scattering of polarized photons off unpolarized
electrons. In order to derive this equation, the authors use polarization
vectors. This is clear since in standard QED covariant polarization tensors
cannot be constructed from the polarization vectors. Hence, we perform a similar
calculation in the modified theory, where we can directly test our polarization
vectors given by Eqs. \eqref{eq:polarization-mode1-parity-violating}
and \eqref{eq:polarization-mode2-parity-violating}.

The second possibility is to compute the cross sections according to Eq.
(5.81) of Ref.~\cite{PeskinSchroeder1995}, where polarization tensors are used.
Note that here Compton scattering of \textit{unpolarized} photons is
considered, hence it is averaged over initial and summed over final photon
polarizations. Only under this condition can polarization tensors be used
in standard QED. However, for parity-odd ``nonbirefringent'' modified Maxwell
theory an analogous computation is also possible for Compton scattering with
polarized photons. Hence, we have to calculate
\begin{subequations}
\begin{align}
\widetilde{\sigma}_{1X}&=\frac{1}{4m\omega_1(\mathbf{k}_1)|\mathbf{v}_{\mathrm{gr},\,1}|}\sum_{\lambda'=1,2}\int \frac{\mathrm{d}^3k_2}{(2\pi)^22\omega_{\lambda'}(\mathbf{k}_2)2E_2}\,\delta\big(\omega_1(\mathbf{k}_1)+m-\omega_{\lambda'}(\mathbf{k}_2)-E_2\big) \notag \\
&\hspace{2.5cm}\,\times\big(|\mathcal{M}(k_1,k_2)|^2\big)^{\mu\nu\varrho\sigma}\big(\Pi_{\mu\varrho}(k_1)|_{\lambda=1}\big)\big(\Pi_{\nu\sigma}(k_2)|_{\lambda'}\big)\Big|_{\substack{k_1^0=\omega_1(\mathbf{k}_1) \\ k_2^0=\omega_{\lambda'}(\mathbf{k}_2)}}\,,
\end{align}
\begin{align}
\widetilde{\sigma}_{2X}&=\frac{1}{4m\omega_2(\mathbf{k}_1)|\mathbf{v}_{\mathrm{gr},\,2}|}\sum_{\lambda'=1,2}\int \frac{\mathrm{d}^3k_2}{(2\pi)^22\omega_{\lambda'}(\mathbf{k}_2)2E_2}\,\delta\big(\omega_2(\mathbf{k}_1)+m-\omega_{\lambda'}(\mathbf{k}_2)-E_2\big) \notag \\
&\hspace{2.5cm}\,\times\big(|\mathcal{M}(k_1,k_2)|^2\big)^{\mu\nu\varrho\sigma}\big(\Pi_{\mu\varrho}(k_1)|_{\lambda=2}\big)\big(\Pi_{\nu\sigma}(k_2)|_{\lambda'}\big)\Big|_{\substack{k_1^0=\omega_2(\mathbf{k}_1) \\ k_2^0=\omega_{\lambda'}(\mathbf{k}_2)}}\,,
\end{align}
\end{subequations}
where the energy of the final electron is denoted as $E_2$.
Note the division by the group velocity of the first and second polarization state (see also Ref.~\cite{Colladay:2001wk}),
respectively, where in the standard theory
$|\mathbf{v}_{\mathrm{gr},\,1}|=|\mathbf{v}_{\mathrm{gr},\,2}|=1$.
The tensor $(|\mathcal{M}(k_1,k_2)|^2\big)^{\mu\nu\varrho\sigma}$ is
given by the trace term of Eq. (5.81) of Ref.~\cite{PeskinSchroeder1995} with
some modifications due to the Lorentz-violating kinematics.
\begin{figure}[h]
\centering
\includegraphics{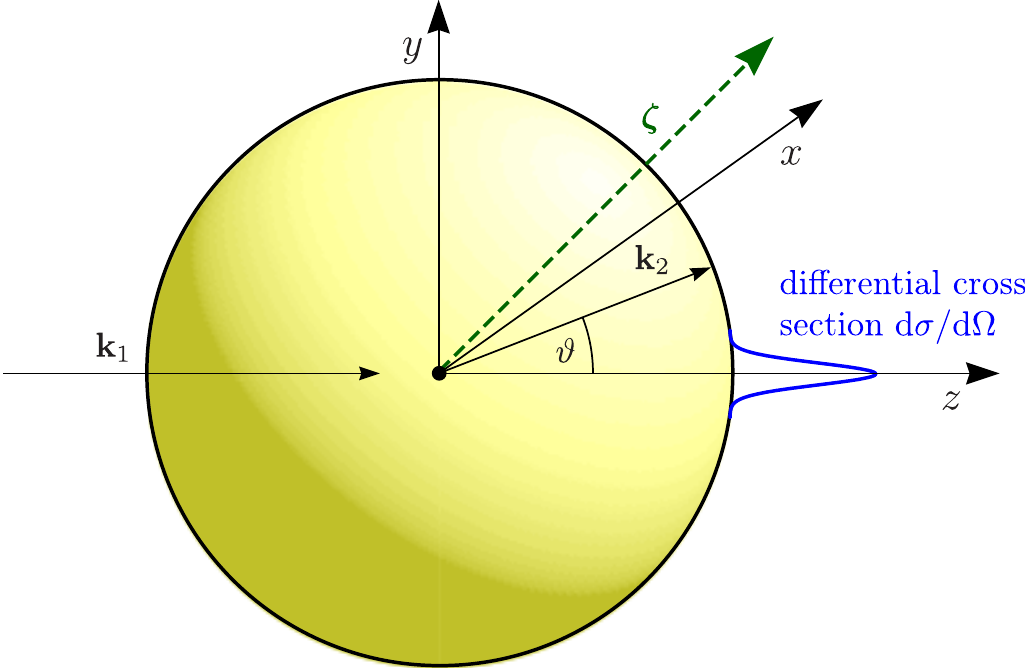}
\caption{Chosen coordinate system for the phase space integration, where the initial
photon three-momentum $\mathbf{k}_1$ lies along the third axis of the coordinate
system. For the outgoing photon momentum $\mathbf{k}_2$, spherical coordinates
are chosen with the azimuthal angle $\vartheta$ corresponding to the angle between
$\mathbf{k}_2$ and the third axis. Cases are treated with the three-vector
$\boldsymbol{\zeta}$ having equal or unequal components.}
\label{fig:phase-space-integration}
\end{figure}
The purely algebraic part of the calculation, that includes computation of traces,
contraction of indices and inserting kinematical relations, is performed with
\verb|Form| \cite{Vermaseren:2000nd}.
The subsequent phase space calculation is done numerically with \verb|C++|, since
the resulting matrix element squared contains hundreds of terms. The limit of zero
Lorentz violation has to be taken with care and ``long double'' precision does not
suffice here. Therefore, the GNU Multiple Precision Arithmetic Library \verb|GMP|
\cite{GMP:2011} is used with its \verb|C++| interface described in Sec. 12 of the
reference previously mentioned.

The first idea was to choose the coordinate system such that $\boldsymbol{\zeta}$ lies
along the third axis. Then the phase space should have been integrated with cylindrical
coordinates $(k_{2,\bot},\varphi,k_{2,\parallel})$. To cover a general situation, where
the initial photon momentum $\mathbf{k}_1$ points in an arbitrary direction, the cylindrical
axes would have to point in that direction as well. As a result of this, the coordinate
frame must be rotated in order to compute the cross section. This treatment has turned out
to be unsuitable. Therefore, a better approach is the following, which is sketched in Fig.
\ref{fig:phase-space-integration}. The phase space
integration is performed with spherical coordinates $(|\mathbf{k}_2|,\vartheta,\varphi)$,
where the initial photon momentum points along the third axis of the coordinate system.
The general case is mimicked by $\boldsymbol{\zeta}$ pointing in an arbitrary direction.
As a special --- but nevertheless very generic --- case we can choose its components to
be equal (however, computations were also done for different
cases as shown in Table \ref{tab:Compton-scattering-results3}):
\begin{equation}
\boldsymbol{\zeta}=\widetilde{\kappa}\begin{pmatrix}
1 \\
1 \\
1 \\
\end{pmatrix}\,,\quad \mathcal{E}=\sqrt{3}\,|\widetilde{\kappa}|\,.
\end{equation}
The integration over $|\mathbf{k}_2|$ is eliminated at once with the energy conservation
equation in the $\delta$-function. Here, we have to keep in mind that
\begin{align}
\delta&\big(\omega_{\lambda}(k_1)+m-\omega_{\lambda'}(k_2)-E_2\big)= \notag \\
&\hspace{2cm}\left|\frac{\partial\big(\omega_{\lambda}(k_1)+m-\omega_{\lambda'}(k_2)-E_2\big)}{\partial k_2}\right|^{-1}\delta\big(k_2-(k_2)^0\big)\,,\quad \lambda,\,\lambda'\in \{1,2\}\,,
\end{align}
where $(k_2)^0$ is the corresponding zero. The analytic solution $(k_2)^0$ is a complicated
function of $\mathcal{E}$ and $m$, so we determine it numerically with Newton's method
inside the \verb|C++| program. The integrations over $\vartheta$ and $\varphi$ are
performed with the Simpson rule, which is sufficient for our purpose. The integration domain,
that includes all physical states, is determined automatically with $(k_2)^0$. If no zero
$(k_2)^0$ exists, then the corresponding angles $\vartheta$ and $\varphi$ lie outside the
domain.

\section{Compton scattering and Thomson limit in standard (quantum) electrodynamics}
\label{subsec:thomson-scattering}

The low-energy limit of the Compton scattering cross section (Thomson limit)
can be calculated classically via the following equation (see e.g. Ref.~\cite{Jackson:1975}):
\begin{equation}
\left(\frac{\mathrm{d}\sigma}{\mathrm{d}\Omega}\right)^{\mathrm{Th}}_{\lambda\lambda'}=\frac{\alpha^2}{m^2}|\overline{\boldsymbol{\varepsilon}_{f,\lambda'}}\cdot \boldsymbol{\varepsilon}_{i,\lambda}|^2\,,
\end{equation}
where $\boldsymbol{\varepsilon}_{i,\lambda}$ is the polarization three-vector of the
incoming and $\boldsymbol{\varepsilon}_{f,\lambda'}$ that of the outgoing electromagnetic
wave. For the initial wave traveling along the $z$-axis we can choose the transverse
polarization vectors as
\begin{equation}
\boldsymbol{\varepsilon}_{i,1}\equiv\begin{pmatrix}
1 \\
0 \\
0 \\
\end{pmatrix}\,,\quad \boldsymbol{\varepsilon}_{i,2}\equiv\begin{pmatrix}
0 \\
1 \\
0 \\
\end{pmatrix}\,.
\end{equation}
In general, the propagation direction of the final wave can be described in spherical
coordinates by the basis vector
$\widehat{\mathbf{e}}_r=(\sin\vartheta\cos\varphi,\sin\vartheta\sin\varphi,\cos\vartheta)$.
Then we can pick the physical polarization vectors to point along the other two basis
vectors $\widehat{\mathbf{e}}_{\vartheta}$ and $\widehat{\mathbf{e}}_{\varphi}$:
\begin{equation}
\label{eq:polarization-vectors-standard-electrodynamics}
\boldsymbol{\varepsilon}_{f,1}\equiv\widehat{\mathbf{e}}_{\vartheta}=\begin{pmatrix}
\cos\vartheta\cos\varphi \\
\cos\vartheta\sin\varphi \\
-\sin\vartheta \\
\end{pmatrix}\,,\quad \boldsymbol{\varepsilon}_{f,2}\equiv\widehat{\mathbf{e}}_{\varphi}=\begin{pmatrix}
-\sin\varphi \\
\cos\varphi \\
0 \\
\end{pmatrix}\,.
\end{equation}
This leads to the polarized Thomson scattering cross sections in standard
electrodynamics:
\begin{equation}
\sigma_{11}^{\mathrm{Th}}=\sigma_{21}^{\mathrm{Th}}=\frac{2\pi\alpha^2}{3m^2}\,,\quad
\sigma_{12}^{\mathrm{Th}}=\sigma_{22}^{\mathrm{Th}}=\frac{2\pi\alpha^2}{m^2}\,.
\end{equation}
If we rotate, for example, the set of initial polarization vectors by angle $\alpha$
and the final ones by angle $\beta$ in their corresponding polarization planes, the
single contributions $\sigma_{\lambda\lambda'}^{\mathrm{Th}}$ will depend on $\beta$.
However, this dependence cancels in $\sigma_{1X}^{\mathrm{Th}}$ and $\sigma_{2X}^{\mathrm{Th}}$
that are defined as follows:
\begin{subequations}
\begin{equation}
\label{eq:result-thomson-scattering-individual-1}
\sigma_{1X}^{\mathrm{Th}}\equiv\frac{1}{2}(\sigma_{11}^{\mathrm{Th}}+\sigma_{12}^{\mathrm{Th}})=\frac{4\pi\alpha^2}{3m^2}\,,
\end{equation}
\begin{equation}
\label{eq:result-thomson-scattering-individual-2}
\sigma_{2X}^{\mathrm{Th}}\equiv\frac{1}{2}(\sigma_{21}^{\mathrm{Th}}+\sigma_{22}^{\mathrm{Th}})=\frac{4\pi\alpha^2}{3m^2}\,,
\end{equation}
\end{subequations}
From $\sigma_{1X}^{\mathrm{Th}}=\sigma_{2X}^{\mathrm{Th}}$ is clear that both initial modes
deliver equal contributions to the total Thomson result of Eq. \eqref{eq:result-thomson-scattering-total}.

This is also the case for MCS-theory. The MCS polarization tensors of Eq.
\eqref{eq:polarization-tensors-mcs-theory-limit} even give equal results
for each individual polarized scattering process in the limit
$m_{\mathrm{\scriptscriptstyle{CS}}}\mapsto 0$:
\begin{equation}
\sigma_{11}^{\mathrm{MCS,Th}}=\sigma_{12}^{\mathrm{MCS,Th}}=\sigma_{21}^{\mathrm{MCS,Th}}=\sigma_{22}^{\mathrm{MCS,Th}}=\frac{4\pi\alpha^2}{3m^2}\,,
\end{equation}
For parity-odd modified
``nonbirefringent'' modified Maxwell theory the individual contributions are not equal
for $\mathcal{E}\mapsto 0$. However, the above expressions from Eqs.
\eqref{eq:result-thomson-scattering-individual-1} and
\eqref{eq:result-thomson-scattering-individual-2} correspond to each
other.

With Eq. (11-13) of Ref.~\cite{JauchRohrlich1976} and the standard polarization vectors
from Eq. \eqref{eq:polarization-vectors-standard-electrodynamics} we obtain the
polarized Compton scattering values given in Table. \ref{tab:Compton-scattering-standard-qed-2}.

\end{appendix}

\section*{Acknowledgments}

It is a pleasure to thank F.R.~Klinkhamer for most helpful discussions. The
author appreciates M.~R\"{u}ckauer's (KIT) help with \verb|GMP|. Furthermore, the
author is indebted to M. Schwarz (KIT) and S. Thambyahpillai (KIT) for proofreading the
article and useful suggestions.

The author also thanks the anonymous referee for useful comments on the manuscript.
This work was supported by the German Research Foundation (\textit{Deutsche
Forschungsgemeinschaft, DFG}) within the Grant No. KL 1103/2-1.

\newpage



\begin{thebibliography}{99}

\bibitem{OPERA:2011zb}
T.~Adam {\it et al.} [OPERA Collaboration],
``Measurement of the neutrino velocity with the OPERA detector in the CNGS beam,''
arXiv:1109.4897 [hep-ex].

\bibitem{Klinkhamer:2011mf}
F.~R.~Klinkhamer,
``Superluminal muon-neutrino velocity from a Fermi-point-splitting model of Lorentz violation,''
arXiv:1109.5671 [hep-ph].

\bibitem{Klinkhamer:2011iz}
F. R. Klinkhamer and G. E. Volovik,
``Superluminal neutrino and spontaneous breaking of Lorentz invariance,''
Pisma Zh.\ Eksp.\ Teor.\ Fiz.\ {\bf 94}, 731 (2011)
[JETP Lett.\  {\bf 94} (2012) 673],
arXiv:1109.6624 [hep-ph].

\bibitem{Schreck:2011ni}
M.~Schreck,
``Multiple Lorentz groups --- a toy model for superluminal muon neutrinos,''
J. Mod. Phys. {\bf 3}, 1398 (2012),
arXiv:1111.7268 [hep-ph].

\bibitem{Contaldi:2011}
C.~R.~Contaldi,
``The OPERA neutrino velocity result and the synchronisation of clocks,''
arXiv:1109.6160 [hep-ph].

\bibitem{Besida:2011fi}
O.~Besida,
``Three errors in the article: 'The OPERA neutrino velocity result and the synchronisation of clocks',''
arXiv:1110.2909 [hep-ph].

\bibitem{vanElburg:2011ze}
R.~A.~J.~van Elburg,
``Measuring time of flight using satellite-based clocks,''
arXiv:1110.2685 [physics.gen-ph].

\bibitem{Wu:1957}
C.~S.~Wu, E.~Ambler, R.~W.~Hayward, D.~D.~Hoppes und R.~P.~Hudson,
``Experimental test of parity conservation in beta decay''
Phys.\ Rev.\ {\bf 105}, 1413 (1957).

\bibitem{Christenson:1964fg}
J.~H.~Christenson, J.~W.~Cronin, V.~L.~Fitch, R.~Turlay,
``Evidence for the $2\uppi$ decay of the $\mathrm{K}_2^{\phantom{2}0}$ meson,''
Phys.\ Rev.\ Lett.\ {\bf 13}, 138 (1964).

\bibitem{ChadhaNielsen1983}
S.~Chadha and H.~B.~Nielsen,
``Lorentz invariance as a low-energy phenomenon,''
Nucl. Phys. B {\bf 217}, 125 (1983).

\bibitem{Wheeler:1957mu}
J.~A.~Wheeler,
``On the nature of quantum geometrodynamics,''
Annals Phys.\ {\bf 2}, 604 (1957).

\bibitem{Hawking:1979zw}
S.~W.~Hawking,
``Space-time foam,''
Nucl.\ Phys.\ B {\bf 144}, 349 (1978).

\bibitem{ColladayKostelecky1998}
D.~Colladay and V.~A.~Kosteleck\'{y},
``Lorentz-violating extension of the standard model,''
Phys. Rev. D {\bf 58}, 116002 (1998),
arXiv:hep-ph/9809521.

\bibitem{Kostelecky:2009zp}
V.~A.~Kosteleck\'{y} and M.~Mewes,
``Electrodynamics with Lorentz-violating operators of arbitrary dimension,''
Phys.\ Rev.\ D {\bf 80}, 015020 (2009),
arXiv:0905.0031 [hep-ph].

\bibitem{Kostelecky:2011gq}
V.~A.~Kosteleck\'{y} and M.~Mewes,
``Neutrinos with Lorentz-violating operators of arbitrary dimension,''
Phys.\ Rev.\ D {\bf 85}, 096005 (2012),
arXiv:1112.6395 [hep-ph].

\bibitem{Cambiaso:2012vb}
M.~Cambiaso, R.~Lehnert and R.~Potting,
``Massive photons and Lorentz violation,''
Phys.\ Rev.\ D {\bf 85}, 085023 (2012),
arXiv:1201.3045 [hep-th].

\bibitem{JordanPauli1928}
P.~Jordan and W.~Pauli,
``Zur Quantenelektrodynamik ladungsfreier Felder,''
Z. Phys. 47, 151 (1928).

\bibitem{KosteleckyLehnert2000}
V.~A.~Kosteleck\'{y} and R.~Lehnert,
``Stability, causality, and Lorentz and \textit{CPT} violation,''
Phys. Rev. D {\bf 63}, 065008 (2001),
arXiv:hep-th/0012060.

\bibitem{AdamKlinkhamer2001}
C.~Adam and F.~R.~Klinkhamer,
``Causality and \textit{CPT} violation from an Abelian Chern--Simons-like term,''
Nucl. Phys. B {\bf 607}, 247 (2001),
arXiv:hep-ph/0101087.

\bibitem{Liberati:2001sd}
S.~Liberati, S.~Sonego and M.~Visser,
``Faster-than-c signals, special relativity, and causality,''
Annals Phys.\ {\bf 298}, 167 (2002),
arXiv:gr-qc/0107091.

\bibitem{Mavromatos:2009xg}
N.~E.~Mavromatos,
``High-energy gamma-ray astronomy and string theory,''
J.\ Phys.\ Conf.\ Ser.\ {\bf 174}, 012016 (2009),
arXiv:0903.0318 [astro-ph.HE].

\bibitem{Casana-etal2009}
R.~Casana, M.~M.~Ferreira, A.~R.~Gomes, and P.~R.~D.~Pinheiro,
``Gauge propagator and physical consistency of the \textit{CPT}-even part of the
  standard model extension,''
Phys. Rev. D {\bf 80}, 125040 (2009),
arXiv:0909.0544 [hep-th].

\bibitem{Casana-etal2010}
R.~Casana, M.~M.~Ferreira, A.~R.~Gomes, and F.~E.~P.~dos Santos,
``Feynman propagator for the nonbirefringent \textit{CPT}-even
electrodynamics of the standard model extension,''
Phys.\ Rev.\ D {\bf 82}, 125006 (2010),
arXiv:1010.2776 [hep-th].

\bibitem{Klinkhamer:2010zs}
F.~R.~Klinkhamer, M.~Schreck,
``Consistency of isotropic modified Maxwell theory: Microcausality and unitarity,''
Nucl.\ Phys.\ B {\bf 848}, 90 (2011),
arXiv:1011.4258 [hep-th].

\bibitem{Klinkhamer:2011ez}
F.~R.~Klinkhamer and M.~Schreck,
``Models for low-energy Lorentz violation in the photon sector: Addendum to `Consistency of isotropic modified Maxwell theory',''
Nucl.\ Phys.\ B {\bf 856}, 666 (2012),
arXiv:1110.4101 [hep-th].

\bibitem{Beall:1970rw}
E.~F.~Beall,
``Measuring the gravitational interaction of elementary particles,''
Phys.\ Rev.\ D {\bf 1}, 961 (1970).

\bibitem{Coleman:1997xq}
S.~R.~Coleman, S.~L.~Glashow,
``Cosmic ray and neutrino tests of special relativity,''
Phys.\ Lett.\ B {\bf 405}, 249 (1997),
hep-ph/9703240.

\bibitem{Phillips:2000dr}
D.~F.~Phillips, M.~A.~Humphrey, E.~M.~Mattison, R.~E.~Stoner, R.~F.~C.~Vessot, R.~L.~Walsworth,
``Limit on Lorentz and \textit{CPT} violation of the proton using a hydrogen maser,''
Phys.\ Rev.\ D {\bf 63}, 111101 (2001),
physics/0008230.

\bibitem{Bear:2000cd}
D.~Bear, R.~E.~Stoner, R.~L.~Walsworth, V.~A.~Kosteleck\'{y}, C.~D.~Lane,
``Limit on Lorentz and \textit{CPT} violation of the neutron using a two species noble gas maser,''
Phys.\ Rev.\ Lett.\ {\bf 85}, 5038 (2000),
physics/0007049.

\bibitem{Carroll-etal1990}
S.~M.~Carroll, G.~B.~Field, and R.~Jackiw,
``Limits on a {L}orentz- and parity-violating modification of electrodynamics,''
Phys. Rev. D {\bf 41}, 1231 (1990).

\bibitem{KosteleckyMewes2002}
V.~A.~Kosteleck\'{y} and M.~Mewes,
``Signals for Lorentz violation in electrodynamics,''
Phys. Rev. D {\bf 66}, 056005 (2002),
arXiv:hep-ph/0205211.

\bibitem{Kostelecky:2001mb}
V.~A.~Kosteleck\'{y}, M.~Mewes,
``Cosmological constraints on Lorentz violation in electrodynamics,''
Phys.\ Rev.\ Lett.\ {\bf 87}, 251304 (2001),
arXiv:hep-ph/0111026.

\bibitem{BaileyKostelecky2004}
Q.~G.~Bailey and V.~A.~Kosteleck\'{y},
``Lorentz-violating electrostatics and magnetostatics,''
Phys. Rev. D {\bf 70}, 076006 (2004),
arXiv:hep-ph/0407252.

\bibitem{Itin:2009aa}
Y.~Itin,
``On light propagation in premetric electrodynamics. Covariant dispersion relation,''
J.\ Phys.\ A {\bf 42}, 475402 (2009),
arXiv:0903.5520 [hep-th].

\bibitem{Hehl:2003}
F.~W.~Hehl, Y.~N.~Obukhov,
\emph{Foundations of Classical Electrodynamics: Charge, Flux, and Metric (Progress in Mathematical Physics)}, 1st ed.
(Birkh\"{a}user, Boston, 2003).

\bibitem{Gomes:2009ch}
M.~Gomes, J.~R.~Nascimento, A.~Y.~.Petrov and A.~J.~da Silva,
``On the aether-like Lorentz-breaking actions,''
Phys.\ Rev.\ D {\bf 81}, 045018 (2010),
arXiv:0911.3548 [hep-th].

\bibitem{KlinkhamerRisse2008b}
F.~R.~Klinkhamer and M.~Risse,
``Addendum: Ultrahigh-energy cosmic-ray bounds
on nonbirefringent modified Maxwell theory,''
Phys. Rev. D {\bf 77}, 117901 (2008),
arXiv:0806.4351 [hep-ph].

\bibitem{Kostelecky:2008ts}
V.~A.~Kosteleck\'{y}, N.~Russell,
``Data Tables for Lorentz and \textit{CPT} Violation,''
Rev.\ Mod.\ Phys.\ {\bf 83}, 11 (2011),
arXiv:0801.0287 [hep-ph].

\bibitem{Heitler1954}
W.~Heitler,
\textit{The Quantum Theory of Radiation}, 3rd ed.
(Oxford University Press, London, 1954).

\bibitem{JauchRohrlich1976}
J.~M.~Jauch and  F.~Rohrlich,
\textit{The Theory of Photons and Electrons}, 2nd ed.
(Springer, New York, USA, 1976).

\bibitem{Veltman1994}
M.~J.~G.~Veltman,
\textit{Diagrammatica: The path to Feynman rules}
(Cambridge University Press, Cambridge, England, 1994).

\bibitem{Brillouin1960}
L.~Brillouin,
\textit{Wave Propagation and Group Velocity}
(Academic, New York, USA, 1960).

\bibitem{ItzyksonZuber1980}
C.~Itzykson and J.~B.~Zuber, \textit{Quantum Field Theory}
(McGraw-Hill, New York, USA, 1980).

\bibitem{PeskinSchroeder1995}
M.~E.~Peskin and D.~V.~Schroeder, \textit{An Introduction to Quantum Field
Theory}
(Addison--Wesley, Reading, USA, 1995).

\bibitem{Kaufhold:2005vj}
C.~Kaufhold and F.~R.~Klinkhamer,
``Vacuum Cherenkov radiation and photon triple-splitting in a
Lorentz-noninvariant extension of quantum electrodynamics,''
Nucl.\ Phys.\ B {\bf 734}, 1 (2006),
arXiv:hep-th/0508074.

\bibitem{Fluegge1958}
\textit{Korpuskeln und Strahlung in Materie II. Corpuscles and Radiation in Matter II}, vol. 34,
in \emph{Handbuch Der Physik. Encyclopedia of Physics}, edited by S. Fl\"{u}gge (Springer, Berlin $\cdot$ G\"{o}ttingen $\cdot$ Heidelberg, 1958).

\bibitem{Bocquet:2010ke}
J.-P.~Bocquet, D.~Moricciani, V.~Bellini, M.~Beretta, L.~Casano, A.~D'Angelo, R.~Di Salvo, A.~Fantini {\it et al.},
``Limits on light-speed anisotropies from Compton scattering of high-energy electrons,''
Phys.\ Rev.\ Lett.\ {\bf 104}, 241601 (2010),
arXiv:1005.5230 [hep-ex].

\bibitem{Exirifard:2010xm}
Q.~Exirifard,
``Cosmological birefringent constraints on light,''
Phys.\ Lett.\ B {\bf 699}, 1 (2011),
arXiv:1010.2054 [gr-qc].

\bibitem{Colladay:2001wk}
D.~Colladay and V.~A.~Kosteleck\'{y},
``Cross sections and Lorentz violation,''
Phys.\ Lett.\ B {\bf 511}, 209 (2001),
hep-ph/0104300.

\bibitem{Vermaseren:2000nd}
J.~A.~M.~Vermaseren,
``New features of \verb|FORM|,''
arXiv:math-ph/0010025.

\bibitem{GMP:2011}
T.~Granlund and the \verb|GMP| development team,
``The GNU multiple precision arithmetic library, edition 5.0.2,''
\verb|http://gmplib.org|.

\bibitem{Jackson:1975}
J.~D.~Jackson,
\textit{Classical Electrodynamics}, 2nd ed.,
(John Wiley \& Sons Inc, New York, USA, 1975).

\end{thebibliography}
\end{document}